\documentclass[12pt, oneside, a4paper]{report}
\usepackage[thref, framed, amsthm, thmmarks]{ntheorem}
\usepackage{amsmath, amsfonts, amssymb, cancel}
\usepackage[T1]{fontenc}
\usepackage{lmodern}
\usepackage{framed}
\usepackage{graphicx}
\usepackage{grffile}
\usepackage{subfig}
\usepackage{makeidx}
\usepackage{fullpage}
\usepackage{tikz}
\usepackage{siunitx}
\usepackage[titletoc]{appendix}
\usepackage{booktabs}
\usepackage[nottoc]{tocbibind}
\usepackage{algorithm,algpseudocode}
\usepackage{longtable, multirow, array}
\usepackage[font=small]{caption}
\usepackage{fancyhdr}
\usepackage{cite}
\usepackage{CJKutf8}
\usepackage[overlap, CJK]{ruby}

\usepackage{afterpage}
\usepackage[hidelinks, bookmarksnumbered=true, pdftitle=Control of Continuous Quantum Systems with Many Degrees of Freedom based on Convergent Reinforcement Learning
, pdfauthor=Zhikang Wang]{hyperref}
\makeindex
\DeclareMathOperator*{\argmax}{argmax}
\DeclareMathOperator*{\argmin}{argmin}
\theoremstyle{definition}
\newframedtheorem{theorem}{Theorem}[section]

\begin{document}

	\begin{titlepage}
		\begin{center}
	\LARGE
	Doctoral Dissertation\\
	{\begin{CJK}{UTF8}{ipxm}
			博士論文
	\end{CJK}}\\

	\vspace*{1.7cm}
	\Large
	{Control of Continuous Quantum Systems with Many Degrees of Freedom based on Convergent Reinforcement Learning}\\[0.6cm]
	\large
	{\begin{CJK}{UTF8}{ipxm}
		(収束的な強化学習に基づく連続的な多自由度量子系の制御)
		\end{CJK}}\\
	\Large
	\vspace{2.1cm}
	A Dissertation Submitted for
	the Degree of Doctor of Science\\
	December, 2022\\[0.4cm]
	{\begin{CJK}{UTF8}{ipxm}
		令和4年12月博士（理学）申請
	\end{CJK}}\\

	\vfill
	\Large
    Department of Physics, Graduate School of Science,\\
    The University of Tokyo\\[0.3cm]
    {\begin{CJK}{UTF8}{ipxm}
    		東京大学大学院　理学系研究科　物理学専攻
    \end{CJK}}
	\vspace{1.5cm}
	
\LARGE

{\begin{CJK}{UTF8}{ipxm}
		\ruby{王}{ワン}　\ruby{智康}{ヅーカン}
\end{CJK}}\\[0.3cm]
Wang Zhikang\\
	\vspace{1.5cm}
\end{center}
	\end{titlepage}
	
	\chapter*{Abstract}\label{abstract}
	\addcontentsline{toc}{chapter}{\numberline{}Abstract}
	With advances in digital technology in recent years, parallel computation utilizing GPUs has achieved remarkable efficiency, which has made it possible to use large-scale machine learning algorithms with large datasets, and deep learning, which is machine learning utilizing deep neural networks trained with millions to billions of data and parameters, has become increasingly popular and found its use for various tasks including facial recognition, language translation and combinatorial problems, etc., achieving considerably many record-breaking results. As a powerful generic tool to solve problems, deep learning has been applied to physical problems as well, and has become an important alternative tool for scientists to solve a number of important problems. 

With the development of experimental quantum technology over the past several decades, quantum control has attracted increasing attention due to the realization of controllable artificial quantum systems, which have opened up a new regime of complex quantum systems that can be engineered. Quantum control has found its use in controlled quantum-chemical processes and artificial quantum systems including quantum dots, superconducting qubits, trapped ions, and cavity optomechanical systems, etc., which are of considerable importance for future technology as candidates for sensors and quantum-computational devices. However, because quantum-mechanical systems are often too difficult to analytically deal with, heuristic strategies and generic numerical algorithms which search for proper control protocols are adopted. Therefore, deep learning, especially deep reinforcement learning, is a promising generic candidate solution for the control problems. Although there have already been a few successful examples of applications of deep reinforcement learning to quantum control problems, most of the existing reinforcement learning algorithms intrinsically suffer from instabilities and unsatisfactory reproducibility, and as a consequence, they typically require a large amount of fine-tuning and a large computational budget, both of which limit their applicability to quantum control problems and require expertise in machine learning. 

To resolve the issue of instabilities of the reinforcement learning algorithms, in this thesis, we first investigate the non-convergence issue of Q-learning, which is one of the most efficient reinforcement learning strategies. Then, we investigate the weakness of existing convergent approaches that have been proposed, and we develop a new convergent Q-learning algorithm, which we call the convergent deep Q network (C-DQN) algorithm, as an alternative to the conventional deep Q network (DQN) algorithm. We prove the convergence of the C-DQN algorithm, and since the algorithm is scalable and computationally efficient, we apply it to a standard reinforcement learning benchmark, the \textit{Atari 2600} benchmark, to demonstrate its effectiveness. We show that when the DQN algorithm fail, the C-DQN algorithm still learns successfully. Then, we apply the algorithm to the measurement-feedback cooling problems of a quantum-mechanical quartic oscillator and a trapped quantum-mechanical rigid body. We establish the physical models and analyse their properties, and we show that although both the C-DQN algorithm and the DQN algorithm can learn to cool the systems, the C-DQN algorithm tends to behave more stably, and when the task is difficult and the DQN algorithm suffers from instabilities, the C-DQN algorithm can achieve a better performance. As the performance of the DQN algorithm can have a large variance and lack consistency from trial to trial, the C-DQN algorithm can be a better choice for researches on complicated physical control problems. The system of the trapped quantum-mechanical rigid body that we investigate is of theoretical interest and experimental relevance, and it has applications in sensing devices and fundamental physical research as well. We therefore also contribute to the study of the control of a quantum-mechanical rigid body, which is expected to be experimentally realized in near future in systems of trapped nanoparticles. We hope our research helps cool the system of a trapped rigid body to a quantum regime, and helps the development of the use of machine learning technologies and the development of better control of the microscopic quantum world in general.

	\chapter*{Acknowledgements}\label{acknowledgements}
	\addcontentsline{toc}{chapter}{\numberline{}Acknowledgements}
	I gratefully thank Yuto Ashida for bringing my attention to the topic of the quantum-mechanical rigid body. I am especially grateful to professor Masahito Ueda for his persistent patience and the kind discussions and help on my research, my writing, and my personal issues. Without their help the research would not be possible. I would also like to thank for the financial support of the Global Science Graduate Course (GSGC) program at the University of Tokyo and the support of the computational resources of the Institute for Physics of Intelligence at the University of Tokyo, both of which the project has relied on. I am grateful to the altruistic online communities including the Stack Exchange community and the Pytorch community for their valuable documents, comments, discussions and technical helps, which have taught me how to develop the computational program. I would also like to thank Ziyin Liu for discussion on deep learning and programming, and I thank professor Mio Murao for discussions and her valuable advice. 

I wish to express my gratitude to my parents, my friends, and my church, who have supported and encouraged me through the period of COVID though we could not meet each other in person. I am thankful for their trust, which has encouraged me to complete the work.
	\afterpage{\null\newpage}
	\tableofcontents
	
	\chapter{Introduction}\label{Introduction}
	\pagenumbering{arabic}
	\section{Background}
\subsection{Control in Quantum Mechanics}
Over the past several decades, quantum control has attracted increasing attention due to rapid developments of experimental techniques for quantum mechanical systems \cite{quantumControlSurvey,quantumControlScience,ColdAtomQuantumControl}, which have made it possible to directly measure and manipulate the quantum states. With the increased controllability of quantum systems, quantum control has found its use in controlled quantum-chemical processes \cite{quantumChemistryControl1, quantumChemistryControl2, quantumChemistryControl3}, engineered quantum dots, superconducting qubits, trapped ions, cavity optomechanical systems and many more \cite{QuantumDots,QuantumSpinDots,QuantumSimulation,TrappedIon,NVCenter,Superconducting,OpticalQuantumComputation,CavityOptomechanics} for quantum simulation, information processing and sensing. In general, one wishes to achieve a certain goal, or to maximize or minimize some prescribed target, in the process of control. For example, one usually wishes to minimize decoherence and to maximize purity of the quantum state for a quantum computing device, and one may wish to reduce the temperature and minimize the energy for a sensing device so that noise is reduced. In many cases, one considers the maximization of the fidelity of the state with respect to a target state after the process of control, typically starting from a given initial state, and the maximization is done with respect to the control variables, or the control sequence or control pulses, typically external fields in experiments, which affects the system Hamiltonian and affects the time evolution of the state, so that the state can be driven towards the target state. 

Formally, a control problem that is deterministic is to find the trajectory of the optimal control variable $u^*$ \cite{LinearQuadraticControl}:
\begin{equation}\label{control}
	u^*=\argmin_u \left[\left(\int_{0}^{T} L\left (u,x,t\right )\, dt\right)  + L_T\left (u_T,x_T\right )\right],\quad dx=f(x,u)dt,
\end{equation}
where $u$ represents the control variable, $x$ is the controlled state, both of which change in time, and $f$ describes the time evolution of the state $x$. The control loss that is to be minimized is in general given by the time integral of a loss function $L$ that depends on the state $x$ and the control variable $u$ during the control, plus a loss function $L_T$ that depends on the state $x_T$ at the end of the control and also the control variable $u_T$. The control process starts at time $0$ and terminates at time $T$, and the minimization in Eq.~(\ref{control}) is done with respect to the control variable, the function $u$, usually with a given initial state $x(0)=x_0$.

Control problems include a wide range of cases. If the control variable $u$ only depends on time and does not explicitly depend on the state $x$, the control is called \textit{open-loop} control. For example, given an initial state $x_0$, in Eq.~(\ref{control}), the time evolution is deterministic and both the state and the control variable can be regarded as functions of time only, and it is not needed to measure the state $x$ in order to carry out the control $u$ during the control process. However, in cases where randomness, such as noise, is present, a measurement-feedback control scheme is often preferred and the control variable $u$ should depend on the outcome of the measurement of the state and therefore should depend on $x$. Such control is called \textit{closed-loop} control, because one decides the control variable $u$ based on observation of the controlled state $x$, while in open-loop control it is not needed to observe the state.

For open-loop control problems, given a fixed initial state, one is allowed to simulate and repeat the time evolution of $x$ with different $u$, so that the optimal, or at least a locally optimal, control protocol $u(t)$ can be found. In quantum control literature, such kind of method includes QOCT \cite{QuantumOptimalControlTheory}, CRAB \cite{CRAB} and GRAPE \cite{GRAPE}, which are numerical algorithms based on gradients to maximize the fidelity of the state at the end of the control process. However, these algorithms suffer from the presence of local optima and sometimes humans should come up with better strategies to overcome this problem \cite{humanOutperformNumericalMethods,geneticQuantumControl}.

Generally speaking, compared to classical control problems, there are far fewer results known for quantum control. A classical system usually only requires several real numbers in order to fully describe the state; however, a quantum mechanical state can require a large number of complex numbers to describe due to its large Hilbert space, and the dynamics is also much more complicated. These issues make the analysis of quantum control considerably more complicated than its classical counterpart, and rigorous results for quantum control problems are few, except for several simple cases \cite{spinOptimal, HarmonicOscillatorControl, refereeCitation}. Most quantum control strategies are therefore designed on a case-by-case basis, based on approximations and assumptions specific to the systems involved, or relying on special properties of the systems \cite{superconductingQubitAnharmonicityPulseControl, shortcutsToAdiabaticity, SpinEcho, DynamicDecoupling}. Especially, most of these strategies can only deal with open-loop control. For closed-loop control with stochastic measurement outcomes, so far there is no generally applicable known approach. It is therefore desirable to explore generic methods to deal with quantum control problems, and this is why machine learning, as a generic solution to various problems, has become a promising candidate to solve the control problems.
\subsection{Deep Learning}
With advances in digital technology in recent years, especially with the development of GPU, parallel computation has become much more efficient, and the use of large-scale machine learning algorithms with large datasets has become available. Since the well-known breakthrough in 2012 on image classification problems using multi-layer neural networks, a.k.a.~deep neural networks \cite{ImageNetClassification}, \textit{deep learning}, which is simply machine learning utilizing deep neural networks, has attracted much interest and become increasingly popular and found its use for various tasks including facial recognition, speech recognition, language translation, gaming and solving combinatorial problems, and image generation based on textual description \cite{facialRecognition,ZeroResourceSpeech, transformer, RLDota2,Starcraft2, chess-like, diffusionNet}. Deep learning has achieved considerably many record-breaking results, and the artificial intelligence (AI) can often outperform humans. However, deep learning is typically not explainable and it works as a black box. The number of learned parameters in a deep neural network often ranges from millions to billions, and the field of deep learning is rapidly developing. 

Machine learning equipped with deep neural networks, which serves as a powerful generic tool to solve various problems, has also been applied to physical problems including predictions and classification problems in astrophysics, condensed matter physics and high-energy physics \cite{neutronStar, phaseTransition, tauDecay, quarticTomographyNN}, and design of materials \cite{materialDesign}, and design of open-loop control protocols \cite{GateOptimization, SpinControlDeepLearning, alphaZeroSearchControl, ErrorCorrectionDeepLearning}. The advanced machine learning technique has become an important alternative tool for scientists to solve problems. In general, machine learning makes use of a large number of data that are either collected or simulated, and it produces a model to make predictions or decisions based on the data that are learned. Further details and the formalism of deep learning are addressed in Section \ref{deep learning subsection}. 
\section{Outstanding Issues}
In this thesis, we focus on the application of deep learning, or deep reinforcement learning, to quantum control problems. Despite the success of deep learning in recent years, there are many outstanding issues concerning the application of deep reinforcement learning to quantum physics. 

First, most of the current deep reinforcement learning algorithms suffer from significant instability and randomness in the training process, and they often require a large amount of manual fine-tuning of the training hyperparameters so that the learning proceeds satisfactorily. The instability in training results in a lot of subtleties and makes it difficult for one to reproduce published results \cite{deepReinforcementLearningReproducibility}, and therefore, the instability undermines the reliability and consistency of results when deep reinforcement learning is applied to scientific problems. The final performance of the AI in the learning algorithm can vary significantly from trial to trial (for example, see Fig.~\ref{quartic cooling DQN CDQN} and Ref.~\cite{deepReinforcementLearningDoesntWorkYet}), which makes it difficult for one to fairly evaluate the performance and to improve or debug the learning algorithm. Therefore, in this thesis, we aim to solve this problem by investigating the issue in detail and by newly developing a convergent reinforcement learning algorithm based on the Q-learning reinforcement learning algorithm, so that the instability can be removed. We test our convergent algorithm on standard deep reinforcement learning benchmarks and apply the algorithm to quantum control problems to demonstrate its advantage for physical problems. 

Secondly, regarding quantum control, most existing applications of deep reinforcement learning to quantum control are only concerned with open-loop control, and are concerned with discrete systems such as spins and qubits \cite{GateOptimization, SpinControlDeepLearning, ErrorCorrectionDeepLearning}. However, quantum systems with continuous variables can be more complicated, and they can be of significant importance in many experimental scenarios. In this thesis, we focus on quantum control of continuous-space systems, and we focus on measurement-feedback closed-loop control, in which case the AI tracks the quantum state to make decision on how to control. We consider the case where information of the quantum state is obtained via continuous measurement, and the AI takes statistical information of the state as its input and it outputs the control variable. The physical system that we consider is a quantum-mechanical rigid body, which is a nonlinear multi-dimensional system that is of practical importance and can be realized using a trapped nanorotor \cite{quantumRotorNature}, and we consider stabilization and cooling of the rigid body.

Lastly, modern developments of deep learning always come together with field-specific techniques and modifications of the deep learning algorithm, such as convolutional neural network structures and data augmentation techniques in image classification \cite{residualNet, cutout}, and complicated transformer network structures in language processing models \cite{transformer}. These field-specific techniques are indispensable in order to achieve performance that is comparable to the state-of-the-art; without the techniques, the performance would be significantly qualitatively worse. The field-specific techniques make use of essential properties in the problems, such as symmetries, to help the deep neural networks to focus on essential points and to better deal with the tasks. However, concerning physical applications, we know little about how modifications of the deep learning algorithm could be made in order to improve the performance. There is much room for adjustment and design when one applies deep learning to physical problems. We will also discuss this issue for our case when we apply deep reinforcement learning in Chapter \ref{control rigid body}.
\section{Outline}
The rest of the thesis is organised as follows.

In Chapter \ref{continuous quantum measurement}, we review the general theory of measurement in quantum mechanics, and by taking the continuous limit, we derive the stochastic time-evolution equation of a quantum state that is subjected to continuous measurement. A specific realization of the stochastic time evolution is known as a \textit{quantum trajectory} of the state. The equations derived in this chapter are used in Chapter \ref{control rigid body} to model the relevant continuous measurements and used in the numerical experiments.

In Chapter \ref{deep reinforcement learning}, we review deep reinforcement learning, especially deep Q-learning. We introduce the framework of reinforcement learning, and we introduce how deep neural networks are used in combination with reinforcement learning algorithms, leading to deep reinforcement learning algorithms. We focus on one of the most efficient reinforcement learning algorithms, Q-learning, and introduce the deep Q network (DQN) algorithm.

In Chapter \ref{convergent DQN}, we first address the non-convergence issue in Q-learning with function approximation, and we briefly summarise some existing works and results that attempt to solve the non-convergence issue. Then, we point out and investigate several serious weaknesses of the conventional approach to the non-convergence problem, and we demonstrate using simple tabular tasks. The weaknesses of the conventional approach can prevent the AI from learning effectively, and therefore, we do not follow the conventional approach and we propose a new reinforcement learning algorithm instead, and we prove that our proposed algorithm is convergent by design. We test our algorithm, convergent DQN (C-DQN) algorithm, on a standard deep reinforcement learning benchmark \textit{Atari 2600} \cite{AtariEnvironment}, and we show that the algorithm is indeed convergent, especially in cases where the conventional DQN algorithm is unstable and diverges. The convergence property of our algorithm allows one to use large discount factors and long-term horizons, and as a result, C-DQN outperforms DQN on several difficult tasks where long-term planning is needed, in which case DQN diverges in our numerical experiments. This chapter is based on the following publication \cite{CDQN}: 
\begin{itemize}
	\item Zhikang T. Wang and Masahito Ueda. Convergent and efficient deep Q network algorithm, In \textit{International Conference on Learning Representations}, 2022.
\end{itemize}

In Chapter \ref{control rigid body}, we consider measurement-feedback cooling problems for a quantum quartic oscillator and a quantum rigid body. We first apply the C-DQN algorithm to measurement-feedback cooling of a quantum quartic oscillator and compare the results with those of the DQN algorithm. We show that although the DQN algorithm does not diverge on this task, it suffers from instabilities in the process of learning, and it has a large variance in its final performance. In contrast, the C-DQN algorithm learns stably and quickly approaches the highest performance achieved by the DQN algorithm, and the C-DQN algorithm has almost no variance in its final performance, showing the stability and the consistency of the results of the C-DQN algorithm. Then, we apply the DQN and the C-DQN algorithms to the problem of measurement-feedback cooling of a trapped quantum-mechanical rigid body. We first introduce the background and the motivation for controlling a quantum-mechanical rigid body, and we follow existing works to derive the general form of the rotational Hamiltonian of a rigid body. Then, we introduce our model of the trapped quantum-mechanical nanorod, and we derive its Hamiltonian and time-evolution equation in terms of the physical coordinates that are straightforwardly measurable, and we discuss the properties and the assumptions involved. Lastly, we compare the results of the C-DQN algorithm and the DQN algorithm on the measurement-feedback cooling problem of the system, and we compare them with the standard LQG control strategy. We show that both the DQN and the C-DQN algorithms can successfully learn to cool the system and have performances comparable to that of the LQG control. However, when the energy of the system is high and the dynamics is sufficiently nonlinear, the performance of the LQG control is lower than those of the DQN and the C-DQN algorithms, and we find that the C-DQN algorithm tends to have a more stable control strategy compared with DQN. This chapter is based on the following publication \cite{quantumCartpole} and unpublished original work:
\begin{itemize}
	\item Zhikang T. Wang, Yuto Ashida and Masahito Ueda. Deep reinforcement learning control of quantum cartpoles. \textit{Physical Review Letters}, 125(10):100401, 2020.
\end{itemize}

In Chapter \ref{conclusion}, we summarise the results and discuss future perspectives.

	\chapter{Continuous Measurement on Quantum Systems}\label{continuous quantum measurement}
	In this chapter, we review measurement in quantum mechanics and the continuous limit of the measurement. We review a general measurement model in Section \ref{measurement model}, and in Section \ref{continuous measurement} we discuss the continuous measurement, and in Section \ref{COM measurement} we discuss a quantum measurement of the center of mass of multiple particles.

\section{Model of Measurement}\label{measurement model}
A quantum measurement is described by a set $\{M_m\}$ of linear operators $M_m$, where $m$ corresponds to a specific outcome of the measurement, and the postmeasurement state after obtaining an outcome $m$ is given by
\begin{equation}
	|\psi'\rangle=\dfrac{M_m|\psi\rangle}{\sqrt{\langle\psi|M_m^{\dagger}M_m|\psi\rangle}},
\end{equation}
where $|\psi\rangle$ is the initial state and $\langle\psi|M_m^{\dagger}M_m|\psi\rangle$ is the probability of obtaining an outcome $m$ \cite{NielsenChuang}. The set of the measurement operators $\{M_m\}$ is required to satisfy the completeness condition
\begin{equation}\label{completeness}
	\sum_{m}M^{\dagger}_{m}M_{m}=I,
\end{equation}
where $I$ is the identity operator, so that the probabilities of different measurement outcomes adds up to 1, i.e.~$\sum_{m}\langle\psi|M_m^{\dagger}M_m|\psi\rangle=1$, and $M_m^{\dagger}M_m$ must have non-negative eigenvalues. For a general mixed state described by the density matrix $\rho$, the postmeasurement state and the probability of an outcome are given by 
\begin{equation}
	\rho'=\dfrac{M_m\rho M^{\dagger}_m}{\text{tr}\left(M_{m} \rho M^{\dagger}_{m}\right)},
\end{equation}
and
\begin{equation}
	\text{tr}\left(M_{m} \rho M^{\dagger}_{m}\right).
\end{equation}

The simplest measurement is the projection measurement, described by a set of projection operators $\{P_m\}$, satisfying 
\begin{equation}
	P_m^{2} = P^{\dagger}_m = P_m,\quad  \sum_m P_m=I,\quad P_m P_n = \delta_{mn}P_m, 
\end{equation}
with 
\begin{equation}
	\delta_{mn}=\left\{
\begin{array}{rl}
	0 & \text{if } m \ne n;\\
	1 & \text{if } m = n.
\end{array} \right.
\end{equation}
The projection operator $P_m$ projects the state onto the space that corresponds to an outcome $m$, which is orthogonal to the space of other measurement outcomes, and therefore, repeated measurements always yield the same result. In the following, we show that by introducing a meter state $\rho_e$ that interacts with the state of the system $\rho$, we can perform a general measurement $\{M_m\}$ on $\rho$ by performing a projection measurement $\{P_m\}$ on $\rho_e$, and this strategy is referred to as the indirect measurement scheme.

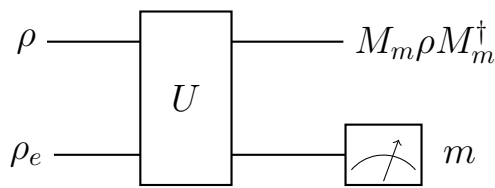
\begin{figure}[t]
	\centering
\begin{tikzpicture}
\draw[thick] (0,0) rectangle (1.2,2.3);
\node (U) at (0.6,1.15) {{\large $U$}};
\node (e) at (-1.5,0.4) {{\large $\rho_e$}};
\node (rho) at (-1.5,1.9) {{\large $\rho$}};
\draw[thick] (e) -- (0,0.4);
\draw[thick] (rho) -- (0,1.9);
\draw[thick] (2.7,0) rectangle (3.7,0.8);
\draw[thin] (3.62,0.2) arc (40:140:0.55);
\draw[thin,->] (3.2,0.05) -- (3.4,0.6);
\draw[thick] (1.2,0.4) -- (2.7,0.4);
\node (m) at (4.2,0.4) {{\large $m$}};
\node (rhom) at (3.75,1.9) {{\large $M_m\rho M^{\dagger}_m$}};
\draw[thick] (1.2,1.9) -- (rhom);
\end{tikzpicture}
\caption{Indirect measurement scheme. We let a meter interact with the measured state, and then, we measure the meter to obtain a measurement outcome $m$.}
\label{fig:indirect measurement}
\end{figure}
Given a state $\rho$ that we wish to measure, we prepare a meter state $\rho_e = |\psi_e\rangle\langle\psi_e|$ for which the initial state is known and pure. Then, we let $\rho$ and $\rho_e$ interact via a unitary evolution $U$, and we measure the state of the meter using projection measurement, as schematically illustrated in figure \ref{fig:indirect measurement}. When we obtain a measurement outcome $m$ on the meter state, the unnormalized postmeasurement state for the premeasurement state $\rho$ is given by
\begin{equation}\label{indirect measurement1}
\tilde{\rho}_m=\text{tr}_e \left((I\otimes P_m)U(\rho \otimes \rho_e)U^{\dagger}(I\otimes P_m)\right)=\text{tr}_e \left((I\otimes P_m)U(\rho \otimes |\psi_e\rangle\langle\psi_e|)U^{\dagger}\right),
\end{equation}
where $\text{tr}_e(\cdot)$ denotes the trace over the meter. Provided that the rank of the projection operator $P_m$ is one, i.e.~$P_m=|\psi_m\rangle\langle\psi_m|$, the above result becomes
\begin{equation}\label{indirect measurement2}
\tilde{\rho}_m=\langle\psi_m|U\left(|\psi_e\rangle\langle\psi_e| \otimes \rho\right) U^{\dagger}|\psi_m\rangle\,.
\end{equation}
By defining $M_{m}\equiv\langle\psi_m|U|\psi_e\rangle$, we obtain
\begin{equation}\label{indirect measurement3}
\tilde{\rho}_m=M_{m}\,\rho M^{\dagger}_{m}\,,
\end{equation}
and the normalized postmeasurement state $\rho_m$ is given by \begin{equation}\label{indirect measurement result}
	\rho_m=\dfrac{M_m\rho M^{\dagger}_m}{\text{tr}\left(M_{m} \rho M^{\dagger}_{m}\right)},
\end{equation}
which is equivalent to the case where a measurement operator $M_{m}$ is applied to the measured state $\rho$. The completeness condition Eq.~(\ref{completeness}) can be derived from the completeness of the projection operators $\{P_m\}$, i.e.
\begin{equation}
	\sum_m M_{m}^{\dagger}M_{m} = \sum_m \langle\psi_e|U^{\dagger}|\psi_m\rangle\langle\psi_m|U|\psi_e\rangle = \langle\psi_e|U^{\dagger}U|\psi_e\rangle=1.
\end{equation}
Conversely, given a set of measurement operators $\{M_{m}\}$ satisfying the completeness condition, one can construct a unitary matrix $U$ so that $\langle\psi_m|U|\psi_e\rangle=M_{m}$ is satisfied, which allows us to perform the measurement $\{M_{m}\}$ using the indirect measurement scheme \cite{NielsenChuang}. This result shows that general measurements can be realized by using projection measurements and the indirect measurement scheme.

\section{Continuous Measurement}\label{continuous measurement}
Continuous measurement, or monitoring, is modelled as a weak quantum measurement repeatedly performed every infinitesimal time interval $dt$, and it can be realized in several different ways. 
\subsection{Unconditional Evolution}
When we do not take the outcomes of the measurements into account, the state is a statistical ensemble of different postmeasurement states and it is generally a mixed state, and its evolution is deterministic and does not depends on the measurement outcomes. We denote the evolution, or the mapping, by $\rho\to\mathcal{E}(\rho)$, with $\mathcal{E}(\rho)=\sum_{m}M_m\rho M^{\dagger}_m$. We require this evolution to be continuous in time, i.e.
\begin{equation}\label{continuous}
0<\left\|\lim_{dt\to 0}\dfrac{\mathcal{E}(\rho)-\rho}{dt}\right\|<\infty,
\end{equation}
where $\mathcal{E}$ and $M_m$ depends on $dt$.

For simplicity, we first consider a binary measurement $\{M_0,M_1\}$ with two measurement outcomes $0$ and $1$. To meet the requirements $M^{\dagger}_{0}M_{0}+M^{\dagger}_{1}M_{1}=I$ and $\left(\mathcal{E}(\rho)-\rho\right)\sim dt$, we set $M_0=I-\frac{R\,dt}{2}$, where $R$ is a Hermitian operator, so that
\begin{equation}\label{binary default}
M_0 \rho M^{\dagger}_0=(I-R\,dt)\rho + O(dt^{2}),
\end{equation}
and $M_1$ should satisfy
\begin{equation}\label{binary non-default}
M_1 = L_1\,\sqrt{dt},\quad L^{\dagger}_1L_1=R.
\end{equation}
We therefore have $M_1^{\dagger}M_1 = R\,dt$ and $M^{\dagger}_{0}M_{0}+M^{\dagger}_{1}M_{1}=I$ \cite{QuantumMeasurement}. Similarly, it is possible to introduce more operators $M_m$ into the set of measurement operators $\{M_m\}$. Then, they should satisfy
\begin{equation}\label{multiple measurement outcomes}
M_i=L_i\,\sqrt{dt},\quad i=1,2,\cdots,m\ ,\quad M_0=I-\frac{dt}{2}\sum^{m}_{i=1}L^{\dagger}_iL_i,
\end{equation}
which leads to the Lindblad equation that describes the time evolution of the state $\rho$: 
\begin{equation}\label{Lindblad}
\frac{d\rho}{dt}=-\frac{i}{\hbar}[\hat{H},\rho]+\sum_{i}\gamma_i\left(L_i\rho L^{\dagger}_i-\frac{1}{2}\{L^{\dagger}_iL_i,\rho\}\right),
\end{equation}\\
where a self-evolution term $[\hat{H},\rho]$ has been added to the time evolution equation with a Hamiltonian $\hat{H}$, and $\{\cdot,\cdot\}$ is the anticommutator satisfying $\{A,B\}\equiv AB+BA$, and $\gamma_i$ is a factor describing the measurement strength. This equation can also be derived from the indirect measurement scheme, considering repeated projection measurements on a meter which continuously interacts with the measured system \cite{OpenQuantumSystemsAngelRivas}.

\subsection{Quantum Trajectory with Jumps}\label{Quantum Trajectory section}
When the measurement outcomes are observed, the time evolution of the quantum state conditioned on the measurement outcomes follows a quantum trajectory. This quantum trajectory is a stochastic process and is not necessarily continuous.

The probability of obtaining the measurement outcome $m$ in an infinitesimal time interval $dt$ is given by
\begin{equation}\label{measurement outcome 1}
p_m(dt)=\text{tr}(L_m\rho L^{\dagger}_m)\,dt\,,
\end{equation}
which is proportional to $dt$. Therefore, in an infinitesimal time interval, measurement outcomes other than the outcome $0$ ($M_0=I-\frac{dt}{2}\sum^{m}_{i=1}L^{\dagger}_iL_i$) only appear with vanishingly small probabilities. We therefore take the limit that measurement outcomes other than $0$ are sparse in time and two or more of them do not occur in the same time interval $dt$. Denoting the number of obtaining measurement outcome $i$ in $dt$ as $dN_i(t)$, the random value $dN_i(t)$ obeys the following relations:
\begin{equation}\label{measurement events}
dN_i(t)= 0\ \text{or}\ 1,\quad dN_idN_j=\delta_{ij}\,dN_i,\quad E\left[dN_i(t)\right]=\text{tr}\left(L_i\rho(t)L^{\dagger}_i\right)dt\, .
\end{equation}
We say that a quantum jump occurs when $dN_i(t)$ is equal to $1$, since the state jumps and changes suddenly due to the measurement backaction.

Using the random variables $\{dN_i(t)\}$, we can write down the stochastic differential equation to describe the stochastic time evolution of the system. For a pure state $|\psi\rangle$, the time evolution equation is given by
\begin{equation}\label{jump evolution pure}
\begin{split}
d|\psi(t)\rangle&=\left[\sum_i dN_i(t)\left(\frac{L_i}{\sqrt{\langle L^{\dagger}_i L_i\rangle}}-1\right)+[1-\sum_i dN_i(t)]\left(\frac{M_0}{\sqrt{\langle M_0^{\dagger}M_0\rangle}}-1\right)\right]|\psi(t)\rangle\\
&=\left[\sum_i dN_i(t)\left(\frac{L_i}{\sqrt{\langle L^{\dagger}_i L_i\rangle}}-1\right)+1\cdot\left(\frac{1-\frac{dt}{2}\sum^{m}_{i=1}L^{\dagger}_iL_i}{\sqrt{\langle 1-dt\sum^{m}_{i=1}L^{\dagger}_iL_i\rangle}}-1\right)\right]|\psi(t)\rangle\\
&=\left[\sum_i dN_i(t)\left(\frac{L_i}{\sqrt{\langle L^{\dagger}_i L_i\rangle}}-1\right)-dt\left(\frac{1}{2}\sum^{m}_{i=1}L^{\dagger}_iL_i-\frac{1}{2} \sum^{m}_{i=1}\langle L^{\dagger}_iL_i\rangle\right)\right]|\psi(t)\rangle,
\end{split}
\end{equation}
where $\langle\cdot\rangle$ denotes the expectation value. In the above equation, we have expanded the terms up to the lowest order in $dt$, and have replaced the term $[1-\sum_i dN_i(t)]$ by $1$ because non-zero events of $dN_i$ are vanishingly rare. If we take the Hamiltonian $\hat{H}$ of the system into account, the time evolution equation becomes
\begin{equation}\label{jump evolution with Hamiltonian}
d|\psi(t)\rangle=\left[\sum_i dN_i(t)\left(\frac{L_i}{\sqrt{\langle L^{\dagger}_i L_i\rangle}}-1\right)-dt\left(\frac{i}{\hbar}\hat{H}+\frac{1}{2}\sum^{m}_{i=1}L^{\dagger}_iL_i-\frac{1}{2} \sum^{m}_{i=1}\langle L^{\dagger}_iL_i\rangle\right)\right]|\psi(t)\rangle,
\end{equation}
which is a nonlinear stochastic Schr\"odinger equation (SSE). For a mixed state, the corresponding equation is given by
\begin{align}\label{jump evolution with Hamiltonian mixed}
d\rho=-dt\frac{i}{\hbar}(\hat{H}_{\text{eff}}\rho-\rho \hat{H}^{\dagger}_{\text{eff}})&+dt\sum^{m}_{i=1}\langle L^{\dagger}_iL_i\rangle\rho+\sum^{m}_{i=1}dN_i\left(\frac{L_i\rho L^{\dagger}_i}{\langle L^{\dagger}_iL_i\rangle}-\rho\right),\\ \hat{H}_{\text{eff}}&:=\hat{H}-\frac{i\hbar}{2}\sum_{i=1}^{m}L^{\dagger}_i L_i.
\end{align}
Note that the above equations are nonlinear, because the terms involving expectation values $\langle\cdot\rangle$ also depend on the current state $\rho$ or $|\psi\rangle$.
 
\subsection{Diffusive Quantum Trajectory}\label{continuous measurement subsection}
When the quantum state is continuously measured, instead of experiencing discrete quantum jumps, the state may evolve stochastically and continuously in a diffusive manner. Such behaviour appears if the measurement operators $\{M_{m}\}$ in an infinitesimal time interval $dt$ are only slightly different from the identity.

A typical example is low-resolution position measurement through, for example, shining off-resonant probe light at an atom and probing scattered light from it. Because the resolution is extremely low, one effectively combines and takes the average of the results of many repetitions of the same measurement to obtain a single outcome, and due to the central limit theorem, the measurement operator is effectively described by a Gaussian function \cite{GaussianMeasurement, GaussianMeasurement2}
\begin{equation}\label{Gaussian measurement}
	M_q=\left(\frac{\gamma dt}{\pi}\right)^\frac{1}{4}e^{-\frac{\gamma dt}{2}(\hat{x}-q)^2},
\end{equation}
where $\hat{x}$ is the position operator of the measured particle, $\gamma$ is the measurement strength, and $q$ is the measurement outcome. The standard deviation of the Gaussian function, i.e.~the resolution of the measurement, is proportional to $\frac{1}{\sqrt{dt}}$, which shows the correct scaling property respecting the amount of information received within the time interval $dt$. The completeness condition of the measurement is automatically satisfied due to the property of the Gaussian function
\begin{equation}\label{Gaussian measurement completeness}
	\int_{-\infty}^{\infty}M^{\dagger}_qM_q\, dq=1\,.
\end{equation}
The probability of obtaining measurement outcome $q$ is given by 
\begin{equation}
	\begin{split}
			\langle\psi|M^{\dagger}_qM_q|\psi\rangle &=  \left(\frac{\gamma dt}{\pi}\right)^\frac{1}{2}\langle\psi|e^{-{\gamma dt}(\hat{x}-q)^2}|\psi\rangle \\
			&=\left(\frac{\gamma dt}{\pi}\right)^\frac{1}{2}e^{-{\gamma dt}(\langle\hat{x}\rangle-q)^2}  ,
	\end{split}
\end{equation}
and therefore, the measurement outcome $q$ follows the Gaussian distribution $q\sim\mathcal{N}\left (\langle\hat{x}\rangle,\dfrac{1}{2\gamma\, dt}\right )$, and the order of magnitude of $q$ is $O(\frac{1}{\sqrt{dt}})$.

To obtain the time evolution equation, we note 
\begin{equation}
	M_q|\psi\rangle = \left(\frac{\gamma dt}{\pi}\right)^\frac{1}{4}e^{-\frac{\gamma dt}{2}(\hat{x}-q)^2}|\psi\rangle,
\end{equation}
and therefore the postmeasurement state is given by \cite{DIOSI1988419}
\begin{equation}\label{position post measurement calculation}
	\begin{split}
	\frac{M_q|\psi\rangle}{\sqrt{\langle\psi|M_q^{\dagger}M_q|\psi\rangle}}&= e^{\frac{\gamma dt}{2}(\langle\hat{x}\rangle-q)^2}e^{-\frac{\gamma dt}{2}(\hat{x}-q)^2}|\psi\rangle \\
	&= e^{\frac{\gamma dt}{2}(-\hat{x}^{2}+\langle\hat{x}\rangle^{2}+2q(\hat{x}-\langle\hat{x}\rangle))}|\psi\rangle \\
	&=e^{-\frac{\gamma dt}{2}(\hat{x}-\langle\hat{x}\rangle)^{2}+\gamma dt(q-\langle\hat{x}\rangle)(\hat{x}-\langle\hat{x}\rangle)}|\psi\rangle.
	\end{split}
\end{equation}
Since we have $q\sim\mathcal{N}\left (\langle\hat{x}\rangle,\dfrac{1}{2\gamma\, dt}\right )$, $dt(q-\langle\hat{x}\rangle)$ is a Gaussian variable with mean zero and variance $\dfrac{dt}{2\gamma}$. Using the It\^o calculus \cite{MeasurementItoFormalism}, defining $dW$ as the Wiener increment of a Wiener process satisfying $\mathbb{E}[dW]=0$ and $\mathbb{E}[dW^{2}]=dt$, Eq.~(\ref{position post measurement calculation}) can be expressed as
\begin{equation}
	\begin{split}
	\frac{M_q|\psi\rangle}{\sqrt{\langle\psi|M_q^{\dagger}M_q|\psi\rangle}}&= e^{-\frac{\gamma dt}{2}(\hat{x}-\langle\hat{x}\rangle)^{2}+\sqrt{\frac{\gamma}{2}}dW(\hat{x}-\langle\hat{x}\rangle)}|\psi\rangle\\
	&\approx \left (1-\frac{\gamma dt}{2}(\hat{x}-\langle\hat{x}\rangle)^{2}+\sqrt{\frac{\gamma}{2}}dW(\hat{x}-\langle\hat{x}\rangle) + \frac{\gamma dt}{4}(\hat{x}-\langle\hat{x}\rangle)^{2}\right )|\psi\rangle \\
	&= \left (1-\frac{\gamma dt}{4}(\hat{x}-\langle\hat{x}\rangle)^{2}+\sqrt{\frac{\gamma}{2}}dW(\hat{x}-\langle\hat{x}\rangle) \right )|\psi\rangle ,\\
\end{split}
\end{equation}
where we have used the Taylor series of the exponential function up to the second order, and have used the It\^o rule $\mathbb{E}[dW^{2}]=dt$. Taking the Hamiltonian of the system into account, the time evolution equation is given by
\begin{equation}\label{position time evolution equation}
	d|\psi\rangle=\left[\left(-\frac{i}{\hbar}\hat{H}-\frac{\gamma}{4}(\hat{x}-\langle\hat{x}\rangle)^2\right)dt+\sqrt{\dfrac{\gamma}{2}}(\hat{x}-\langle\hat{x}\rangle)dW\right]|\psi\rangle.
\end{equation}
For a general mixed state $\rho$, the time evolution equation is given by
\begin{equation}\label{position time evolution equation mixed}
d\rho=-\frac{i}{\hbar}[\hat{H},\rho]dt-\frac{\gamma}{4}[\hat{x},[\hat{x},\rho]]dt+\sqrt{\frac{\gamma}{2}}\{\hat{x}-\langle \hat{x}\rangle,\rho\}dW.
\end{equation}

\section{Measurement of the Center of Mass}\label{COM measurement}
To derive the measurement of the center of mass, or the average position, of several particles, we consider identical particles on a lattice and we introduce the measurement operator \cite{COMmeasurement}
\begin{equation}
	M_{q}\propto \sum_{m}f(q-md)\hat{b}_m^{\dagger}\hat{b}_m
\end{equation}
where $\hat{b}_m$ is the annihilation operator on site $m$, $f(\cdot)$ is the point spread function, $d$ is the spacing of the lattice, and $q$ is the measurement outcome. At the low-resolution limit, for simplicity, we use the Gaussian point spread function 
\begin{equation}
	f(x)=\left(\frac{1}{\sigma^{2}\pi}\right)^\frac{1}{4}e^{-\frac{x^2}{2\sigma^{2}}},
\end{equation}
where the resolution $\sigma$ is large.

To measure the center of mass of multiple particles, the measurement must not resolve the positions of the particles separately, and therefore, for any two sites $m$ and $n$ with non-zero particles, we have $\sigma \gg (m-n)d$. Without loss of generality, we may use $\sigma\gg md$. Then, the measurement operator $M_{q}$ can be expressed as
\begin{equation}\label{COM measurement begin}
	\begin{split}
	M_{q} &\propto \sum_{m}e^{-\frac{(q-md)^2}{2\sigma^{2}}}\hat{b}_m^{\dagger}\hat{b}_m\\
	&= \sum_{m}e^{-\frac{q^{2}-2mdq+(md)^2}{2\sigma^{2}}}\hat{b}_m^{\dagger}\hat{b}_m\\
	&\approx \sum_{m}e^{-\frac{q^{2}-2mdq}{2\sigma^{2}}}\hat{b}_m^{\dagger}\hat{b}_m\,.\\
	\end{split}
\end{equation}
Because the measurement outcome $q$ and the resolution $\sigma$ are of the same order, using $\sigma\gg md$, Eq.~(\ref{COM measurement begin}) can be expanded as
\begin{equation}
	\begin{split}
	M_{q} &\propto  \sum_{m}e^{-\frac{q^{2}}{2\sigma^{2}}}e^{\frac{mdq}{\sigma^{2}}}\hat{b}_m^{\dagger}\hat{b}_m\\
	&\approx\sum_{m}e^{-\frac{q^{2}}{2\sigma^{2}}}\left (1+\frac{mdq}{\sigma^{2}}\right )\hat{b}_m^{\dagger}\hat{b}_m\\
	&=e^{-\frac{q^{2}}{2\sigma^{2}}}\left (N+\frac{q}{\sigma^{2}}\sum_{m}md\hat{n}_{m}\right )\\
	&={N}e^{-\frac{q^{2}}{2\sigma^{2}}}\left (1+\frac{q}{\sigma^{2}}\sum_{m}\frac{md\hat{n}_{m}}{N}\right )\\
	&\approx{N}e^{-\dfrac{\left (q-\frac{\sum_{m}\hat{n}_{m}md}{N}\right )^{2}}{2\sigma^{2}}},\\
\end{split}
\end{equation}
where $N$ is the total number of particles, and $\hat{n}_{m}\equiv\hat{b}_m^{\dagger}\hat{b}_m $ is the number operator at site $m$. Therefore, the measurement $M_{q}$ that collectively measures all the particles with a low resolution effectively measures the center of mass, i.e.~$\frac{\sum_{m}\hat{n}_{m}md}{N}$. It is clear that the same argument also applies to the case of continuous space, where the measurement effectively measures the average position of several particles, i.e.~$\frac{\sum_{m} \hat{x}_{m}}{N}$. The time evolution equation is thus given by
\begin{equation}\label{COM time evolution equation}	d|\psi\rangle=\left[\left(-\frac{i}{\hbar}\hat{H}-\frac{\gamma}{4}(\hat{x}_{\text{COM}}-\langle\hat{x}_{\text{COM}}\rangle)^2\right)dt+\sqrt{\dfrac{\gamma}{2}}(\hat{x}_{\text{COM}}-\langle\hat{x}_{\text{COM}}\rangle)dW\right]|\psi\rangle,
\end{equation}
where $\hat{x}_{\text{COM}}\equiv\dfrac{\sum_{m}\hat{x}_{m}}{N}$ is the center-of-mass operator.
	\chapter{Deep Reinforcement Learning}\label{deep reinforcement learning}
	In this chapter, we introduce deep reinforcement learning, and specifically, we focus on deep Q-learning. In Section \ref{reinforcement learning}, we introduce the basics of reinforcement learning, and in Section \ref{Q-learning section} we introduce Q-learning, and in Section \ref{deep learning DQN}, we discuss how current deep learning technology is used with Q-learning.

\section{Reinforcement Learning}\label{reinforcement learning}
\subsection{Problem Setting}
Reinforcement learning generally deals with a Markov decision process (MDP), where the state $s_t$ of an environment (or, the state of a system) at time step $t$ makes a transition to the next state $s_{t+1}$ depending on the current state $s_t$ as well as the current action $a_t$ of the agent, producing a reward $r_t$. The reward $r_t$ and the resultant state $s_{t+1}$ and can be either deterministic or probabilistic. There can also be some predefined terminal states $s_T$ so that the process terminates when $s_T$ is reached. The goal of reinforcement learning is to learn a policy $a_t\sim\pi(s_t)$ to determine action $a_t$ based on the current state $s_t$, so that the return $\sum_{t=0}^{T}r_{t}$, i.e., the sum of all future rewards, is maximized by following the policy $\pi$ to decide the actions. For realistic problems, examples of rewards to maximize include scores in games, +1 representing a winning and -1 representing a loss in competitive games such as chess, fidelity in quantum control problems, and the amount of energy decrease in cooling problems. 

In practice, a discounted return $\sum_{t=0}^{T}\gamma^t r_{t}$ is often considered instead of the original return $\sum_{t=0}^{T}r_{t}$, where $\gamma$ is the discount factor satisfying $\gamma<1$ and $\gamma\approx1$, so that the expression $\sum_{t=0}^{T}\gamma^t r_{t}$ is convergent for $T\to\infty$ and that the rewards in future steps are discounted exponentially according to how distant they are in the future. This strategy results in an effective time horizon $\frac{1}{1-\gamma}$, beyond which the agent does not plan for more time steps in the future. The factor $\gamma$ is often set to be around $0.99$. Note that the optimal policy that maximizes $\sum_{t=0}^{T}\gamma^t r_{t}$ may be different from the policy that maximizes $\sum_{t=0}^{T}r_{t}$, but for convenience, we often only consider the case of $\sum_{t=0}^{T}\gamma^t r_{t}$. Ideally, the policy that maximizes $\sum_{t=0}^{T}\gamma^t r_{t}$ should be similar to the policy that maximizes $\sum_{t=0}^{T}r_{t}$, and is supposed to perform satisfactorily even when the performance is measured in terms of $\sum_{t=0}^{T}r_{t}$. 

\subsection{The Value Function and the Q Function}
Given a policy $\pi$, the value function $V_\pi(s_t)$ is defined as the expected return starting from a state $s_t$ following the policy $\pi$. The Q function $Q_\pi(s_t, a_t)$ is defined as the expected return starting from a state $s_t$, choosing $a_t$ as the action at time step $t$, while following the policy $\pi$ for the subsequent time steps.
\begin{equation}\label{value and Q definition}
	V_\pi(s_t):=\mathbb{E}_{a_t\sim\pi(s_t),\{(s_{t+i}, a_{t+i})\}_{i=1}^{T-t}}\left [\sum_{i=0}^{T-t}\gamma^i r_{t+i}\right ],\quad Q_\pi(s_t, a_t):=\mathbb{E}_{\{(s_{t+i}, a_{t+i})\}_{i=1}^{T-t}}\left [\sum_{i=0}^{T-t}\gamma^i r_{t+i}\right ].
\end{equation}
The expectation in Eq.~(\ref{value and Q definition}) is taken over the trajectories $\{(s_{t+i},a_{t+i})\}_{i=1}^{T-t}\,$, with $a_{t+i}\sim\pi(s_{t+i})$ for $1\le i\le T-t$. Clearly they satisfy
\begin{equation}
	V_\pi(s_t)=\mathbb{E}_{a_t\sim\pi(s_t)}\left [Q_\pi(s_t, a_t)\right].
\end{equation}

The $Q$ function has an important recursive property given by the following equation
\begin{equation}\label{Q recursive}
	Q^{\pi}(s_t,a_t)=r_t+\gamma\,\mathbb{E}_{s_{t+1},\,a_{t+1}\sim\pi{(s_{t+1})}}\left[Q^{\pi}(s_{t+1},a_{t+1})\right],
\end{equation}
which can be shown directly using the definition in Eq.~(\ref{value and Q definition}).

If a policy maximizes the Q function, we say that the policy is optimal, and we denote the maximized Q function and the policy by $Q^*$ and $\pi^*$, respectively. The optimality implies that $Q^*$ satisfies the Bellman equation \cite{ReinforcementlearningAnintroduction}
\begin{equation}\label{Bellman equation}
	Q^*(s_t,a_t)=r_t+\gamma \mathbb{E}_{s_{t+1}}\left[\max_{a'}Q^*(s_{t+1},a') \right],
\end{equation}
and the policy $\pi^*$ is greedy with respect to $Q^*$, i.e.~$\pi^*(s)=\argmax_{a'}Q^*(s,a')$, always choosing the action that corresponds to the largest Q value. This is because if the policy does not choose the action $\argmax_{a'}Q^*(s,a')$, then, by changing its action to $\argmax_{a'}Q^*(s,a')$, it will be able to obtain more reward by definition. Therefore, the optimality implies that the optimal policy must always choose $\argmax_{a'}Q^*(s,a')$ as its action, and the Bellman equation---Eq.~(\ref{Bellman equation})---follows. It is known that the optimal Q function $Q^*$ that satisfies the Bellman equation is unique \cite{ReinforcementlearningAnintroduction}. The optimal policy $\pi^*$ is the policy that we look for in a reinforcement learning task, which maximizes $\sum_{t=0}^{T}\gamma^t r_{t}$.

\section{Q-learning}\label{Q-learning section}
Q-learning uses the recursive relation in Eq.~(\ref{Bellman equation}) to learn $Q^*$. In cases where the spaces of $s_t$ and $a_t$ are discrete and small enough to enumerate, we can arbitrarily initialize the Q value for every $(s_t,a_t)$ pair and write them down in a table, and apply the iteration rule
\begin{equation}\label{Q table learning}
	Q(s_t,a_t)\gets r_t+\gamma \mathbb{E}_{s_{t+1}}\left[\max_{a'}Q(s_{t+1},a') \right].
\end{equation}
Such an iteration always converges because the absolute norm $||Q_1 - Q_2||_{l_1}$ contracts during the iteration, and the Q function converges to the optimal Q function $Q^*$, from which we can derive the optimal policy $\pi^*(s)=\argmax_{a'}Q^*(s,a')$. The convergence of the iteration can also be understood by recursively substituting the term $\max_{a'}Q(s_{t+1},a')$ in Eq.~(\ref{Q table learning}) by the left-hand side of the equation, from which we can see that the prefactor of the Q function decreases exponentially by $\gamma^n$, where $n$ is the number of substitutions, implying that the final result of the iteration is independent of the initial condition of $Q(s_t,a_t)$ and only depends on the rewards. In many cases where random sampling is involved, the following iteration rule, or learning rule, is used instead
\begin{equation}\label{Q table learning with learning rate}
	\Delta Q(s_t,a_t) = \alpha \left (r_t+\gamma \max_{a'}Q(s_{t+1},a') -Q(s_t,a_t)\right ),
\end{equation}
where $\alpha$ is a small constant called the \textit{learning rate}. This strategy is called Q-table learning.

Compared with other reinforcement learning strategies, especially policy gradient strategies, which simply increases the probability of choosing an action if the action leads to larger future rewards, the Q-learning strategy is relatively more efficient and it learns the structure of the task quickly by directly backpropagating information from $s_{t+1}$ to $s_t$, reaching for the optimal policy. 

In many real-world cases where $(s_t,a_t)$ is not enumerable and cannot be written into a table, Q-table learning becomes impossible, and some function approximation is used to represent the Q function that is to be learned. With learnable parameters $\theta$, the Q function denoted by $Q_\theta$ may be learned via the basic Q-learning rule
\begin{equation}\label{Q learning with parameters}
	\Delta \theta = \alpha \nabla_\theta Q_\theta(s_t,a_t)\left (r_t+\gamma \max_{a'}Q_\theta(s_{t+1},a') -Q_\theta(s_t,a_t)\right ),
\end{equation} 
which can be interpreted as changing the value of $Q_\theta(s_t,a_t)$ so that $Q_\theta(s_t,a_t)$ approaches the target $r_t+\gamma \max_{a'}Q_\theta(s_{t+1},a')$, following the gradient. However, this iteration may not converge in practice. This is because the term $\max_{a'}Q_\theta(s_{t+1},a')$ is $\theta$-dependent and can change simultaneously with $Q_\theta(s_t,a_t)$, and therefore the iteration may diverge. This can be a serious issue for realistic tasks, because the adjacent states $s_t$ and $s_{t+1}$ often have similar representations, which makes $\nabla_\theta Q_\theta(s_t,\cdot)$ closely related to $\nabla_\theta Q_\theta(s_{t+1},\cdot)$. There have been many attempts to resolve this problem in practical applications and to stabilize the learning. One important strategy is to keep a copy $\tilde{\theta}$ of $\theta$, and use $\tilde{\theta}$ to compute the term $\max_{a'}Q_\theta(s_{t+1},a')$ in Eq.~(\ref{Q learning with parameters}), and update $\tilde{\theta}$ to $\theta$ periodically. In this way, the learning target $r_t+\gamma\max_{a'}Q_\theta(s_{t+1},a')$ does not change simultaneously with $\theta$, and the learning can be stabilized. This strategy is an important ingredient in the deep Q network (DQN) algorithm and $\tilde{\theta}$ is called the target network \cite{DQN}, which will be discussed in the next section. 

\section{Deep Q-Learning}\label{deep learning DQN}
\subsection{Deep Learning}\label{deep learning subsection}
Learning usually refers to the process of building a model to relate different pieces of information so as to give predictions on things that were previously unpredictable. Machine learning automatizes this process by making use of machine learning algorithms. Typically, machine learning utilizes a dataset $\{(\boldsymbol{x}_i,\boldsymbol{y}_i)\}=\mathcal{S}$ containing example questions $\boldsymbol{x}_i$ and the corresponding answers $\boldsymbol{y}_i$, and by learning from the dataset using some algorithms, the AI program is expected to give the correct answer $\boldsymbol{y}'$ when it is given a question $\boldsymbol{x}'$ which is not in $\mathcal{S}$ but is similar to the learned data $\boldsymbol{x}_i$.

To model the relation between $\boldsymbol{x}$ and $\boldsymbol{y}$, deep learning uses parametrized functions $f$ which are known as deep neural networks, or artificial neural networks, in analogy to the biological neural networks. A deep neural network $f$ is defined as 
\begin{equation}\label{feedforward neural network}
	f(\boldsymbol{x})=\textbf{M}_n\left(f_{n-1}\circ f_{n-2} \circ f_{n-3}\cdots\circ f_1\left(\boldsymbol{x}\right)\right)+\boldsymbol{b}_n\,,
\end{equation}
\begin{equation}\label{feedforward neural network2}
	f_i(\boldsymbol{x})=\sigma\left(\textbf{M}_i\boldsymbol{x}+\boldsymbol{b}_i\right),
\end{equation}
where $\textbf{M}_i$ are matrices known as weights, $\boldsymbol{b}_i$ are vectors known as biases, and $\sigma$ is a nonlinear function applied to the vectors element-wise and is called the \textit{activation function} \cite{DeepLearningBook}. In practice, the activation function $\sigma$ is often chosen to be the rectified linear unit (ReLU) function, i.e.~$\sigma=\max(0,x)$.
\begin{figure}[htb]
	\centering
	\includegraphics[width=0.45\linewidth]{"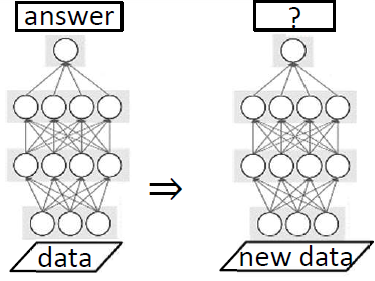"}
	\caption{A schematic illustration of deep learning. We train the neural network to fit into existing data-answer pairs, and use the trained neural network to give predictions on new data.}
	\label{fig:neural network}
\end{figure}

The function $f$ can be considered as having $n$ layers, and each layer consists of $w$ neurons representing the $w$ real values in the vector $\boldsymbol{x}$, and neurons in adjacent layers $i$ and $i+1$ are connected by a linear mapping $\textbf{M}_{i+1}\boldsymbol{x}+\boldsymbol{b}_{i+1}$, with an activation function that is applied to every neuron. This structure is in analogy to biological neural networks, and can potentially represent complicated functions. With sufficiently wide layers, i.e.~large $w$, the function $f$ is known to be universal in the sense that it can approximate any function with arbitrary precision. The neural network $f$ is also known as a feedforward neural network, because the information in it only goes in a single direction from bottom to top, as schematically shown in Fig.~\ref{fig:neural network}. The parameters to be learned are the matrices $\textbf{M}_i$ and biases $\boldsymbol{b}_i$. 

To learn the parameters in $f$, we define a loss function to measure the difference between the prediction of $f$, i.e.~$f(\boldsymbol{x}_i)$, and the correct answer $\boldsymbol{y}_i$. By minimizing this difference, the prediction of $f$ can get close to the correct answer and $f$ may be used to give predictions on new data. The simplest loss function for real-valued $\boldsymbol{y}_i$ is the mean squared error (MSE)
\begin{equation}
	L_{\text{MSE}}=\frac{1}{|\mathcal{S}|}\sum_{(\boldsymbol{x}_i,\boldsymbol{y}_i)\in\mathcal{S}} \left | f(\boldsymbol{x}_i) - \boldsymbol{y}_i\right |^{2}.
\end{equation}
By calculating the gradient of $L_{\text{MSE}}$ with respect to the parameters in $f$, we can adjust the parameters by following the direction of the gradient to minimize the loss function. This training procedure is called \textit{gradient descent}. Denoting the parameters by $\theta$, the basic gradient descent does following iteration:
\begin{equation}\label{gradient descent}
	\Delta \theta = - \alpha \nabla_\theta L_{\text{MSE}},
\end{equation}
where $\alpha$ is the learning rate. The learning rate $\alpha$ is often annealed and decreased gradually during the training process to control the speed and the precision of learning. Besides the basic gradient descent as given by Eq.~(\ref{gradient descent}), many variations also exist, involving additional momentum terms and normalization techniques to speed up the training process \cite{DeepLearningBook, Adam,RMSprop}, which have been widely adopted in deep learning practice.

The calculation of the gradient $\nabla_\theta L_{\text{MSE}}$ can be done through a procedure called backpropagation of gradient. As the neural network $f$ is composed of linear mappings and activation functions $\sigma=\max(0,x)$ only, the gradient with respect to the internal parameters of $f$ can be computed easily using matrix computation, which is typically carried out on modern GPUs (graphical processing units) which can do large-scale matrix computation quickly. Also, instead of evaluating $\nabla_\theta L_{\text{MSE}}$ using the whole dataset $\mathcal{S}$, one typically samples a minibatch from $\mathcal{S}$ for each iteration, and evaluate $\nabla_\theta L_{\text{MSE}}$ on the sampled minibatch only, and the resulting training procedure is therefore called \textit{stochastic gradient descent}. 

\subsection{Deep Q Network Algorithm}\label{DQN algorithm section 3}
The Q-learning strategy that employs a deep neural network to represent the Q function is called deep Q-learning, and the most widely used deep Q-learning algorithm is the deep Q network (DQN) algorithm \cite{DQN}. In the DQN algorithm, the neural network inputs a representation of the state $s_t$ and produces several outputs as the evaluated Q function values for different choices of the action $a_t$. Note that the action $a_t$ here is an element from a finite set and it cannot be a continuous variable. The training loss is defined to effectively achieve the Q-learning rule in Eq.~(\ref{Q learning with parameters}) and it is given by
\begin{equation}
	L_{\text{DQN}}=\mathbb{E}_{(s_t,a_t,r_t,s_{t+1})}\left [\left| Q_{\theta}(s_t,a_t)-\left(r(s_t,a_t)+\gamma\,\max_{a_{t+1}}Q_{\tilde{\theta}}(s_{t+1},a_{t+1})\right)\right|^{2}\right ],
\end{equation}
where $\theta$ is the network parameters, and $Q_{\tilde{\theta}}$ is the target network which is a copy of $Q_{\theta}$, and $\tilde{\theta}$ is updated to $\theta$ periodically. The period of the target network update is typically several thousand iterations of gradient descent. $(s_t,a_t,r,s_{t+1})$ is a sampled piece of experience from the replay memory. The replay memory is a dataset containing the experience data that are observed and accumulated, when the AI follows its policy to interact with the environment. 

\begin{figure}[htb]
	\centering
	\includegraphics[width=0.7\linewidth]{"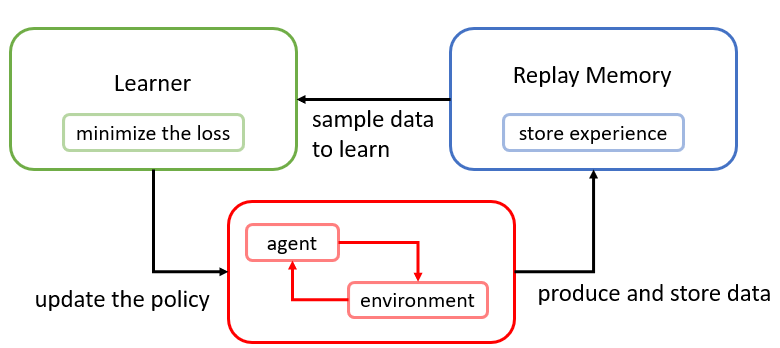"}
	\caption{A schematic diagram of the reinforcement learning system. These three parts of the reinforcement learning system can work simultaneously and be parallelized.}
	\label{fig:reinforcementlearningsystem}
\end{figure}

Unlike other types of machine learning, reinforcement learning does not have a preexisting dataset, and through interactions with the environment, it accumulates experience to build up the dataset to learn. The reinforcement learning system is summarized and schematically shown in Fig.~\ref{fig:reinforcementlearningsystem}. To explore various possibilities of different actions and policies for the task, when the agent interacts with the environment, for some small probability $\epsilon$, the agent takes a random action; otherwise the agent chooses $\argmax_{a'}Q_\theta(s,a')$ as its action. Such a policy is called the $\epsilon$-greedy policy, and is important for Q-learning to proceed in realistic tasks.

Here we summarize the procedure of deep reinforcement learning. First, we initialize a neural network using random parameters. Second, we use this neural network as $Q_{\theta}$ and use policy $\argmax_{a'}Q(s,a')$ for the agent to take actions to interact with the environment, but with a small probability $\epsilon$, we replace the action by a random one. Third, we sample from what the agent has experienced, specifically, $(s_t,a_t,r,s_{t+1})$ data tuples, and calculate the loss $L_{\text{DQN}}$, and use gradient descent to minimize the loss. Finally, we repeat the second and third steps until the AI performs well enough. Note that the second and the third steps above can be parallelized and be executed simultaneously.

In real scenarios, a large amount of technical modification is applied to the above learning scheme to improve the learning system in various aspects so as to make the learning more efficient. These techniques include the prioritized sampling strategy of experience data \cite{prioritizedSampling}, using double Q-networks to separately decide actions and to calculate the $Q$ values \cite{DoubleDQN}, a dueling network structure to learn the average of $Q$ and the $Q$ values for each action separately \cite{DuelDQN}, etc. These techniques can improve the performance and stability of the DQN algorithm, and they are often used in combination to facilitate learning. We also adopt these techniques in our research. These techniques, together with many other modifications, have been combined together in Ref.~\cite{RainbowDQN} and the resulting algorithm is called \textit{Rainbow DQN}. However, we do not use Rainbow DQN, because the algorithm is too complicated and not suitable for our research.

	\chapter{Convergent Deep Q-Learning}\label{convergent DQN}
	In this chapter, we introduce our work, the convergent deep Q network (C-DQN) algorithm \cite{CDQN}, which is a generalization of the conventional DQN to overcome the issue of convergence of the algorithm. In Section \ref{CDQN background}, we introduce the background of the convergence issue in Q-learning, and in Section \ref{inefficiency of RG} we discuss the inefficiency problems of the existing approaches to the non-convergence issue. We introduce our C-DQN algorithm in Section \ref{CDQN}, and we present results of the numerical experiments of C-DQN in Section \ref{CDQN experiments}, and in Section \ref{CDQN outlook} we present the conclusions and the outlook of this section.

\section{Background}\label{CDQN background}
\subsection{Non-Convergence in Q-Learning}
Although Q-learning remains one of the most efficient methods in reinforcement learning, the problem of non-convergence in Q-learning, or more generally, in temporal difference (TD) learning, has been pointed out decades before by Baird \cite{ResidualLearningInitial} and Tsitsiklis and Van Roy  \cite{TsitsiklisNonlinearDivergence}. As discussed in Section \ref{Q-learning section}, Q-table learning uses the following learning rule
\begin{equation}\label{Q-table learning (CDQN)}
	\Delta Q(s_t,a_t) = \alpha \left (r_t+\gamma \max_{a'}Q(s_{t+1},a') -Q(s_t,a_t)\right ),
\end{equation}
where $\alpha$ is the learning rate and $\gamma$ is the discount factor, and the function value $Q(s_t,a_t)$ for every $(s_t,a_t)$ pair is an independent entry in the Q-table that is updated by the learning rule. On the other hand, for realistic tasks for which $(s_t,a_t)$ cannot be enumerated, function approximation is adopted and the Q function is parametrized by parameters. In this case, parameter $\theta$ determines the Q function denoted by $Q_{\theta}$, and $(s_t,a_t)$ is considered as the input of the function and an output of the function is the corresponding Q value. The Q-learning rule is given by 
\begin{equation}\label{Q-learning (CDQN)}
	\Delta \theta = \alpha \nabla_\theta Q_\theta(s_t,a_t)\left (r_t+\gamma \max_{a'}Q_\theta(s_{t+1},a') -Q_\theta(s_t,a_t)\right ).
\end{equation}
However, this iteration procedure does not necessarily converge. This is because although $Q_\theta(s_t,a_t)$ is learned to approach the value of $r_t+\gamma \max_{a'}Q_\theta(s_{t+1},a')$, the value of $r_t+\gamma \max_{a'}Q_\theta(s_{t+1},a')$ itself is also dependent on the learned parameter $\theta$ and it changes simultaneously with $Q_\theta(s_t,a_t)$. As a simple argument, if the condition 
\begin{equation}\label{assumed condition for divergence}
	\gamma \nabla_\theta Q_\theta(s_t,a_t)\cdot \nabla_\theta \max_{a'}Q_\theta(s_{t+1},a') > {||\nabla_\theta Q_\theta(s_t,a_t)||}^2
\end{equation}
is always satisfied, then, define the \textit{Bellman error}, or, the \textit{Bellman residual}
\begin{equation}\label{Bellman error}
	\delta_t:=r_t+\gamma \max_{a'}Q_\theta(s_{t+1},a')- Q_\theta(s_t,a_t),
\end{equation}
using Eq.~(\ref{Q-learning (CDQN)}), we have 
\begin{equation}\label{Bellman error diverge}
	\begin{split}
		\Delta \delta_t&=\nabla_\theta\delta_t \cdot\Delta\theta+ O(\Delta\theta^{2})\\
		&=\alpha \delta_t\left (\gamma\nabla_\theta Q_\theta(s_t,a_t)\cdot \max_{a'}Q_\theta(s_{t+1},a') -  {||\nabla_\theta Q_\theta(s_t,a_t)||}^2\right ) + O(\Delta\theta^{2}).
	\end{split}
\end{equation}
It can be seen from Eq.~(\ref{Bellman error diverge}) and (\ref{assumed condition for divergence}) that up to the first order in $\Delta\theta$, the Bellman error $\delta_t$ exponentially diverges, which means that the difference between $Q_\theta(s_t,a_t)$ and $r_t+\gamma \max_{a'}Q_\theta(s_{t+1},a')$ only increases, and the learning can never achieve its goal.

To alleviate this problem, the deep Q network (DQN) algorithm, which uses a deep neural network as ${\theta}$, introduces a target network ${\tilde{\theta}}$, and it effectively applies the following learning rule \cite{DQN}
\begin{equation}\label{DQN Q-learning rule}
	\Delta \theta = \alpha \nabla_\theta Q_\theta(s_t,a_t)\left (r_t+\gamma \max_{a'} \underline{Q_{\tilde{\theta}}(s_{t+1},a')}  -Q_\theta(s_t,a_t)\right ), 
\end{equation}
where we have used an underline to highlight the difference from Eq.~(\ref{Q-learning (CDQN)}). As the term $r_t+\gamma \max_{a'}Q_{\tilde{\theta}}(s_{t+1},a')$ does not changes simultaneously with $Q_\theta(s_t,a_t)$, given a fixed $\tilde{\theta}$, parameter $\theta$ can be learned stably without divergence. However, as a drawback, when the target network $\tilde{\theta}$ is fixed, the reward information can only propagate for a single time step from state $s_{t+1}$ to state $s_t$. For example, the Q function $Q_\theta(s_t,\cdot)$ at time step $t$ learns from the target network $Q_{\tilde{\theta}}(s_{t+1},\cdot)$ at time step $t+1$, and the Q function $Q_\theta(s_{t+1},\cdot)$ at time step $t+1$ learns from the target network $Q_{\tilde{\theta}}(s_{t+2},\cdot)$ at time step $t+2$, and the information cannot propagate for more than one time step. Therefore, to allow effective learning, in the DQN algorithm, $\tilde{\theta}$ is updated to $\theta$ periodically during the training process. Typically, $\tilde{\theta}$ is updated once every few thousand iterations of $\theta$ updates. For the $i$-th target network $\tilde{\theta}_i$, the loss function of the DQN algorithm is given by
\begin{equation}\label{DQN loss single}
	{\ell}_{\textit{DQN}}(\theta;\tilde{\theta}_i) =\left(Q_\theta(s_t,a_t)-r_t-\gamma \max_{a'}Q_{\tilde{\theta}_i}(s_{t+1},a') \right) ^2.
\end{equation}
Note that if the target network $\tilde{\theta}$ is immediately updated to $\theta$ whenever $\theta$ is learned and updated, the learning rule would reduce to the conventional Q-learning rule in Eq.~(\ref{Q-learning (CDQN)}). 

\begin{figure}[htb]
	\centering
	\includegraphics[width=0.5\linewidth]{"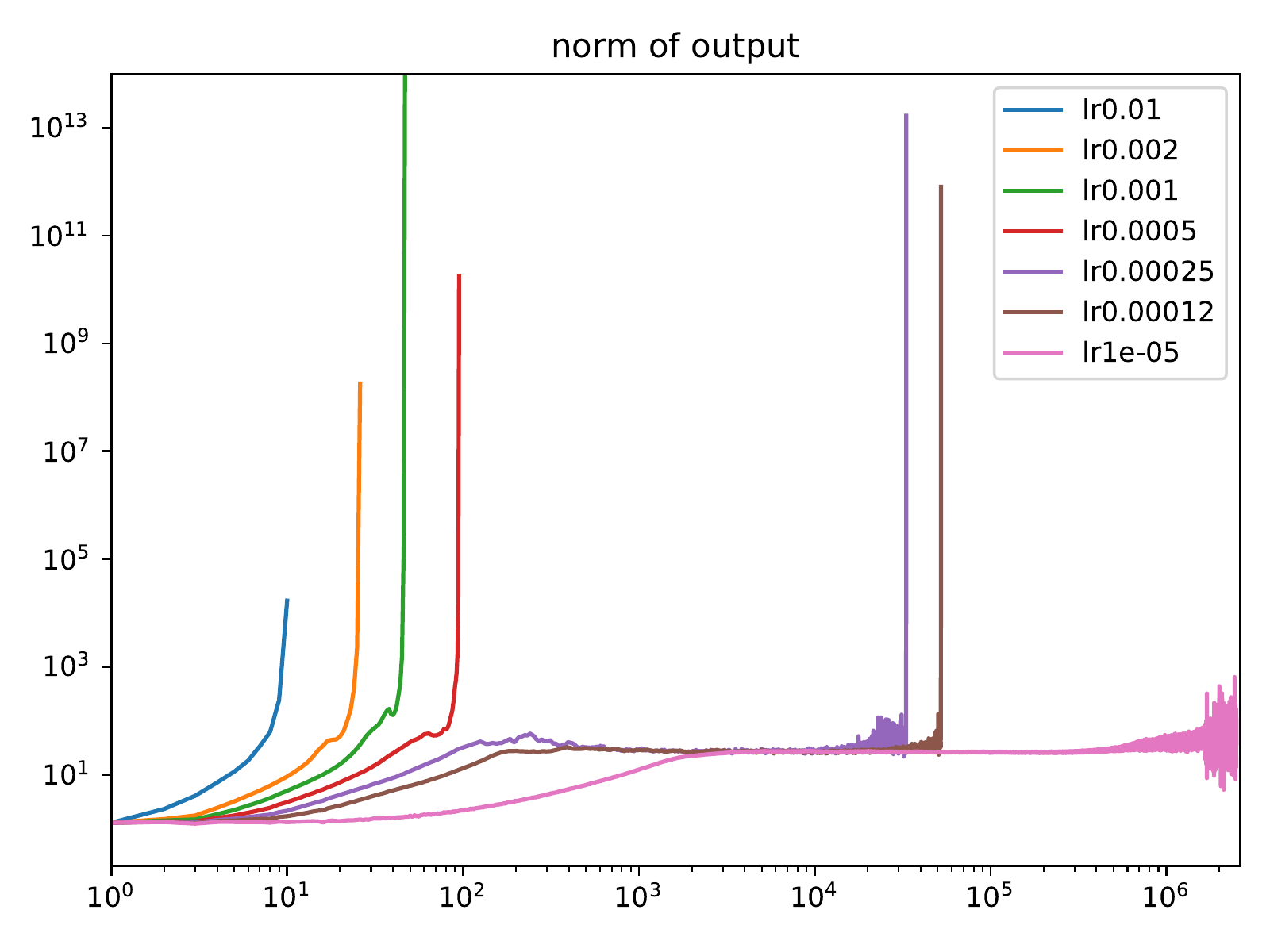"}
	\caption{Norm of the output of the function $Q_{\theta}$ during training in the DQN algorithm with different learning rates, when the target network is not used and the replay memory, i.e.~the dataset, is fixed. The horizontal axis shows the number of updates, and the vertical axis shows the average norm of $Q_{\theta}(s_t, \cdot)$ during training, and {lr} in the legend is an abbreviation of \textit{learning rate}, and the results for different learning rates are shown. The optimization algorithm is stochastic gradient descent (SGD), and the task is \textit{MsPacman} in the \textit{Atari 2600} benchmark. It can be seen that $Q_{\theta}$ is prone to diverge and is always unstable no matter how small the learning rate is. It should be noted that such exponential divergence usually cannot be seen directly in conventional DQN training procedure, because conventional DQN adopts adaptive gradient descent methods such as Adam \cite{Adam} instead of SGD, which effectively normalizes the update step $\Delta \theta$ to a constant magnitude and therefore disallows exponential divergence. Nevertheless, in such cases, the norm of $Q_{\theta}$ can still increase polynomially and the learning can fail similarly.}
	\label{Q diverge}
\end{figure}

Empirically, the use of the target network $\tilde{\theta}$ and the DQN loss given in Eq.~(\ref{DQN loss single}) significantly stabilizes the learning process, and the deep reinforcement learning AI successfully learns most of the simple arcade games in the \textit{Atari 2600} benchmark and achieves scores that are comparable to human performance \cite{DQN, AtariEnvironment}. As a confirmation, we checked that the algorithm indeed diverges if the target network is not used---specifically, the norm of the output of the function $Q_{\theta}$ goes to infinity, as shown in Fig.~\ref{Q diverge}.

Although the use of target network can significantly alleviate the problem of divergence in practice, the technique is essentially ad hoc, and it does not come with any theoretical guarantee. As a consequence, there are still many benchmark tasks where the DQN algorithm exhibits instability and divergence, such as the examples in Ref.~\cite{DoubleDQN}, where the author proposed further technical improvements to alleviate the problem. Hasselt \textit{et al.}~\cite{DeadlyTriad} investigated the effect of different settings of the learning algorithm on the instability, and it has been found that if one uses larger neural networks, the algorithm would be more vulnerable to instability. In Ref.~\cite{TCLoss, ConstrainedTDUpdate}, it has been shown that DQN does not perform well when the discount factor $\gamma$ is large, for example, when we have $\gamma\gtrsim0.999$, which is consistent with the fact that Q-learning is prone to diverge when the discount factor $\gamma$ is too large \cite{ReinforcementlearningAnintroduction}. Such limitation on $\gamma$ shortens the effective time horizon $\frac{1}{1-\gamma}$ of planning, and therefore prohibits the DQN agent from learning tasks that require long-term planning.

\begin{figure}[tb]
	\centering
	\includegraphics[width=0.52\linewidth]{"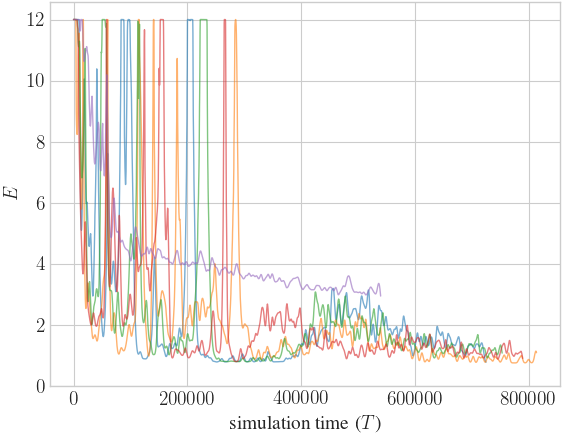"}
	\caption{Training curves of the DQN algorithm on the task of measurement feedback cooling of a quantum quartic oscillator in numerical simulation \cite{quantumCartpole}, where each curve represents a repetition of the same experiment with a different random seed. The horizontal axis shows the simulation time of the system and represents the training time, and the vertical axis shows the energy of the cooled system, and a lower energy indicates higher performance. }
	\label{DQN quartic}
\end{figure}

The non-convergence issue has various adverse effects when one attempts to apply DQN to realistic problems. First, it makes the training result sensitive to fine-tuned hyperparameters and increases the difficulty of training. For example, if the training fails, it is hard to know whether the failure is because the task is too difficult for DQN to learn or simply because the training setting is not tuned well enough. Secondly, since the training does not necessarily converge, the performance during training may not improve consistently and the performance can fluctuate throughout the whole training process, in which case it is difficult to know how long the training process should be, when the learning is completed and when the training process is supposed to be stopped. This issue significantly increases the difficulty of applying DQN to complicated and newly encountered tasks. (For example, see the results on \textit{Video Pinball} in Fig.~5 of Ref.~\cite{RainbowDQN}). Lastly, the instability of DQN increases the randomness in training and undermines the reliability of its results, as several repetitions of the same numerical experiment may end up with qualitatively different final performances, especially for tasks that are relatively difficult to learn, as shown in Fig.~\ref{DQN quartic}. This issue makes it difficult to reproduce existing results, harms the consistency of results and deteriorates the robustness and the reliability of deep reinforcement learning in general.

\subsection{Existing Works on the Non-Convergence Issue}\label{RG algorithm section}
To deal with the non-convergence issue, most of the existing works have been focusing on the use of the following loss function and its alternatives:
\begin{equation}\label{Bellman eq loss}
	\ell_{\text{Bellman}}(\theta)=\left| Q_{\theta}(s_t,a_t)-r_t-\gamma \mathbb{E}_{s_{t+1}|(s_t,a_t)}\left[\max_{a'}Q_{\theta}(s_{t+1},a') \right]\right |^{2},
\end{equation}
which is the mean squared error of the two sides of the Bellman equation  in Eq.~(\ref{Bellman equation}). The symbol $\mathbb{E}_{s_{t+1}|(s_t,a_t)}[\cdot]$ denotes the expectation value over $s_{t+1}$ conditioned on $s_t$ and $a_t$. The argument is that if the loss function is reduced to zero, then the Bellman equation will be satisfied and therefore we have $Q_{\theta}=Q^*$, i.e.~the optimal Q function. The training therefore proceeds through the minimization of this loss function, and it is expected that the performance of the agent improves as the loss decreases. The convergence of the training procedure simply follows the convergence of the optimization of the loss function.

The \textit{residual gradient} (RG) algorithm proposed by Baird \cite{ResidualLearningInitial} uses the squared Bellman residual, or, the mean squared Bellman error (MSBE), as the loss function
\begin{equation}\label{MSBE single}
	{\ell}_{\textit{MSBE}}(\theta)=\left|Q_\theta(s_t,a_t)-r_t-\gamma \max_{a'}Q_{\theta}(s_{t+1},a') \right| ^2.
\end{equation} 
It avoids the evaluation of the expectation term in Eq.~(\ref{Bellman eq loss}), because in practical situations, typically, only data samples $(s_{t}, a_{t}, r_{t}, s_{t+1})$ are available and one cannot evaluate the term $\mathbb{E}_{s_{t+1}|(s_t,a_t)}[\cdot]$ precisely, unless the underlying model of the reinforcement learning task is fully known and the model can be easily evaluated. If the underlying task is deterministic, we have ${\ell}_{\textit{MSBE}}\equiv \ell_{\text{Bellman}}$ and the minimization of ${\ell}_{\textit{MSBE}}$ fulfills the purpose of learning. On the other hand, if the task is stochastic, we have 
\begin{equation}\label{MSBE bias}
	\begin{split}
			\mathbb{E}_{s_{t+1}}[{\ell}_{\textit{MSBE}}(\theta)]={} & \mathbb{E}_{s_{t+1}}\left [\left (Q_\theta(s_t,a_t)-r_t-\gamma \max_{a'}Q_\theta(s_{t+1},a')\right )^2\right ]\\
			={} & \left (Q_\theta(s_t,a_t)-r_t\right )^2 -2\gamma\left (Q_\theta(s_t,a_t)-r_t\right ) \mathbb{E}_{s_{t+1}}\left[\max_{a'}Q_\theta(s_{t+1},a')\right] \\
			&+ \mathbb{E}_{s_{t+1}}\left [\left (\gamma \max_{a'}Q_\theta(s_{t+1},a')\right )^2\right ]\\
			={} & \left (Q_\theta(s_t,a_t)-r_t-\gamma \mathbb{E}_{s_{t+1}}\left[\max_{a'}Q_\theta(s_{t+1},a')\right]\right )^2 + \gamma^2\text{Var}_{s_{t+1}}(\max_{a'}Q_\theta(s_{t+1},a'))\\
			={} & \ell_{\text{Bellman}}(\theta) + \gamma^2\text{Var}_{s_{t+1}}[\max_{a'}Q_\theta(s_{t+1},a')],
	\end{split}
\end{equation}
where $\text{Var}_{s_{t+1}}[\cdot]$ denotes the variance regarding state $s_{t+1}$. That is, the expectation value of the loss ${\ell}_{\textit{MSBE}}$ includes an additional bias term $\gamma^2\text{Var}_{s_{t+1}}(\max_{a'}Q_\theta(s_{t+1},a'))$ which is absent in the original loss function $\ell_{\text{Bellman}}$, and we have $\mathbb{E}[{\ell}_{\textit{MSBE}}]>\mathbb{E}[{\ell}_{\text{Bellman}}]$. Therefore, the minimization of ${\ell}_{\textit{MSBE}}$ in the stochastic setting does not precisely converges to the optimal Q function $Q^{*}$, and the RG algorithm attempts to reduce the variance of $Q_{\theta}(s_{t+1}, \cdot)$, if a state $s_{t}$ with an action $a_{t}$ can make transitions to several different states $s_{t+1}$. However, it has been argued that this issue may not have serious consequences on the learning process, and that the agent may still learn to perform well, depending on the properties of the underlying task \cite{ResidualLearningInitial}. 

\subsubsection{Improvements on the Residual Gradient (RG) Algorithm}
Most recent works on convergent Q learning, or convergent TD learning, have been focusing on how to improve the RG algorithm so that the loss $\ell_{\text{Bellman}}$ is effectively minimized, instead of the loss ${\ell}_{\textit{MSBE}}$ \cite{GradientTDFirst, GradientTDLinear, GradientTDGeneral, SBEEDResidualLearning, GradientTDbyKernelToCorrelateData, GradientTDwithRegularization, GradientTreeBackup}. One way to effectively minimize the original loss $\ell_{\text{Bellman}}$ is to use double sampling to obtain an unbiased estimate of the gradient $\nabla_{\theta}\ell_{\text{Bellman}}$
\begin{equation}\label{double sampling}
	\begin{split}
		\mathbb{E}\left [\nabla_{\theta}\ell_{\text{Bellman}}\right ]={}&\nabla_\theta \left |Q_\theta(s_t,a_t)-r_t-\gamma\mathbb{E}_{s_{t+1}}[\max_{a'}Q_\theta(s_{t+1},a') ]\right |^2\\
		={}&\nabla_{\theta} \left |\left (Q_\theta(s_t,a_t)-r_t-\gamma\mathbb{E}_{s_{t+1}}[\max_{a'}Q_\theta(s_{t+1},a') ]\right )\right.\\
		&\left. \times \left (Q_\theta(s_t,a_t)-r_t-\gamma\mathbb{E}_{s_{t+1}}[\max_{a'}Q_\theta(s_{t+1},a') ]\right )\right |\\
		={}&2 \left (Q_\theta(s_t,a_t)-r_t-\gamma\underbrace{\mathbb{E}_{s_{t+1}}[\max_{a'}Q_\theta(s_{t+1},a')]}_{\text{evaluated using sample 1}}\right )\\
		&\times\left (\nabla_\theta Q_\theta(s_t,a_t)-\gamma\underbrace{\mathbb{E}_{s_{t+1}} [ \nabla_\theta \max_{a'}Q_\theta(s_{t+1},a') ]}_{\text{evaluated using sample 2}}\right ),
	\end{split}
\end{equation}
where the two expectation terms $\mathbb{E}_{s_{t+1}}[\cdot]$ must be evaluated independently to obtain an unbiased estimate of $\mathbb{E}\left [\nabla_{\theta}\ell_{\text{Bellman}}\right ]$, and therefore two independent samples of $s_{t+1}$ are needed, as highlighted in the above equation. This is sometimes referred to as the double sampling issue, which is the main focus of recent works on convergent Q learning or TD learning \cite{GradientTDFirst, GradientTDLinear, GradientTDGeneral, SBEEDResidualLearning, GradientTDbyKernelToCorrelateData, GradientTDwithRegularization, GradientTreeBackup}. All of those works have effectively used either a separate set of parameters or kernel methods to learn the first term in Eq.~(\ref{double sampling}), i.e.,
\begin{equation}
	Q_\theta(s_t,a_t)-r_t-\gamma\mathbb{E}_{s_{t+1}}[\max_{a'}Q_\theta(s_{t+1},a')],
\end{equation}
as a function of $s_t$ and $a_t$. Then, they combine the learned results and samples of the second term
\begin{equation}
	\nabla_\theta Q_\theta(s_t,a_t)-\gamma\mathbb{E}_{s_{t+1}} [ \nabla_\theta \max_{a'}Q_\theta(s_{t+1},a') ]
\end{equation}
to compute the gradient that is needed to train the parameter $\theta$. However, most of the proposed methods can only deal with simple cases where $Q_{\theta}$ is a linear function of the parameter $\theta$, or they require the second-order information, i.e.~the Hessian matrix, of $Q_{\theta}$ \cite{GradientTDGeneral}, which is computationally intensive for deep neural networks and is therefore difficult to use in deep reinforcement learning. Reference \cite{CharacterizingDivergenceInTD} considered the divergence of the Bellman error in Eq.~(\ref{Bellman error}) in Q-learning, and proposed using natural gradient descent to overcome the problem; however, it still requires second-order information and is difficult to apply. So far there have been no convergent methods that have been demonstrated to be successful on complicated tasks using deep neural networks.

\section{Inefficiency of Conventional Approaches}\label{inefficiency of RG}
In this section, we show that even in cases where the underlying task is deterministic, it is inefficient to use the loss function ${\ell}_{\textit{MSBE}}$ to learn, and therefore, the reason why the RG algorithm and the related convergent methods have not been successful so far is that the strategy of minimizing the loss is intrinsically inefficient, and the double sampling issue is not the essential reason for failure. In the following, we use simple deterministic tabular problems as toy tasks, in which case the methods in Ref.~\cite{GradientTDFirst, GradientTDLinear, GradientTDGeneral, SBEEDResidualLearning, GradientTDbyKernelToCorrelateData, GradientTDwithRegularization} all reduce to the RG algorithm ideally, and we show that the RG algorithm actually fails to learn the toy tasks efficiently.

\subsection{Ill-Conditionedness of the Loss Function}
Consider a tabular problem, where the space of state-action pairs $(s_t,a_t)$ is small and the state-action pairs are enumerable. The Q function value of each state-action pair can be written down into a table and treated as a single scalar parameter of the function $Q$. Suppose that the agent starts from state $s_0$ and interacts with the environment until it encounters a terminal state $s_N$ where the process terminates, producing a sequence of experience data $\{(s_t,a_t,r_t,s_{t+1})\}_{t=0}^{N-1}$, following the greedy policy, i.e., $a_t=\argmax_{a'}Q(s_t,a')$. Then, we consider using reinforcement learning to learn from this sequence of data. Setting the discount factor $\gamma$ to be $1$, the MSBE loss is given by
\begin{equation}\label{trajectory MSBE}
	L_{\textit{MSBE}}:=\frac{\sum_t\ell_{\textit{MSBE}}}{N}=\frac{1}{N}\left[\sum_{t=0}^{N-2}\left(Q(s_t,a_t) - r_t - Q(s_{t+1},a_{t+1})\right)^2 + \left(Q(s_{N-1},a_{N-1}) - r_{N-1} \right)^2 \right],
\end{equation}
where $Q(s_N,\cdot)$ is equal to zero and therefore omitted, as $s_{N}$ is a terminal state. Although Eq.~(\ref{trajectory MSBE}) has a simple quadratic form, if we regard $Q(s_t,a_t)$ for each $t$ as a single parameter and denote it by $Q_{t}$, the Hessian matrix of $L_{\textit{MSBE}}$ with respect to parameters $\{Q_t\}_{t=0}^{N-1}$ is ill-conditioned, and the loss $L_{\textit{MSBE}}$ cannot be optimized efficiently using gradient descent methods. The condition number $\kappa$ of the Hessian matrix is given by $\kappa=\frac{\left |\lambda_\text{max}\right |}{\left |\lambda_\text{min}\right |}$, where $\lambda_\text{max}$ and $\lambda_\text{min}$ denote the largest and the smallest eigenvalues of the Hessian matrix. When $\kappa$ is exceedingly large, the loss function is said to be ill-conditioned and the curvature of the loss function is sharp in the parameter space, and it is difficult for gradient descent methods to make progress in optimizing the loss. We find that the condition number $\kappa$ of the Hessian of $L_{\textit{MSBE}}$ in Eq.~(\ref{trajectory MSBE}) grows approximately as $O(N^2)$. To simplify $L_{\textit{MSBE}}$ and derive an analytic expression of $\kappa$, we add an additional term $Q_0^2$ to $L_{\textit{MSBE}}$, so that it becomes
\begin{equation}\label{expansion of ill-conditioned loss function}
	\begin{split}
		L_{\textit{MSBE}}'=\frac{1}{N}[Q_0^2&+\sum_{t=0}^{N-2}\left( Q_t - r_t - Q_{t+1}\right)^2 + \left(Q_{N-1} - r_{N-1} \right)^2 ]\\
		=\frac{1}{N}[Q_0^2&+\sum_{t=0}^{N-2}\left( Q_t^2 - 2 r_t Q_t + 2 r_tQ_{t+1}- 2Q_tQ_{t+1} + r_t^2+ Q_{t+1}^2\right) \\
		& + \left(Q_{N-1}^2 - 2 r_{N-1}Q_{N-1} + r_{N-1}^2\right) ].
	\end{split}
\end{equation}
Then, straightforwardly we have
\begin{equation}
	\begin{split}
	\frac{\partial^2 L_{\textit{MSBE}}'}{\partial Q_t^2}=\frac{4}{N},\qquad t\in\{0,1,..., N-1\},\\
	\frac{\partial^2 L_{\textit{MSBE}}'}{\partial Q_t\,\partial Q_{t+1}}=-\frac{2}{N},\qquad t\in\{0,1,..., N-2\},
\end{split}
\end{equation}
and the other second-order derivatives of $L_{\textit{MSBE}}'$ are zero. The Hessian matrix of $L_{\textit{MSBE}}'$ is given by
\begin{equation}\label{trajectory hessian matrix}
	H_{1}=\frac{1}{N}\begin{pmatrix}
		4 & -2 &  &\\
		-2 & 4 & \ddots & \\
		& \ddots & \ddots & -2\\
		&  & -2 & 4\\
	\end{pmatrix}_{N\times N}\ .
\end{equation}
The eigenvectors and eigenvalues of this matrix can be explicitly found due to its special structure. First, we have
\begin{equation}\label{hessian matrix simplification}
	N\cdot H_{1}= 4I-2H',\qquad H':=\begin{pmatrix}
		0 & 1 &  &\\
		1 & 0 & \ddots & \\
		& \ddots & \ddots & 1\\
		&  & 1 & 0\\
	\end{pmatrix}_{N\times N}\ ,
\end{equation}
where $I$ is the identity matrix. Then, it suffices to find the eigenvectors and eigenvalues of $H'$. The eigenvectors of $H'$ are known to have the form of standing waves, given by 
\begin{equation}\label{eigenvector 1}
	\left (\sin\frac{k\pi}{N+1}, \sin\frac{2k\pi}{N+1},... , \sin\frac{Nk\pi}{N+1}\right )^T \quad \text{ for } k\in\{1,2,... ,N\},
\end{equation}
which can be confirmed through the trigonometric identity
\begin{equation}
	\sin\frac{(i-1)k\pi}{N+1}+\sin\frac{(i+1)k\pi}{N+1}=\sin\frac{ik\pi}{N+1}\cdot\left(2\cos\frac{k\pi}{N+1}\right) 
\end{equation}
and
\begin{equation}
	\sin\frac{0\cdot k\pi}{N+1}=\sin\frac{(N+1)k\pi}{N+1}=0.
\end{equation}
Therefore for an eigenvector in Eq.~(\ref{eigenvector 1}), after being multiplied by $H'$, each element of the vector becomes effectively multiplied by $2\cos\frac{k\pi}{N+1}$, and the eigenvalue is thus equal to $2\cos\frac{k\pi}{N+1}$. Therefore, using Eq.~(\ref{hessian matrix simplification}), the eigenvalue of the Hessian matrix $H_{1}$ is given by
\begin{equation}\label{eigenvalue 1}
	\lambda_k=\frac{4-4\cos\frac{k\pi}{N+1}}{N},
\end{equation}
and the condition number is given by
\begin{equation}\label{condition number 1}
	\kappa = \frac{\left |\lambda_\text{max}\right |}{\left |\lambda_\text{min}\right |} = \frac{1-\cos\frac{N\pi}{N+1}}{1-\cos\frac{\pi}{N+1}}.
\end{equation}
When $N$ is large, we have
\begin{equation}
	\kappa\approx\frac{1-\left (\cos\pi+\frac{1}{2}\left (\frac{\pi}{N+1}\right )^{2}\right )}{1-\left(\cos0 -\frac{1}{2}\left (\frac{\pi}{N+1}\right )^{2}\right) }\approx\frac{4\left (N+1\right )^{2}}{\pi^{2}}\sim O(N^2),
\end{equation}
that is, the condition number grows quadratically with respect to the length $N$ of the data sequence. As the data sequence is obtained by the agent interacting with the environment until termination, the length of the sequence, which is the total number $N$ of time steps in the environment, can be regarded as the size of the underlying problem of the reinforcement learning task. Therefore, we can say that the condition number $\kappa$ grows quadratically with respect to the size of the reinforcement learning problem in terms of the number of time steps. In practical problems, we typically have $N\sim300$ to $10000$ and $\kappa>10^{4}$.

The the ill-conditioned property of the loss has several important implications. Because the convergence rate of gradient descent is known to be $O(\kappa)$\footnote{Although batch gradient descent (i.e., to compute the gradient using the whole dataset at each iteration) can be accelerated up to a convergence rate of $O(\sqrt{\kappa})$ using the momentum strategy, the same does not apply to the commonly used stochastic gradient descent, because the strategy involves a large momentum factor $(\frac{\sqrt{\kappa}-1}{\sqrt{\kappa}+1})^2$ \cite{PolyakMomentum} which would dramatically increase the noise in stochastic gradient descent.} \cite{PolyakMomentum}, the property of the condition number $\kappa$ implies that the required learning time for the RG algorithm is quadratic in the problem size, i.e., $O(N^2)$, in terms of computational cost. In contrast, because Q-learning is based on the learning rule
\begin{equation}
	Q(s_t,a_t)\gets r_t+\gamma \max_{a'}Q(s_{t+1},a'),
\end{equation}
which straightforwardly propagates information from state $s_{t+1}$ to $s_{t}$, it is clear that Q-learning only requires $O(N)$ time to converge, as there are totally $N$ different states from $s_0$ to $s_{N-1}$. Therefore, the RG algorithm is considerably less computationally efficient compared with Q-learning. 

As another consequence of the ill-conditioned property, a small loss $L_{\textit{MSBE}}$ does not necessarily imply a short distance between the learned Q function $Q_{\theta}$ and the optimal Q function $Q^*$ which is the target of learning. This may explain why it has been observed that $L_{\textit{MSBE}}$ fails to be a useful indicator of performance \cite{IsBellmanResidualABadProxy}. 

So far we have dealt with the case of $\gamma=1$. For $\gamma<1$, suppose that the states $\{s_t\}_{t=0}^{N-1}$ form a cycle, i.e., $s_{N-1}$ makes a transition to $s_0$, the loss is given by
\begin{equation}
	L_{\textit{MSBE}}=\frac{1}{N}\left [\sum_{t=0}^{N-2}\left(Q_t - r_t - \gamma Q_{t+1}\right)^2 + \left(Q_{N-1} - r_{N-1} - \gamma Q_{0} \right)^2 \right ].
\end{equation}
The second-order derivatives are given by
\begin{equation}
	\begin{split}
		\frac{\partial^2 L_{\textit{MSBE}}}{\partial Q_t^2}=2+2\gamma^2,\qquad &t\in\{0,1,... , N-1\},\\
		\frac{\partial^2 L_{\textit{MSBE}}}{\partial Q_t\partial Q_{t+1}}=-2\gamma,\qquad &t\in\{0,1,... , N-1\},\qquad Q_N\equiv Q_0,
	\end{split}
\end{equation}
and the Hessian matrix of the loss is cyclic. Assuming that $N$ is even, the eigenvectors are of the form of periodic waves, given by
\begin{equation}
	\begin{split}
	&\left (\sin\frac{2k\pi}{N}, \sin\frac{4k\pi}{N},... , \sin\frac{2Nk\pi}{N}\right )^T \text{ and } \\
&\left (\cos\frac{2k\pi}{N}, \cos\frac{4k\pi}{N},... ,\cos\frac{2Nk\pi}{N}\right )^T \text{ for } k\in\left \{1,2,... ,\frac{N}{2}\right \},
\end{split}
\end{equation}
with the corresponding eigenvalues given by 
\begin{equation}
	\lambda_k = 2(1+\gamma^2)-4\gamma\cos\frac{2k\pi}{N}.
\end{equation}
This result can be confirmed by following a similar argument and using the trigonometric relations
\begin{equation}
	\sin\frac{2(i-1)k\pi}{N}+\sin\frac{2(i+1)k\pi}{N}=\sin\frac{2ik\pi}{N}\cdot\left (2\cos\frac{2k\pi}{N}\right )
\end{equation}
and 
\begin{equation}
	\cos\frac{2(i-1)k\pi}{N}+\cos\frac{2(i+1)k\pi}{N}=\cos\frac{2ik\pi}{N}\cdot\left (2\cos\frac{2k\pi}{N}\right ).
\end{equation}
For the condition number $\kappa$, at the limit of large $N$, we have
\begin{equation}\label{condition number 2}
	\kappa=\frac{2(1+\gamma^2)-4\gamma\cos\pi}{2(1+\gamma^2)-4\gamma\cos\frac{2\pi}{N}}\approx\frac{2(1+\gamma^2)+4\gamma}{2(1+\gamma^2)-4\gamma}= \frac{(1+\gamma)^2}{(1-\gamma)^2}.
\end{equation}
Using $\gamma\approx1$, we have 
\begin{equation}
	\kappa\approx\frac{4}{\left(1-\gamma\right)^2}\sim O\left (\frac{1}{\left(1-\gamma\right)^2} \right ).
\end{equation}
As $\dfrac{1}{1-\gamma}$ is the time horizon of the planning of the agent, the term $\dfrac{1}{1-\gamma}$ may be regarded as an effective size of the reinforcement learning problem, and therefore, we see that $\kappa$ is still quadratic in the size of the problem. Usually, we have $\gamma\gtrsim0.99$ and $\kappa>10^{4}$, and we conclude that the loss $L_{\textit{MSBE}}$ is ill-conditioned. 

\begin{figure}[tb!]
	\centering
	\includegraphics[width=0.45\linewidth]{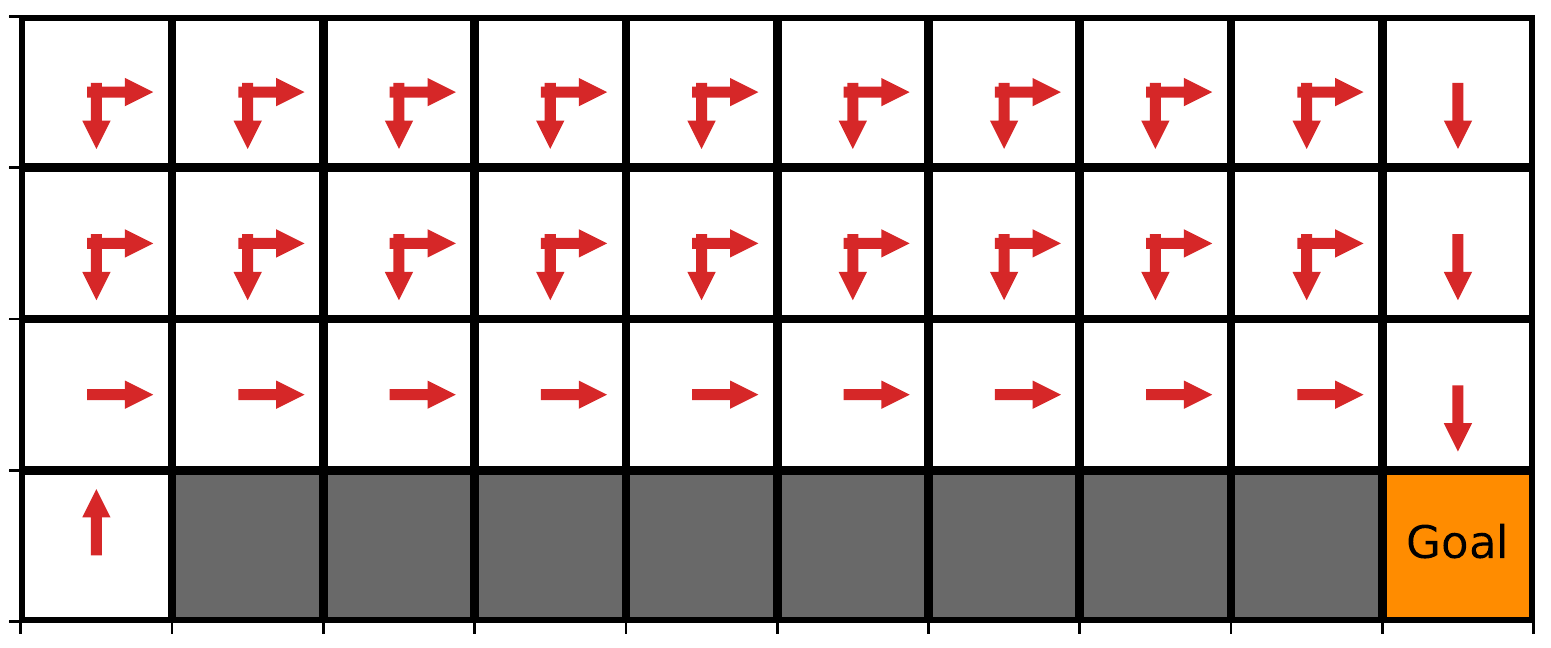}
	\caption{\label{cliff walking 1}The cliff walking task. The agent starts at the lower left corner and is supposed to move to the goal at the lower right corner as quickly as possible, while bypassing the grey region. The red arrows indicate the optimal policy in this problem at each state. The system shown here has the height of 4 and the width of 10.}
\end{figure}

\paragraph{Cliff walking}By way of illustration, we consider the \textit{cliff walking} problem in Ref.~\cite{ReinforcementlearningAnintroduction} as a toy task, which is a tabular problem, illustrated in Fig.~\ref{cliff walking 1}. In this system, the agent starts at the block in the lower left corner of the grid, which is the initial state, and the agent is allowed to move to nearby blocks. When it moves to a white block, it obtains a reward of $-1$; when it moves to a grey block which represents the cliff, it obtains a reward of $-100$ and the process terminates; when it moves to the goal at the lower right corner, the process terminates with a reward of zero. The optimal policy is therefore to move to the lower right corner as soon as possible while bypassing the cliff. 

\begin{figure}[tb!]
	\centering
	\begin{tabular}[b]{c}
		\includegraphics[width=0.73\linewidth]{"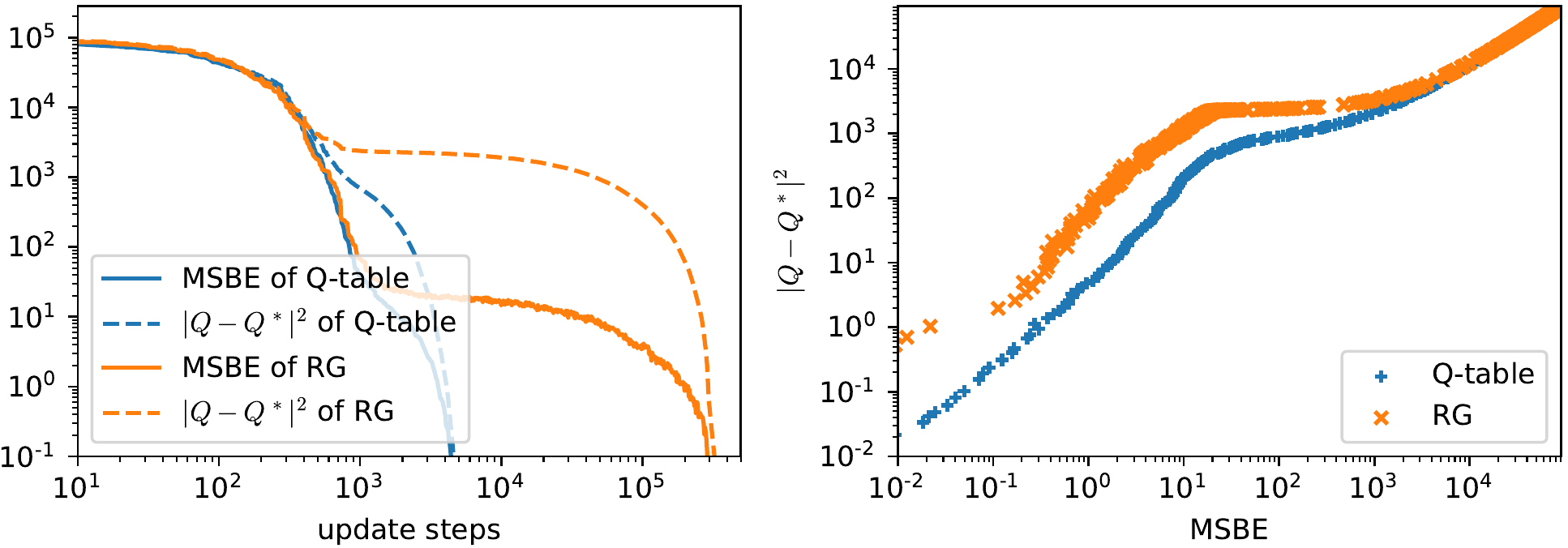"}\\
		\includegraphics[width=0.73\linewidth]{"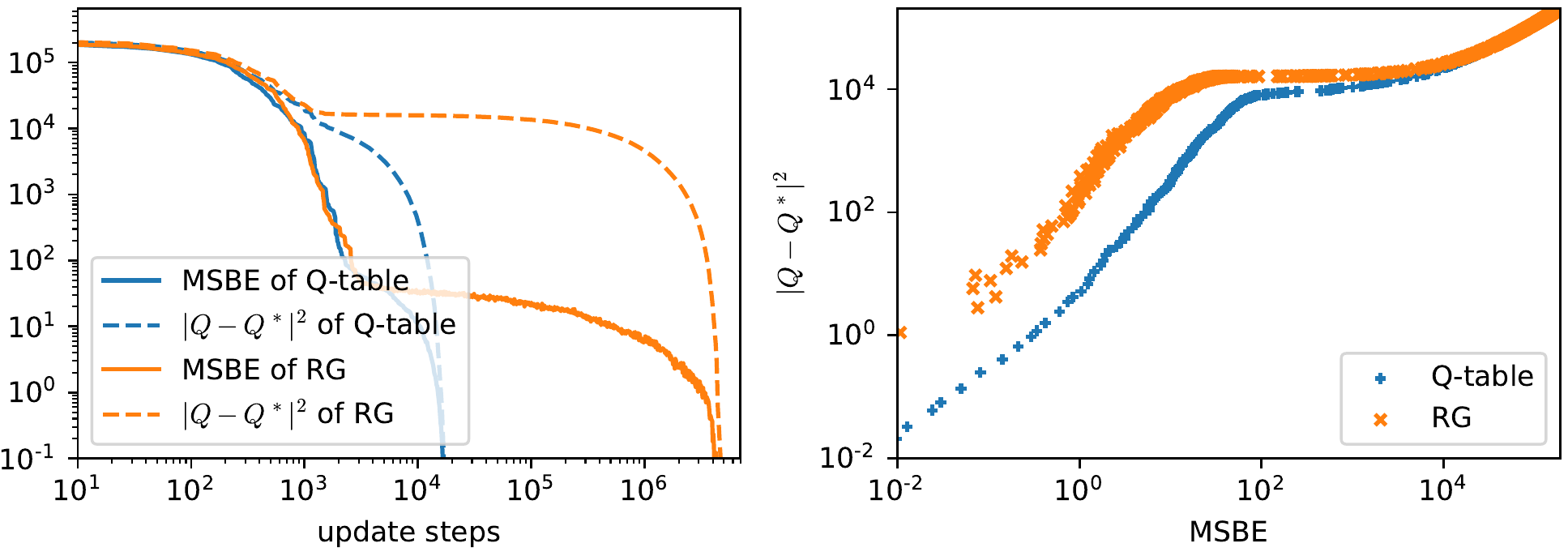"}
	\end{tabular}
	\caption{\label{cliff walking ill-conditionness}Results of training on the cliff walking task, learning from a randomly sampled data in the state-action space at each iteration. The upper two panels show the results of the system with width 10 and $\gamma=0.9$, and the lower two show the results of the system with width 20 and $\gamma=0.95$. $|Q-Q^*|^2$ is the squared distance between the learned Q function and the optimal Q function $Q^*$,\protect\footnotemark \ and MSBE represents the value of the loss function $\sum \ell_{\textit{MSBE}}$. The plots on the left show the progress of learning regarding the number of iteration steps, and the plots on the right show the relation between $|Q-Q^*|^2$ and $\sum \ell_{\textit{MSBE}}$ using the data obtained in the training process. The learning rate is set to be $0.5$, and the data are averaged over 10 repetitions of the same experiments.}
\end{figure}%
To learn the cliff walking task, first, we treat the Q function value for every state-action pair to be a learnable parameter and initialize them to be zero. For each iteration, we randomly choose a state-action pair $(s_t,a_t)$ and find its next state $s_{t+1}$ and its reward $r_t$, and then, we use the obtained sample $(s_t,a_t, r_t, s_{t+1})$ to update the Q function $Q(s_t,a_t)$ via Eq.~(\ref{Q table learning}), which corresponds to Q-table learning, or to minimize the associated loss $\ell_{\textit{MSBE}}$ following the gradient, which corresponds to the RG algorithm. As shown in the plots on the left in Fig.~\ref{cliff walking ill-conditionness}, the RG algorithm learns considerably more slowly than Q-table learning. In addition, as shown in the plots on the right in Fig.~\ref{cliff walking ill-conditionness},\footnotetext{We define $\left |Q-Q^*\right |^2$ by $\sum_{(s,a)}\left |Q(s,a)-Q^*(s,a)\right |^2$, where the summation is taken over all possible state-action pairs $(s,a)$ in the system.} for a fixed value of $L_{\textit{MSBE}}$, the Q function learned by the RG algorithm has a larger distance to the optimal solution $Q^*$ compared with that of Q-table learning, which implies that Q-table learning actually approaches the optimal solution $Q^*$ more efficiently. Note that the data in Fig.~\ref{cliff walking ill-conditionness} are plotted in a log-log scale.

To investigate how the size of the problem affects the behaviour of Q-table learning and the RG algorithm, we consider two different sizes of the cliff walking system. We consider a width of 10 of the system with $\gamma=0.9$, and a width of 20 of the system with $\gamma=0.95$. The size of the second problem is therefore twice the first one, and the results are shown respectively in the upper and in the lower plots in Fig.~\ref{cliff walking ill-conditionness}. Note that the agent learns from randomly sampled data in the state-action space and doubling the size of the system also reduces the sampling efficiency by half. Therefore, as the learning time of Q-learning is linear $O(N)$ and that of RG is quadratic $O(N^{2})$ in the system size without consideration of the sampling, when the problem size is doubled, their learning time should respectively become 4 times and 8 times as long as the previous one. It can be confirmed in Fig.~\ref{cliff walking ill-conditionness} that the experimental results are consistent with this prediction, and therefore the results support our analysis above concerning the scaling properties of the learning algorithms.

\subsection{Tendency of Maintaining the Average Q Value}\label{learning behaviour problem of average}
In the previous section, we considered the case where the learning algorithm learns from data that are collected through random sampling. However, in real cases, the data used for learning are typically collected by the agent using its incompletely learned policy, which can lead to additional difficulties. In the following we show that due to the policy learned by the RG algorithm, besides the ill-conditionedness issue, the RG algorithm has other problems which can also lead to ill-behaved learning dynamics and failure of learning. 

To demonstrate the problems, first we denote $Q_t\equiv Q(s_t,a_t)$ and $Q_{t+1}\equiv \max_{a'}Q(s_{t+1},a')$. Suppose that they initially satisfy $Q_t=\gamma Q_{t+1}$, when a transition from state $s_t$ to $s_{t+1}$ with a non-zero reward $r_t$ is observed, the Q-table learning rule in Eq.~(\ref{Q-table learning (CDQN)}) leads to $\Delta Q_t = r_t$ and $\Delta Q_{t+1}=0$, and therefore we have \begin{equation}
	\Delta\left(Q_t+Q_{t+1} \right)= r_t.
\end{equation}
However, when the RG algorithm is used, the minimization of $L_{\textit{MSBE}}$ following the gradient leads to the following learning rule
\begin{equation}\label{RG learning rule}
	\begin{split}
		\Delta Q(s_t,a_t)  &= \alpha \left (r_t+\gamma \max_{a'}Q(s_{t+1},a') -Q(s_t,a_t)\right ),\\
		\Delta \max_{a'}Q(s_{t+1},a') &= -\gamma \alpha \left (r_t+\gamma \max_{a'}Q(s_{t+1},a') -Q(s_t,a_t)\right ),\\
		\Delta \max_{a'}Q(s_{t+1},a') &\equiv -\gamma \Delta Q(s_t,a_t),
	\end{split}
\end{equation}
and therefore whenever $Q_t$ is changed, $Q_{t+1}$ changes simultaneously by a similar amount, which leads to
\begin{equation}\label{RG learning total change}
	\Delta\left(Q_t+Q_{t+1} \right)= (1-\gamma) r_t.
\end{equation}
Because we have $\gamma\approx1$, $\Delta\left(Q_t+Q_{t+1} \right)$ is usually close to zero, and therefore, the sum of all the Q function values, i.e.~$\sum_{(s,a)}Q(s,a)$, almost does not change, which is different from the case of Q-learning. This can happen because there exists an additional degree of freedom when one tries to modify $Q_t$ and $Q_{t+1}$ in order to satisfy the Bellman equation: Q-learning keeps $Q_{t+1}$ fixed, while the RG algorithm keeps $Q_t+\frac{1}{\gamma} Q_{t+1}$ fixed, except for the case where $s_{t+1}$ is a terminal state and thus $Q_{t+1}$ is constantly zero. This tendency of maintaining the average Q function value has important consequences on the learning behaviour of the RG algorithm as discussed in the following.

\subsubsection{Inefficiency Caused by Loops of States}
\begin{figure}[tb]
	\centering
	\includegraphics[width=0.5\linewidth]{"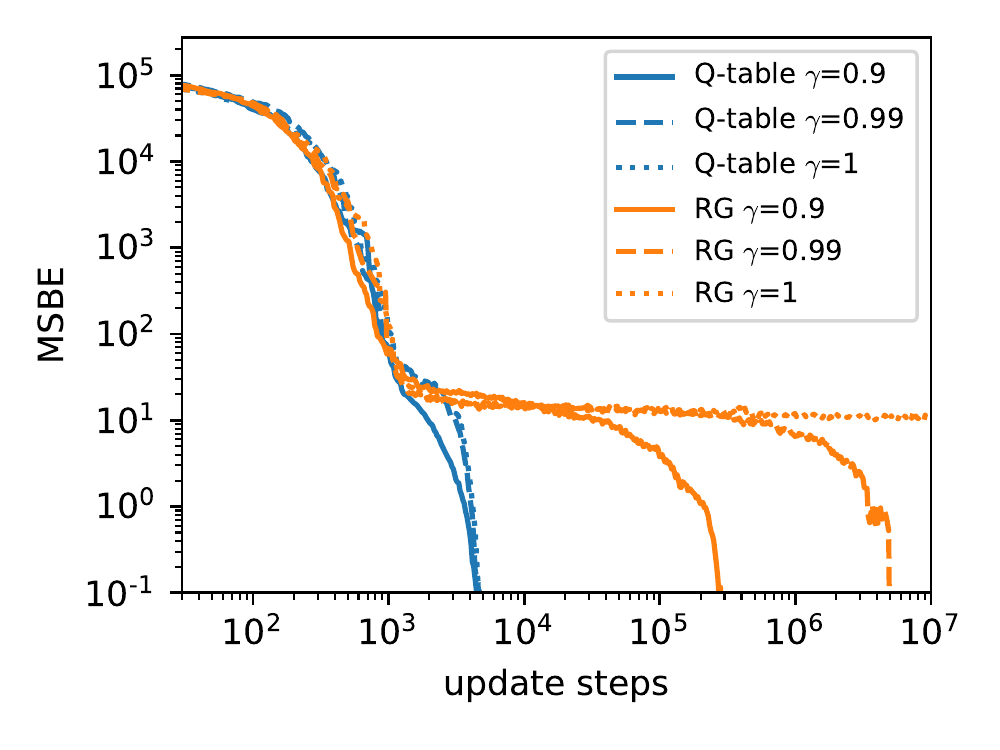"}
	\caption{\label{cliff walking gamma dependence}Results of the cliff walking task in Fig.~\ref{cliff walking ill-conditionness} with different discount factors $\gamma$, with a width of 10 of the system. The data are averaged over 10 repetitions of the experiment. The RG algorithm with $\gamma=1$ does not  converge or continuously makes progress in learning.}
\end{figure}

When the Q function is initialized, if the sum $\sum_{(s,a)}Q(s,a)$ is larger than that of the optimal Q function $\sum_{(s,a)}Q^*(s,a)$, and if it is possible for the transitions among the states to form loops, the learning time of the RG algorithm has an additional scale factor of $\frac{1}{1-\gamma}$ due to Eq.~(\ref{RG learning total change}), even when the size of the problem is finite and much smaller than $\frac{1}{1-\gamma}$. This is because the policy $\argmax_{a'}Q(s,a')$ is likely to make the agent goes into loops, as the transitions following the optimal policy are associated with Q values smaller than what the agent expects, and the agent will tend to stay in the states for which it predicts a high Q value and therefore moves into loops of the states. Then, as terminal states are hardly reached, Eq.~(\ref{RG learning total change}) controls the learning of the sum of $Q$ values, i.e.~$\sum_{(s,a)}Q(s,a)$. The learning target of $\sum_{(s,a)}Q(s,a)$ is $\sum_{(s,a)}Q^*(s,a)$ and the learning time scales as $O (\frac{1}{1-\gamma} )$ according to Eq.~(\ref{RG learning total change}). As shown in Fig.~\ref{cliff walking gamma dependence}, for the cliff walking problem with width 10, the learning time of Q-table learning is roughly the same for different values of $\gamma$, while the time of the RG algorithm scales approximately as $O (\frac{1}{1-\gamma} )$ and it does not learn for $\gamma=1$. This is because given $\gamma=1$, the change of the sum $\sum_{(s,a)}Q_{(s,a)}$ is constantly zero and $\sum_{(s,a)}Q^*(s,a)$ cannot be learned, which leads to failure of learning.

\subsubsection{Deterioration of Policy}
\begin{figure}[tb]
	\centering
	\begin{minipage}[c]{0.4\linewidth}
		\centering
		\subfloat{
			\includegraphics[width=\linewidth]{"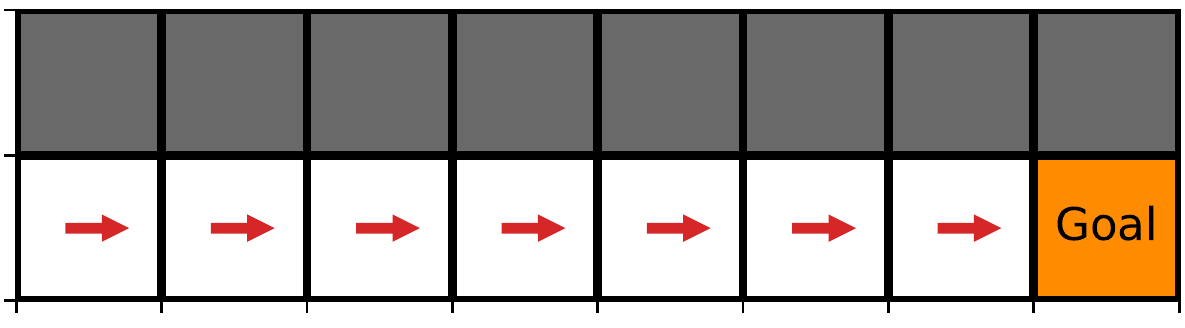"}
		}
	\end{minipage}\ \ 
	\begin{minipage}[c]{0.5\linewidth}
		\centering
		\subfloat{
			\includegraphics[width=\linewidth]{"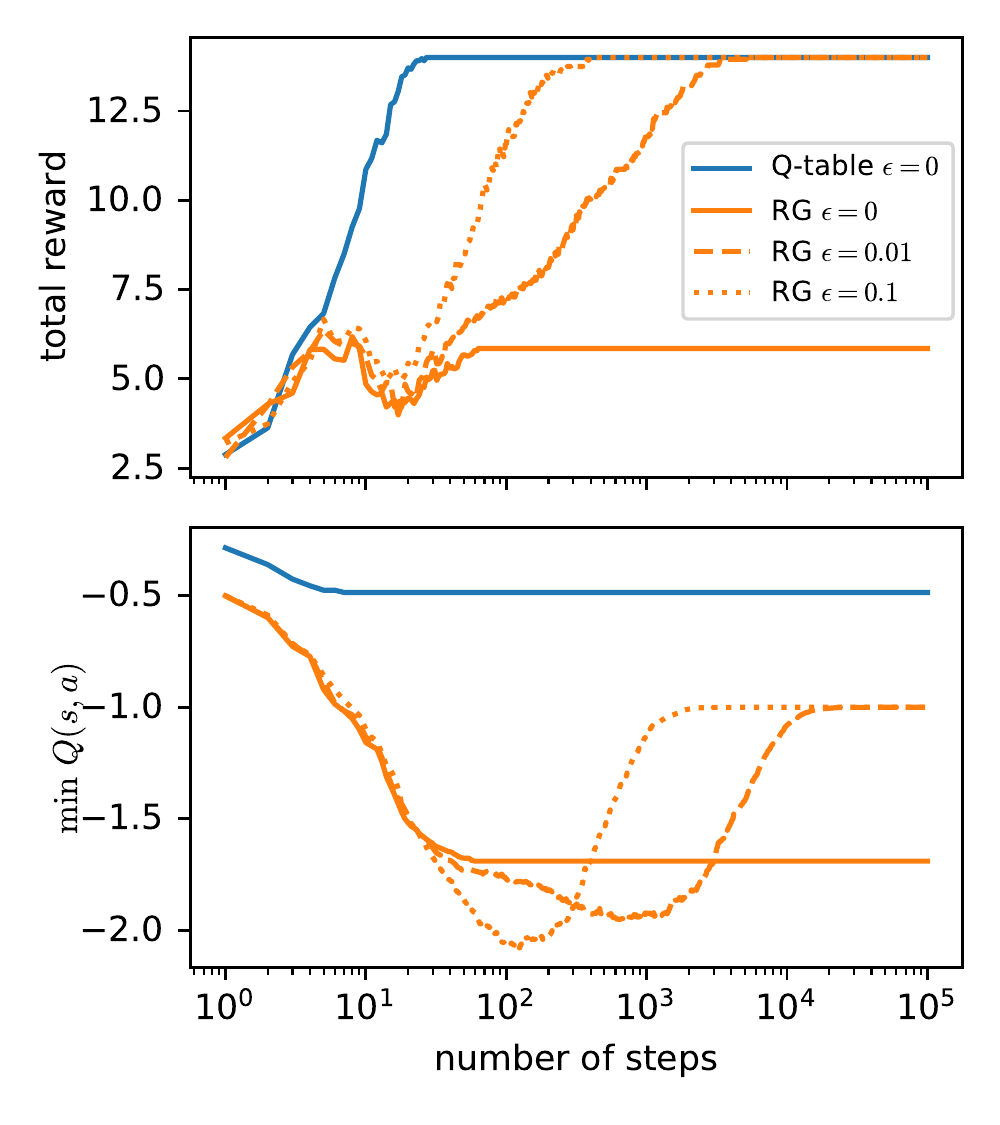"} 
		}
	\end{minipage}
	\caption{Left: one-way cliff walking problem. The agent starts from the state in the lower left corner, and at each step it has two choices of action: it can move to the right and obtain a reward of $2$, or move upwards into the grey block representing the cliff and terminate the process with a reward of $-1$. The process terminates upon reaching the goal in the lower right corner. Right: performance of the learned policy $\argmax_{a'}Q(s,a')$ (top) and the minimum of the learned Q function values $\min\ Q(s,a)$ (bottom) for online Q-table learning and the RG algorithm, following the $\epsilon$-greedy policy with different values of $\epsilon$ in learning, using $\gamma=1$ and a learning rate of $0.5$. Note that the plots are in a log-log scale. The data are obtained by averaging over 100 repetitions of the experiment.}
	\label{one-way cliff walking}
\end{figure}

In practice, the data that the agent learns from are usually collected through interactions with the environment, following the policy $\argmax_{a'}Q(s,a')$ learned by the agent. In such cases, a more common failure mode of the RG algorithm appears if the Q function is initialized to be smaller than $Q^*$, in which case the agent is only able to obtain a small amount of reward and fails to continuously improve its performance. In Fig.~\ref{one-way cliff walking}, we show a typical example of this failure mode, where the Q function is initialized to be zero and the agent learns the observed transitions $(s_t,a_t,r_t,s_{t+1})$ in an online manner following the $\epsilon$-greedy policy\footnote{$\epsilon$-greedy means that with probability $\epsilon$ a random action is used; otherwise $\argmax_{a'}Q(s,a')$ is used.}, i.e., for every step, the agent takes an action $a_t$ at a state $s_t$, receives reward $r_t$ and moves to the next state $s_{t+1}$, and uses the data $(s_t,a_t,r_t,s_{t+1})$ to perform a single iteration of the learning algorithm. In Fig.~\ref{one-way cliff walking}, we see that although Q-table learning finds the optimal policy easily without using a non-zero $\epsilon$, the RG algorithm is not able to find the optimal policy with $\epsilon=0$, and it learns slowly and its speed of learning crucially relies on the value of $\epsilon$. 

The failure of the RG algorithm in Fig.~\ref{one-way cliff walking} occurs because when the RG algorithm learns and the learned Q function values increase for some states, the Q function values must simultaneously decrease for other states, which can cause some of the learned Q values to decrease to a negative number even when no negative reward is observed. When the negative Q values are even smaller than the reward associated with the terminal states, for example, $-1$ in the problem in Fig.~\ref{one-way cliff walking}, the agent regards termination as the action that can bring a higher reward and follows the policy $\argmax_{a'}Q(s,a')$ to choose to terminate. As shown in the lower right panel in Fig.~\ref{one-way cliff walking}, the minimum of the Q values learned by the RG algorithm can indeed decrease to a level that is smaller than the lowest possible reward $-1$ in the system. Then, following the policy $\argmax_{a'}Q(s,a')$, it will always choose to go to the cliff and terminate the process to obtain a reward of $-1$, and as a result, it will fail to explore other possibilities of actions and fail to collect meaningful data, i.e.~the data of moving to the right, to continue its learning. Therefore, its progress of learning crucially relies on the exploration strategy involved, so that diverse data can be collected to allow the agent to learn to correct its behaviour at the states with exceedingly small Q values. In the upper right panel in Fig.~\ref{one-way cliff walking}, it can be seen that the agent learning with $\epsilon=0.1$ indeed makes progress approximately 10 times as fast as the agent learning with $\epsilon=0.01$, showing the importance of exploration. Generally, whenever an unexpected positive reward $r_t$ is encountered and learned by the RG algorithm, according to Eq.~(\ref{RG learning rule}), with an increase in $Q(s_t,a_t)$, $\max_{a'}Q(s_{t+1},a')$ decreases simultaneously by approximately the same amount, and the policy at state $s_{t+1}$, i.e., $\argmax_{a'}Q(s_{t+1},a')$, becomes perturbed, which can cause the agent to choose a worse action at $s_{t+1}$ that results in less reward, and therefore the performance, which is the total reward obtained, may not improve on the whole. This explains the reason why in practice, the performance of the RG algorithm often stays at a low level and does not improve significantly throughout learning. The performance of the RG algorithm therefore relies heavily on the exploration strategy to rediscover the appropriate action at state $s_{t+1}$. However, in practical situations, it is difficult to have efficient exploration, and usually one cannot enhance the exploration without compromising the performance, especially for large-scale problems. The RG algorithm thus encounters difficulties and performs significantly worse than Q-learning for complicated and realistic tasks.

\paragraph{Remark}Although our analysis above have only focused on tabular problems where every Q function value is an independent parameter, in general, the situation is not supposed to be better when function approximations are used. Our arguments explain why most of the successful examples of existing gradient-based convergent methods have tunable hyperparameters that can be used to reduce the methods to conventional TD learning, which involves an update rule that is similar to Q-learning \cite{GradientTDLinear, GradientTDwithRegularization, GradientTreeBackup, SBEEDResidualLearning}. When the methods becomes similar to conventional TD learning without encountering instability issues, the methods usually obtain better efficiency and better quality of policy; when the methods only directly use the gradient of the loss function to learn, the performance can be much worse. This may also explain the reason why the performance of the PCL algorithm \cite{PCL} can deteriorate when the value and the policy neural networks are combined, and the reason why the performance of PCL can be improved with the use of a target network \cite{NAC}. Note that even though the ill-conditionedness issue may be resolved through the use of a second-order optimizer or the Retrace loss \cite{RetraceOperator, Agent57}, the problem in Sec.~\ref{learning behaviour problem of average} cannot be solved, because the optimizer will likely converge to the same solution as the one found by gradient descent, and as a result, the agent will still learn the same policy and have the same learning behaviour. A rigorous and mathematical analysis of the issue discussed in Sec.~\ref{learning behaviour problem of average} is definitely desired, and we leave it for future work.

\section{Convergent Deep Q Network (C-DQN) Algorithm}\label{CDQN}
As discussed above, the learning behaviour of the RG algorithm is significantly worse than conventional Q-learning, and therefore, instead of trying to optimizing the loss $\ell_{\textit{MSBE}}$ or $\ell_{\text{Bellman}}$ directly, we aim to minimally modify the conventional DQN algorithm, so that the algorithm can become convergent but still have learning behaviour similar to Q-learning. 
\subsection{Formulation}
\subsubsection{DQN as Fitted Value Iteration}
We consider the DQN algorithm as fitted value iteration (FVI) \cite{FirstFittedQIteration, FittedValueIteration}. For a transition data $(s_t,a_t,r_t,s_{t+1})$ and a target network $\tilde{\theta}$, the DQN loss of the Q function $Q_{\theta}$ is given by
\begin{equation}\label{DQN loss single CDQN formulation}
	{\ell}_{\textit{DQN}}(\theta;\tilde{\theta}) :=\left(Q_\theta(s_t,a_t)-r_t-\gamma \max_{a'}Q_{\tilde{\theta}}(s_{t+1},a') \right) ^2.
\end{equation}
For a whole dataset $\mathcal{S}$, the DQN loss is given by
\begin{equation}\label{DQN loss dataset}
	\begin{split}
			{L}_{\textit{DQN}}(\theta;\tilde{\theta}) &:=\frac{\sum_{(s_t,a_t,r_t,s_{t+1})} {\ell}_{\textit{DQN}}(\theta;\tilde{\theta}) }{|\mathcal{S}|}\\
			&=\frac{1}{|\mathcal{S}|}\sum_{(s_t,a_t,r_t,s_{t+1})\in\mathcal{S}}\left(Q_\theta(s_t,a_t)-r_t-\gamma \max_{a'}Q_{\tilde{\theta}}(s_{t+1},a') \right) ^2.
	\end{split}
\end{equation}
Or, if the data are considered as randomly sampled, we have
\begin{equation}\label{DQN loss expectation}
		\begin{split}
		{L}_{\textit{DQN}}(\theta;\tilde{\theta}) :={}&\mathbb{E}_{(s_t,a_t,r_t,s_{t+1})}\left [{\ell}_{\textit{DQN}}\right ] \\
		={}&\mathbb{E}_{(s_t,a_t,r_t)}\left [\left(Q_\theta(s_t,a_t)-r_t-\gamma \mathbb{E}_{s_{t+1}|(s_t,a_t)}\left [\max_{a'}Q_{\tilde{\theta}}(s_{t+1},a')\right ] \right) ^2\right . \\
		&\left. + \gamma^2\text{Var}_{s_{t+1}|(s_t,a_t)}[\max_{a'}Q_{\tilde{\theta}}(s_{t+1},a')]\right ],
	\end{split}
\end{equation}
and the gradient $\nabla_{\theta}{L}_{\textit{DQN}}(\theta;\tilde{\theta})$ is given by
\begin{equation}
	\nabla_{\theta}{L}_{\textit{DQN}}(\theta;\tilde{\theta})=2\mathbb{E}_{(s_t,a_t,r_t)}\left[\nabla_{\theta}Q_\theta(s_t,a_t)\left(Q_\theta(s_t,a_t)-r_t-\gamma \mathbb{E}_{s_{t+1}|(s_t,a_t)}\left [\max_{a'}Q_{\tilde{\theta}}(s_{t+1},a')\right ] \right) \right],
\end{equation}
which shows that the gradient of the loss ${L}_{\textit{DQN}}$ indeed corresponds to the learning rule of Q-learning (see Eq.~(\ref{Q table learning})). The conventional DQN algorithm proceeds by performing the iteration 
\begin{equation}\label{DQN update}
	\Delta\theta =-\alpha \nabla_{\theta}L_{\textit{DQN}}(\theta;\tilde{\theta})
\end{equation}
for $N$ times, where $\alpha$ is the learning rate, and performing the update 
\begin{equation}\label{DQN update of target network}
	\tilde{\theta}\gets \theta
\end{equation}
once, and then, repeating the process. Here, $N$ is called the update period of the target network, and we usually have $N\sim 1000$. If we regard the update in Eq.~(\ref{DQN update}) as approximately looking for the minimum of $L_{\textit{DQN}}$, the DQN algorithm can be formulated as 
\begin{equation}
	\begin{split}
			&{\theta}\gets\argmin_\theta L_{\textit{DQN}}(\theta;\tilde{\theta}_i),\\
		&\tilde{\theta}_{i+1} \gets \theta,
	\end{split}
\end{equation} 
or simply
\begin{equation}
	{\tilde{\theta}_{i+1}}=\argmin_\theta L_{\textit{DQN}}(\theta;\tilde{\theta}_i),
\end{equation}
where $\tilde{\theta}_i$ denotes the parameter of the target network at the $i$-update. The parameter $\tilde{\theta}_i$ for a sufficiently large $i$ is therefore the parameter of the fully trained neural network. When the DQN algorithm diverges, the loss $L_{\textit{DQN}}(\theta;\tilde{\theta}_i)$ diverges with $i$, which means that we have $\min_\theta L_{\textit{DQN}}(\theta;\tilde{\theta}_{i+1})>\min_\theta L_{\textit{DQN}}(\theta;\tilde{\theta}_{i})$ for some $i$, that is, the minimization of $L_{\textit{DQN}}(\theta;\tilde{\theta}_{i})$ with respect to $\theta$ eventually increases the loss at the next update, i.e.~$L_{\textit{DQN}}(\theta;\tilde{\theta}_{i+1})$.

\subsubsection{Constructing a Non-Increasing Series of Loss}\label{CDQN proof}
Corresponding to the MSBE loss $\ell_{\textit{MSBE}}(\theta)$ for a single transition data $(s_t,a_t,r_t,s_{t+1})$, we define
\begin{equation}\label{MSBE loss dataset}
	{L}_{\textit{MSBE}}(\theta) :=\frac{1}{|\mathcal{S}|}\sum_{(s_t,a_t,r_t,s_{t+1})\in\mathcal{S}}\left(Q_\theta(s_t,a_t)-r_t-\gamma \max_{a'}Q_{\theta}(s_{t+1},a') \right) ^2,
\end{equation}
for a dataset $\mathcal{S}$, or
\begin{equation}
	{L}_{\textit{MSBE}}(\theta) = \mathbb{E}_{(s_t,a_t,r_t,s_{t+1})}\left [\left(Q_\theta(s_t,a_t)-r_t-\gamma \max_{a'}Q_{\theta}(s_{t+1},a') \right) ^2\right]
\end{equation}
for sampled data. Then, we have the following result:
\begin{theorem}
	The minimum of $L_{\textit{DQN}}(\theta;\tilde{\theta})$ is upper bounded by $L_{\textit{MSBE}}(\tilde{\theta})$.
\end{theorem}
This result can be derived immediately by noticing the important identity
\begin{equation}
	L_{\textit{DQN}}(\tilde{\theta};\tilde{\theta})=L_{\textit{MSBE}}(\tilde{\theta})
\end{equation}
according to Eq.~(\ref{DQN loss dataset}) and (\ref{MSBE loss dataset}), and using
\begin{equation}
	\min_\theta L_{\textit{DQN}}(\theta;\tilde{\theta})\le L_{\textit{DQN}}(\tilde{\theta};\tilde{\theta}),
\end{equation}
which leads to
\begin{equation}
	\min_\theta L_{\textit{DQN}}(\theta;\tilde{\theta})\le L_{\textit{MSBE}}(\tilde{\theta}).
\end{equation} 

Because the minimum of $L_{\textit{DQN}}(\theta;\tilde{\theta})$ is upper bounded by $L_{\textit{MSBE}}(\tilde{\theta})$, when the DQN loss $L_{\textit{DQN}}(\theta;\tilde{\theta}_i)$ diverges with $i$, the MSBE loss $L_{\textit{MSBE}}(\tilde{\theta}_{i})$ also diverges. Specifically, if $L_{\textit{DQN}}(\theta;\tilde{\theta}_i)$ diverges with $i$, although $\tilde{\theta}_{i+1}=\argmin_\theta L_{\textit{DQN}}(\theta;\tilde{\theta}_{i})$ minimizes the $i$-th DQN loss $L_{\textit{DQN}}(\,\cdot\,;\tilde{\theta}_{i})$, $\tilde{\theta}_{i+1}$ also increases the upper bound for the ($i$+1)-th DQN loss, which is $L_{\textit{MSBE}}(\tilde{\theta}_{i+1})$. We therefore want both the current loss $L_{\textit{DQN}}$ and the upper bound $L_{\textit{MSBE}}$ for the loss at the future step to decrease during training, and we define our convergent DQN (C-DQN) loss as
\begin{equation}\label{CDQN loss}
	L_{\textit{CDQN}}(\theta;\tilde{\theta}_i) := \mathbb{E}_{(s_t,a_t,r_t,s_{t+1})}\left[\max\left \{{\ell}_{\textit{DQN}}(\theta;\tilde{\theta}_i), {\ell}_{\textit{MSBE}}(\theta)\right \}\right].
\end{equation}

\begin{theorem}
	The C-DQN loss satisfies $\min_\theta L_{\textit{CDQN}}(\theta;\tilde{\theta}_{i+1})\le \min_\theta L_{\textit{CDQN}}(\theta;\tilde{\theta}_{i})$, with $\tilde{\theta}_{i+1}=\argmin_\theta L_{\textit{CDQN}}(\theta;\tilde{\theta}_i)$.
\end{theorem}
We have
\begin{equation}
	\min_\theta L_{\textit{CDQN}}(\theta;\tilde{\theta}_{i+1})\le L_{\textit{CDQN}}(\tilde{\theta}_{i+1};\tilde{\theta}_{i+1})=L_{\textit{MSBE}}(\tilde{\theta}_{i+1}),
\end{equation}
and
\begin{equation}
	\begin{split}
		L_{\textit{MSBE}}(\tilde{\theta}_{i+1}) &= \mathbb{E}\left[{\ell}_{\textit{MSBE}}(\tilde{\theta}_{i+1})\right]\\
		&\le\mathbb{E}\left[\max\left \{{\ell}_{\textit{DQN}}(\tilde{\theta}_{i+1};\tilde{\theta}_i), {\ell}_{\textit{MSBE}}(\tilde{\theta}_{i+1})\right \}\right]\\
		&=L_{\textit{CDQN}}(\tilde{\theta}_{i+1};\tilde{\theta}_{i})\\
		&= \min_\theta L_{\textit{CDQN}}(\theta;\tilde{\theta}_{i}),
	\end{split}
\end{equation}
which leads to
\begin{equation}
	\min_\theta L_{\textit{CDQN}}(\theta;\tilde{\theta}_{i+1})\le L_{\textit{MSBE}}(\tilde{\theta}_{i+1})\le \min_\theta L_{\textit{CDQN}}(\theta;\tilde{\theta}_{i}).
\end{equation}
Therefore, we obtain the relation $\min_\theta L_{\textit{CDQN}}(\theta;\tilde{\theta}_{i+1})\le \min_\theta L_{\textit{CDQN}}(\theta;\tilde{\theta}_{i})$, which means that the iteration $\tilde{\theta}\gets\argmin_\theta L_{\textit{CDQN}}(\theta;\tilde{\theta})$ is convergent. This convergence holds in the sense that the loss $L_{\textit{CDQN}}(\theta;\tilde{\theta}_{i})$ is both bounded from below and non-increasing with $i$.

When we consider the original formulation of the DQN algorithm as in Eq.~(\ref{DQN update}) and (\ref{DQN update of target network}), where $\tilde{\theta}_{i+1}$ does not exactly minimize $L_{\textit{DQN}}(\,\cdot\, ;\tilde{\theta}_i)$, we still have similar results. At the moment when the target network $\tilde{\theta}$ is updated by $\theta$, we have 
\begin{gather}
		\tilde{\theta}' = \theta,\\
		L_{\textit{CDQN}}(\theta;\tilde{\theta}')= L_{\textit{CDQN}}(\theta;\theta) = L_{\textit{MSBE}}({\theta})\le L_{\textit{CDQN}}(\theta;\tilde{\theta}),
\end{gather}
and therefore, when the target network is updated, the loss is exactly equal to $L_{\textit{MSBE}}(\theta)$, and unlike $L_{\textit{DQN}}$, the convergent DQN loss $L_{\textit{CDQN}}$ does not increase. Specifically, we define
\begin{equation}
	\ell_{\textit{CDQN}}(\theta;\tilde{\theta}):= \max\left \{{\ell}_{\textit{DQN}}(\theta;\tilde{\theta}), {\ell}_{\textit{MSBE}}(\theta)\right \},\qquad L_{\textit{CDQN}}\equiv \mathbb{E}_{(s_t,a_t,r_t,s_{t+1})}\left[\ell_{\textit{CDQN}}\right] ,
\end{equation}
and $\ell_{\textit{CDQN}}$ does not increase for any transition data $(s_t,a_t,r_t,s_{t+1})$ upon the update of the target network. Therefore, provided that $L_{\textit{CDQN}}$ decreases during the optimization process regarding parameter $\theta$, the loss $L_{\textit{CDQN}}$ monotonically decreases throughout the training procedure and converges, since the training is constituted of the optimization with respect to $\theta$ and the replacement of $\tilde{\theta}$ by $\theta$. 

\subsection{Discussion and Limitations of the Convergent DQN Algorithm}\label{CDQN discussion}
The convergence of the iteration $\tilde{\theta}\gets\argmin_\theta L_{\textit{CDQN}}(\theta;\tilde{\theta})$ discussed in the above section relies on the assumption that the dataset $\mathcal{S}$, or the data distribution, that is used to define the loss $L_{\textit{CDQN}}$ is fixed and does not change during training. Although the fixedness of the dataset is usually taken for granted in other fields of machine learning, the situation can be different in reinforcement learning. This is because the data learned by the agent are typically collected by the agent itself, and the data distribution is affected by the learned policy during training, and the distribution can continuously change with the policy. Therefore, although we have guaranteed convergence for the iteration procedure using the C-DQN loss in Section \ref{CDQN proof}, we do not have guaranteed convergence for reinforcement learning discussed in Section \ref{DQN algorithm section 3} as a whole. Nevertheless, we find that the replacement of the DQN loss $L_{\textit{DQN}}$ by the convergent DQN loss $L_{\textit{CDQN}}$ indeed significantly improves the stability of the reinforcement learning algorithm and resolves the issue of divergence. Thus, we obtain our convergent DQN (C-DQN) algorithm, by replacing the loss $L_{\textit{DQN}}$ in the DQN algorithm with $L_{\textit{CDQN}}$ while keeping other ingredients in the DQN algorithm unchanged. 

We empirically find that the loss $L_{\textit{MSBE}}$ in the RG algorithm is typically much smaller than the loss $L_{\textit{DQN}}$ in the DQN algorithm on the same task, and therefore, as $L_{\textit{CDQN}}(\theta;\tilde{\theta})$ takes the maximum between ${\ell}_{\textit{DQN}}(\theta;\tilde{\theta})$ and ${\ell}_{\textit{MSBE}}(\theta)$, we expect that the C-DQN algorithm puts more emphasis on the DQN loss ${\ell}_{\textit{DQN}}(\theta;\tilde{\theta})$, and that it has learning behaviour similar to the DQN algorithm instead of the RG algorithm, which is confirmed in our numerical experiments in Section \ref{CDQN experiments}. As in the DQN algorithm, we use gradient descent methods to minimize the loss $L_{\textit{CDQN}}$, despite that $L_{\textit{CDQN}}$ is not a smooth function of $\theta$. In our experiments, we find that it suffices to use gradient descent methods to minimize the loss $L_{\textit{CDQN}}$ to make learning proceed. Specifically, we define the gradient of $L_{\textit{CDQN}}$ to be
\begin{equation}
	\nabla_{\theta}L_{\textit{CDQN}}=\mathbb{E}_{(s_t,a_t,r_t,s_{t+1})}\left[\nabla_{\theta}\ell_{\textit{CDQN}}\right] ,\qquad \nabla_{\theta}\ell_{\textit{CDQN}}:=
	\begin{cases}
		\nabla_{\theta}{\ell}_{\textit{DQN}} & \text{if } {\ell}_{\textit{DQN}}\ge{\ell}_{\textit{MSBE}};\\
		\nabla_{\theta}{\ell}_{\textit{MSBE}} & \text{if } {\ell}_{\textit{DQN}}<{\ell}_{\textit{MSBE}},
	\end{cases} 
\end{equation}
and we use the gradient $\nabla_{\theta}L_{\textit{CDQN}}$ in gradient descent optimization algorithms.

Because the C-DQN algorithm only minimally modifies the conventional DQN algorithm, it is also compatible with most of the recent improvements and extensions of the DQN algorithm, such as double Q-learning, distributional DQN and soft Q-learning \cite{DoubleDQN,DistributionalDQN, SoftQLearning}, in which case one modifies the loss functions ${\ell}_{\textit{DQN}}$ and ${\ell}_{\textit{MSBE}}$ accordingly. The mean squared error used in the loss functions in our discussion above may also be replaced by other measures of distance, such as the Huber loss (or smooth $\ell1$ loss), as commonly used in DQN algorithms. The Huber loss is given by
\begin{equation}\label{Huber loss definition}
	\ell_{\text{Huber}}(x,y):=\begin{cases}
		\frac{1}{2}\left(x-y\right) ^2 & \text{if } |x-y|<1;\\
		|x-y|-\frac{1}{2} & \text{if } |x-y|\ge1,
	\end{cases}
\end{equation}
and we also use this loss function in our numerical experiments in Section \ref{atari experiments}, i.e.~we use
\begin{equation}\label{Huber loss with data}
	\begin{split}
		{\ell}_{\textit{DQN}}(\theta;\tilde{\theta})&=\ell_{\text{Huber}}\left (Q_\theta(s_t,a_t),\ r_t+\gamma \max_{a'}Q_{\tilde{\theta}}(s_{t+1},a')\right ),\\
		{\ell}_{\textit{MSBE}}(\theta)&=\ell_{\text{Huber}}\left (Q_\theta(s_t,a_t),\ r_t+\gamma \max_{a'}Q_{\theta}(s_{t+1},a')\right ).
	\end{split}
\end{equation}

\subsubsection{Convergence of C-DQN in Stochastic Settings}\label{stochasticity effect on CDQN discussion}
As discussed in Section \ref{RG algorithm section}, when the transitions of states are stochastic, the loss $L_{\textit{MSBE}}$ does not precisely correspond to the Bellman equation because it includes an additional bias term. The loss $L_{\textit{MSBE}}$ can be expressed as
\begin{equation}
	\begin{split}
		L_{\textit{MSBE}}(\theta)&=\mathbb{E}_{s_{t+1}}\left [\left (Q_\theta(s_t,a_t)-r_t-\gamma \max_{a'}Q_\theta(s_{t+1},a')\right )^2\right ]\\
		&=\left (Q_\theta(s_t,a_t)-r_t-\gamma \mathbb{E}_{s_{t+1}}\left[\max_{a'}Q_\theta(s_{t+1},a')\right]\right )^2 + \gamma^2\text{Var}_{s_{t+1}}\left [\max_{a'}Q_\theta(s_{t+1},a')\right ],
	\end{split}
\end{equation}
as shown in Eq.~(\ref{MSBE bias}). On the second line in the above equation, only the first term corresponds to the Bellman equation, and the second term is considered as a bias. Because the C-DQN loss includes $\ell_{\textit{MSBE}}$, when the underlying task is stochastic, the C-DQN algorithm may not converge to the optimal Q function which is the solution to the Bellman equation. In fact, both the minimum of $L_{\textit{MSBE}}$ and stationary points for the DQN algorithm are stationary points for the C-DQN algorithm. In order to show this result, first, we assume that the minimum of $L_{\textit{DQN}}(\,\cdot\,,\tilde{\theta})$ and the minimum of $L_{\textit{MSBE}}(\,\cdot\,)$ are unique and different. Suppose that the DQN algorithm converges at parameter $\theta$, i.e.~$\theta=\tilde{\theta}$ and $\theta=\argmin L_{\textit{DQN}}(\cdot;\tilde{\theta})$, and we have 
\begin{equation}
	\nabla_{\theta}L_{\textit{DQN}}(\theta;\tilde{\theta})=0.
\end{equation}
Because we have $\theta=\tilde{\theta}$, we also have $L_{\textit{DQN}}(\theta;\tilde{\theta})=L_{\textit{MSBE}}(\theta)=L_{\textit{CDQN}}(\theta;\tilde{\theta})$. Then, if the parameter $\theta$ moves infinitesimally by $\delta\theta$ to reduce $L_{\textit{MSBE}}$, we have
\begin{equation}
	L_{\textit{DQN}}(\theta+\delta\theta; \tilde{\theta}_i)>\min L_{\textit{DQN}}(\,\cdot\,; \tilde{\theta})=L_{\textit{DQN}}(\theta;\tilde{\theta})=L_{\textit{MSBE}}(\theta)>L_{\textit{MSBE}}(\theta+\delta\theta),
\end{equation}
and therefore $L_{\textit{DQN}}$ becomes larger than $L_{\textit{MSBE}}$ and the C-DQN algorithm will choose to prioritize the optimization of $L_{\textit{DQN}}$ instead of $L_{\textit{MSBE}}$, and therefore the parameter will return to the minimum of $L_{\textit{DQN}}$. Similar arguments also hold if $\theta$ is the minimum of $L_{\textit{MSBE}}$, and therefore, the C-DQN algorithm can converge to either the stationary points of the DQN and the RG algorithm. More generally, it may converge between the stationary points of the DQN and the RG algorithm. As it tries to minimize the maximum of the loss functions $\ell_{\textit{DQN}}$ and $\ell_{\textit{MSBE}}$, it stops if both of the two loss functions cannot be optimized simultaneously. Interestingly, this issue does not appear to be a serious problem in our numerical experiments as demonstrated in the next section, even though the tasks involve a fairly large amount of noise and subtitles including partially observable states.

\section{Numerical Experiments}\label{CDQN experiments}
\subsection{The Wet-Chicken Benchmark}\label{wet chicken experiment}
To confirm the behaviour of the C-DQN algorithm in a stochastic setting, we use the \textit{wet-chicken} benchmark \cite{WetChickenOriginal,WetChickenPaper, WetChickenReferee} as a toy problem. In the following we describe the problem setting.

In the \textit{wet-chicken} problem, a canoeist starts at position $x=0$ on a river, and his/her goal is to get close to the waterfall at position $x=20$ on the river without reaching the waterfall. Every step he/she can choose to take actions to paddle back, or hold the position, or drift forward, which contributes a movement of -1, 0, or +1 to $x$. There is also random turbulence $z\sim \textit{Uniform}(-2.5,+2.5)$ at each step which also contributes to a change in $x$. The reward in this problem is equal to $x$, and if the canoeist reaches the waterfall at $x=20$, $x$ is reset to be 0. The task does not have a terminal state, and the performance in this task is evaluated as the reward averaged per time step. This problem is known to be highly stochastic, since the random perturbation has a stronger effect than the action of the agent, and the stochasticity can drive a state to states with dramatically different Q values. 

\begin{figure}[tb]
	\centering
	\includegraphics[width=0.9\linewidth]{"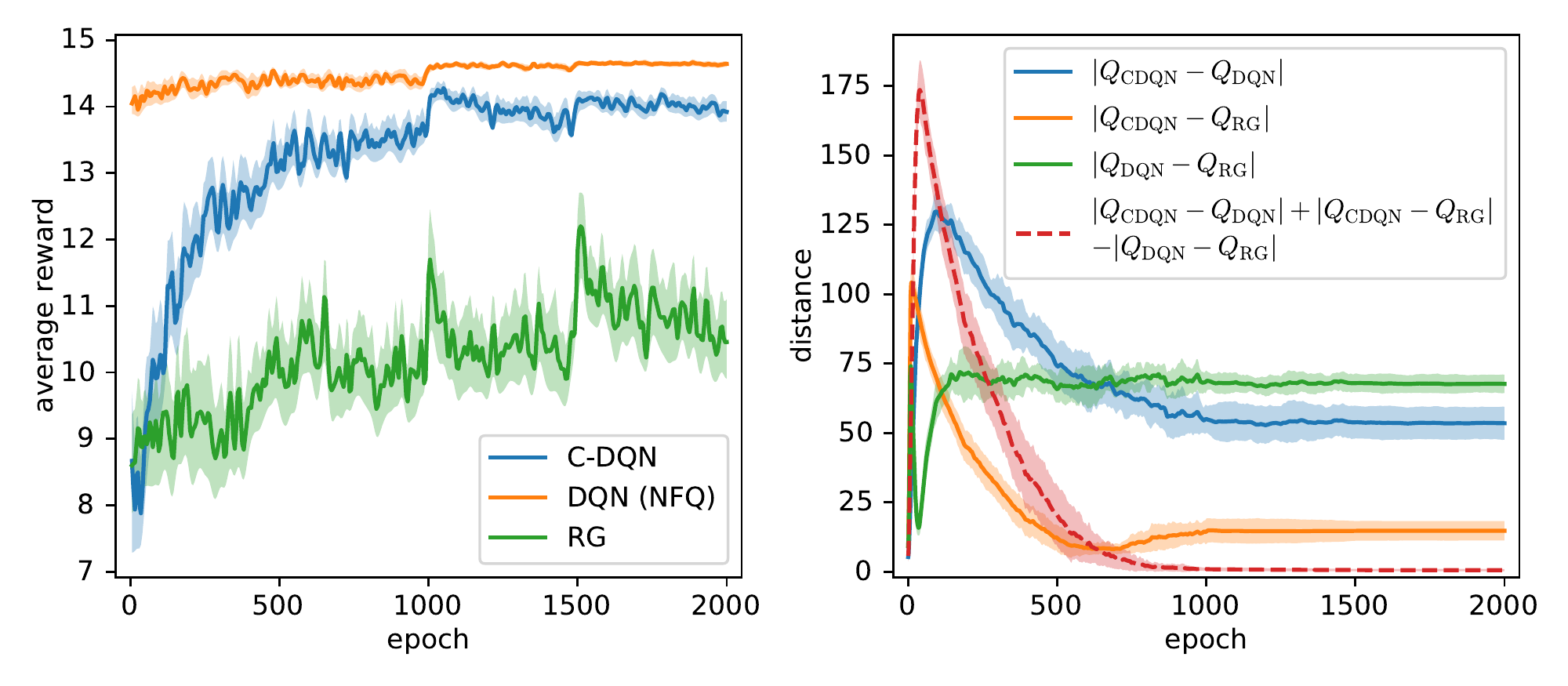"}
	\caption{\label{wet chicken fig}Performance on the wet-chicken benchmark (left) and the distances among the Q functions learned by the different algorithms during training (right). The DQN algorithm learns much more quickly than the other algorithms on this task, and it almost completes learning in the first epoch of training and it reaches a high level of performance. The performance of the RG algorithm does not improve significantly during training, as random actions can already lead to an average reward around $8$ for this task, and the RG algorithm only improves a little compared to the random policy. The C-DQN algorithm learns gradually and approaches the performance of the DQN algorithm, and it learns more slowly than the DQN algorithm on this task. The experiment is repeated for 10 times, and the shaded regions in the left and the right plots show the standard error of the performance and the standard deviation of the corresponding quantities. The small peaks on the learning curves in the left plot at the 1000th and the 1500th epochs are due to the change of the learning rate during training, which perturbs the performance.}
\end{figure}

To learn this task, we use random actions to generate a dataset of 20000 transitions of states, and we train a neural network using this generated dataset following the DQN, the C-DQN and the RG algorithms, and the results are shown in Fig.~\ref{wet chicken fig}. In the left panel of Fig.~\ref{wet chicken fig}, it is shown that the RG algorithm significantly underperforms the DQN algorithm, while the performance of the C-DQN algorithm lies between the DQN and the RG algorithms and is slightly worse than DQN. The result shows that although the task is highly stochastic and the C-DQN algorithm does not converge to the solution that the DQN algorithm finds, the C-DQN algorithm still behaves robustly and has reasonably satisfactory performance. In contrast, the RG algorithm fails to obtain satisfactory performance. 

To obtain further information on the Q functions learned by the different algorithms, we estimate the distances among the Q functions during training. We estimate the distance $|Q_1-Q_2|$ between two Q functions $Q_1$ and $Q_2$ for this task as $\sqrt{\sum_{(x,a)}\left |Q_1(x,a)-Q_2(x,a)\right |^2}$, where the summation over $x$ is taken over the discrete set $\{0,1,2,... ,19\}$. 
In the right panel of Fig.~\ref{wet chicken fig}, the estimated distances among the Q functions learned by the different algorithms are shown. It can be seen that the distance between $Q_{\textit{DQN}}$ and $Q_{\textit{CDQN}}$ increases rapidly in the beginning and slowly decreases later, 
implying that the DQN algorithm learns quickly and the C-DQN algorithm catches up later. Notably, the quantity $|Q_{\textit{CDQN}}-Q_{\textit{DQN}}|+|Q_{\textit{CDQN}}-Q_{\textit{RG}}|-|Q_{\textit{DQN}}-Q_{\textit{RG}}|$ always converges to zero, indicating that the Q function learned by the C-DQN algorithm exactly lies on the line connecting $Q_{\textit{DQN}}$ to $Q_{\textit{RG}}$, which is consistent with our argument in Section \ref{CDQN discussion} that the C-DQN algorithm may converge somewhere between the stationary points of the DQN and the RG algorithms.

\paragraph{Experimental Settings} The neural network $Q_{\theta}$ has 4 hidden layers, with 128 hidden units and using the ReLU as the activation function. The parameters are optimized using the Adam optimizer \cite{Adam} with default hyperparameters. The size of the mini-batch is 200, and the target network is updated per epoch, and therefore the DQN algorithm is essentially the same as the neural fitted Q (NFQ) iteration algorithm \cite{NeuralFittedQIteration}. The training process includes 2000 epochs, and at the 1000th and the 1500th epochs the learning rate is reduced by a factor of 10 and 100. The discount factor $\gamma$ is 0.97. The position $x$ and the reward data are normalized by a factor of $20$ before training, and per 5 epochs we evaluate the performance once to obtain the data in Fig.~\ref{wet chicken fig}. The evaluation includes 200 trials each of which uses 300 time steps. We repeat the entire experiment for 10 times, including the initial data generation process that is used to generate the dataset for training.

\subsection{\textit{Atari 2600} Games}\label{atari experiments}
In this section, we present results of the numerical experiments of the C-DQN algorithm on the standard \textit{Atari 2600} benchmark \cite{AtariEnvironment}. The \textit{Atari 2600} benchmark contains classic \textit{Atari 2600} games that were released around the 1970s and 1980s. Scores in the games are regarded as the reward, and images, or frames, on the screen of the games are used to construct a representation of the state $s_t$ and are used as the input for deep neural networks to predict the Q function. We follow the experimental settings in Ref.~\cite{RainbowDQN} and \cite{DQN} except for several minor differences. Specifically, we adopt the dueling network architecture, prioritized sampling, and double Q-learning techniques where applicable \cite{DuelingDQN, PrioritizedExperienceReplay, DoubleDQN}, for all of the DQN, the C-DQN, and the RG algorithms. Full details of our settings are provided in Ref.~\cite{CDQN}. The codes that can be used to reproduce our experimental results are released online.\footnote{\url{https://github.com/Z-T-WANG/ConvergentDQN}}
\subsubsection{Comparison of C-DQN, DQN and RG}\label{section comparison of CDQN DQN RG}
\begin{figure}[tb]
	\centering
	\includegraphics[width=0.7\linewidth]{"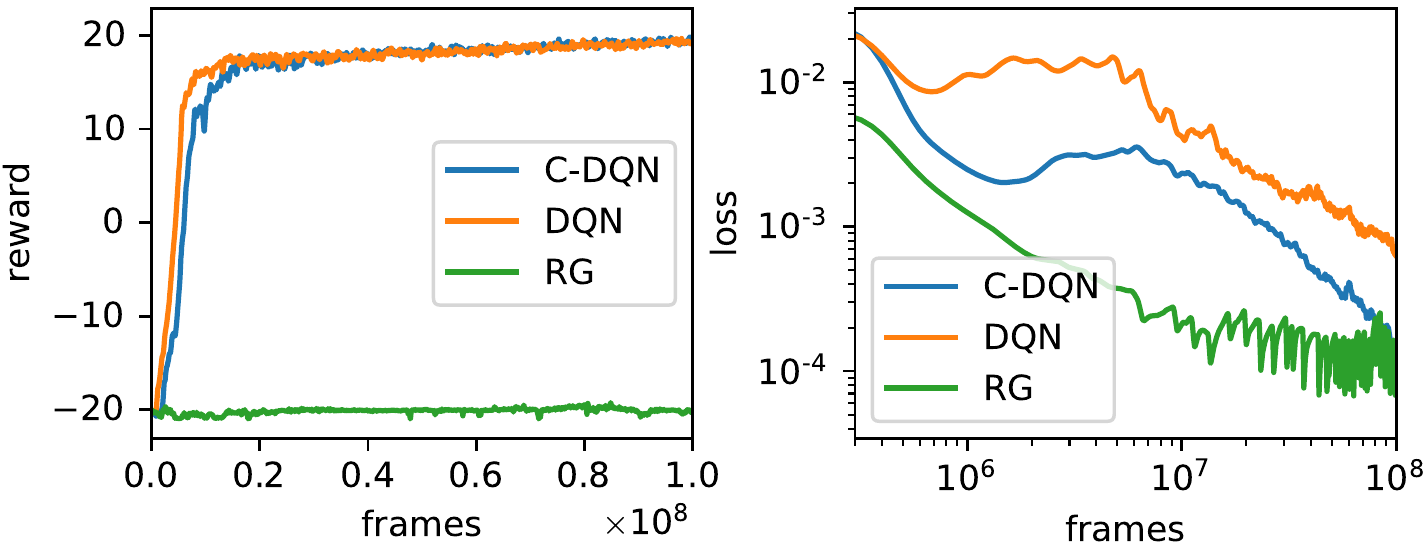"}\\
	\vspace{3mm}
	\includegraphics[width=0.7\linewidth]{"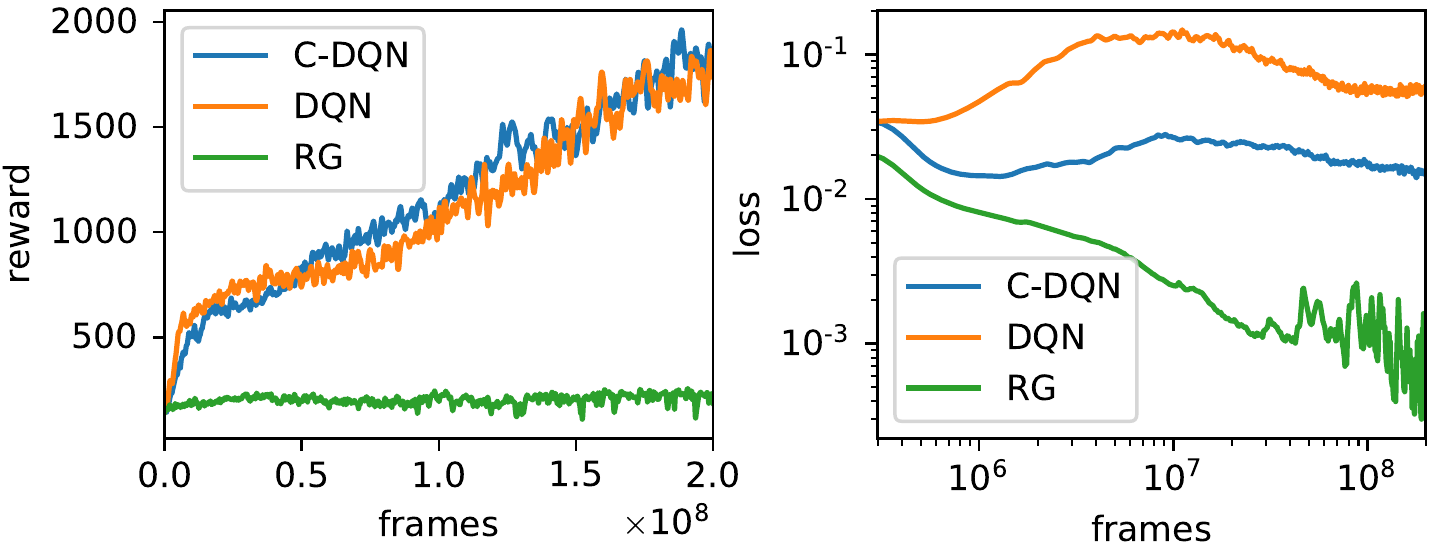"}
	\caption{\label{comparison of DQN CDQN RG}Training performance (left) and training loss (right) on \textit{Atari 2600} games \textit{Pong} (top) and \textit{Space Invaders} (bottom). }
\end{figure}
As the C-DQN, the DQN and the RG algorithms only differ in the loss functions, we use the hyperparameter settings in \cite{RainbowDQN} for all of these algorithms, and we compare their results on two well-known games in the \textit{Atari 2600} benchmark, \textit{Pong} and \textit{Space Invaders}. In Fig.~\ref{comparison of DQN CDQN RG}, we show the learning curves for performance and loss. It can be seen that both the C-DQN and the DQN algorithms can learn the tasks, while the RG algorithm almost does not learn, despite the fact that it has a much smaller loss. The failure of the RG algorithm is consistent with our analysis in Section~\ref{inefficiency of RG}. The results in this experiment confirm that the C-DQN algorithm as a convergent Q-learning method performs well for realistic problems and has performance comparable to the DQN algorithm on standard tasks.

\subsubsection{Learning from Incomplete Trajectories of Experience}
\begin{figure}[tb]%
	\centering
		\includegraphics[width=0.45\linewidth]{"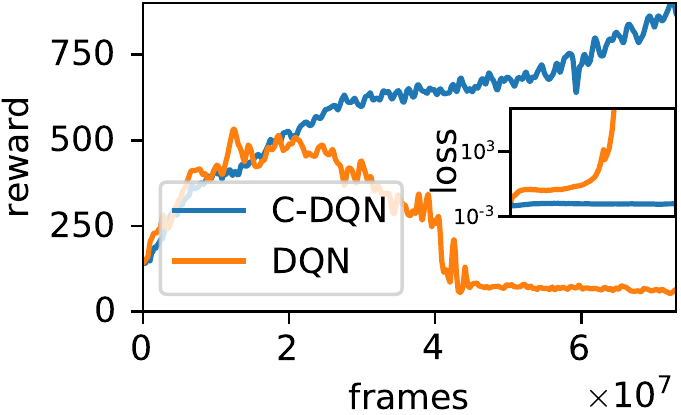"}
		\caption{\label{half discarded space invaders}Training performance and training loss on \textit{Atari 2600} games \textit{Space Invaders} when half of the data are randomly discarded.}
\end{figure}

To demonstrate the instability of the DQN algorithm, we consider training with incomplete trajectories of experience, i.e.~for a transition data $(s_t,a_t,r_t,s_{t+1})$ in the dataset, the subsequent transition $(s_{t+1},a_{t+1},r_{t+1},s_{t+2})$ may not be present in the dataset. This situation makes the DQN algorithm prone to diverge, because while the DQN algorithm learns the value of $Q_\theta(s_{t},a_{t})$ based on the value of $\max_{a'}Q_\theta(s_{t+1},a')$, if data $(s_{t+1},a_{t+1},r_{t+1},s_{t+2})$ is not present in the dataset, the algorithm cannot learn the value of $\max_{a'}Q_\theta(s_{t+1},a')$ directly from data and has to infer it, which increases the possibility of divergence as $Q_\theta(s_{t},a_{t})$ and $\max_{a'}Q_\theta(s_{t+1},a')$ changes simultaneously. To create such a situation, we randomly discard half of the data that are collected by the agent during training, and we also change the loss function from the Huber loss (Eq.~\ref{Huber loss with data}) to the mean squared error so as to allow for divergence in gradient. In our experiments, we find that whether the DQN algorithm diverges or not is generally task-dependent, and it has a larger probability to diverge if the task is more difficult. The result for \textit{Atari 2600} game \textit{Space Invaders} is shown in Fig.~\ref{half discarded space invaders}. It can be seen that while the DQN algorithm diverges, the C-DQN algorithm learns stably and its learning speed is only slightly reduced. This confirms that the C-DQN algorithm is convergent regardless of the properties of the training data.
\begin{figure}[tb]%
	\centering
		\includegraphics[width=0.31\linewidth]{"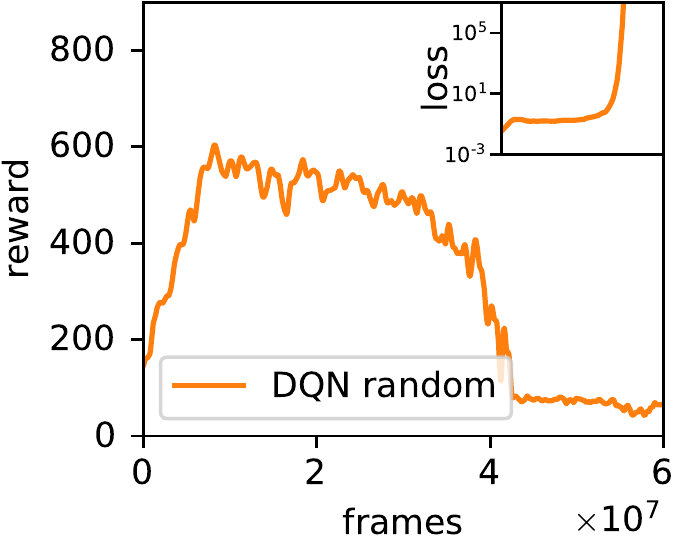"}
		\includegraphics[width=0.67\linewidth]{"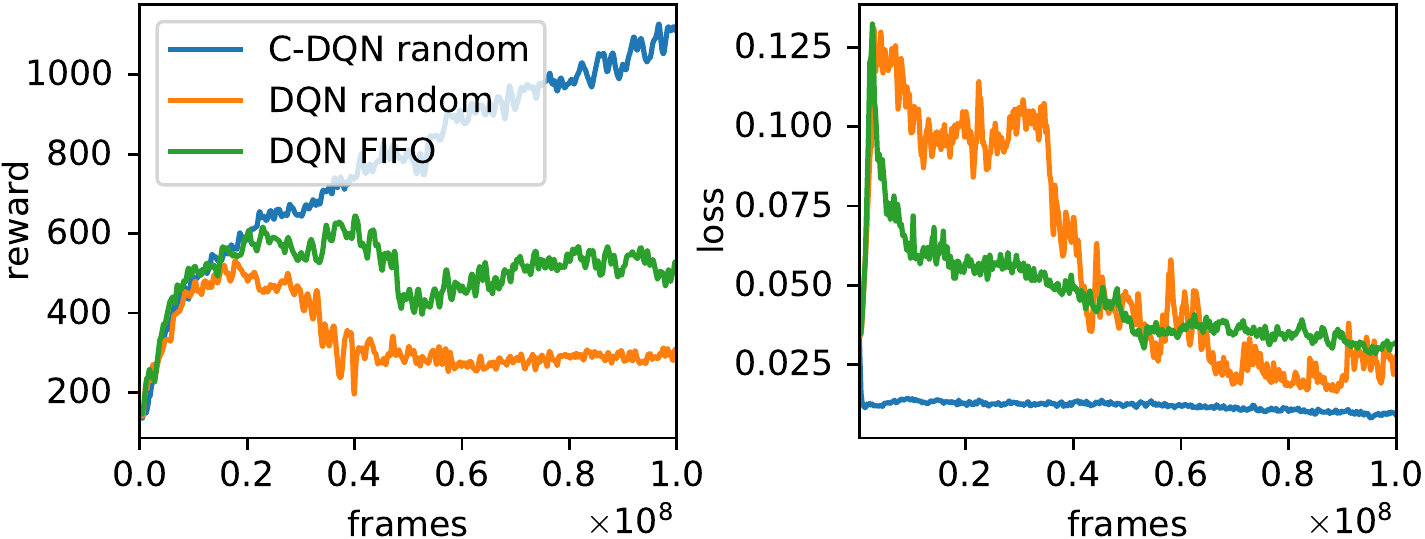"}
		\caption{\label{random replacing replay memory}Training performance and training loss on \textit{Atari 2600} games \textit{Space Invaders} when the replay memory adopts a random replacement strategy (left) and when the size of the replay memory is reduced by a factor of 10 and adopts different strategies (middle and right).}
\end{figure}%

The same situation arises if when the replay memory (i.e.~the dataset) is full, one does not use the first-in-first-out (FIFO) strategy to replace old data by new data, but randomly chooses old transition data to be replaced by new data. In this case, the dataset can also contain incomplete trajectories of data. In Fig.~\ref{random replacing replay memory}, we show that the DQN algorithm can actually diverge in this simple setting. In existing literature on DQN algorithms, the replacement strategy used with the dataset is often ignored, while here we find that it can be an important detail that affects the final results of reinforcement learning. In practice, the FIFO strategy is almost always used; nevertheless, the FIFO strategy makes the data in the dataset less diverse, and it also increases the risk of non-convergence due to the oscillation of the co-evolvement of the learned policy and the dataset. As a consequence, a large size of the replay memory is often necessary in reinforcement learning. In Fig.~\ref{random replacing replay memory}, it can be seen that when we reduce the size of the replay memory by a factor of $10$, the C-DQN algorithm can utilize the random replacement strategy to achieve a higher performance, while the DQN algorithm cannot. Note that the DQN algorithm does not show divergence in this experiment. We conjecture that it is because the replay memory is small, it contains more recent data and more complete trajectories of experience, which alleviates divergence. As the C-DQN algorithm can be trained stably with data that comes from an arbitrary distribution, the result opens up a new possibility of reinforcement learning of only storing and learning important data, which is not possible with the conventional DQN algorithm.

\subsubsection{Difficult Games in \textit{Atari 2600}}\label{section difficult atari games}
Here we consider difficult games in the \textit{Atari 2600} benchmark, which require the use of large discount factors $\gamma$. Although the DQN algorithm becomes unstable and often diverges when $\gamma$ becomes increasingly close to $1$, the convergence property of the C-DQN algorithm does not depend on $\gamma$ and it can work with any $\gamma$ in principle. Nevertheless, we notice that a large $\gamma$ does not necessarily lead to better performance, because a large $\gamma$ requires the agent to learn to predict rewards that are distant in future time steps, which is often not necessary and is irrelevant for learning the task. Therefore, a large $\gamma$ can reduce the learning efficiency. We also notice that if we have $\gamma\ge0.9999$, the order of magnitude of the term $(1-\gamma)Q_\theta$ becomes close to the inherent noise in the gradient descent optimization algorithm due to the finite learning rate, and the learning process can stagnate. Therefore, we do not expect $\gamma$ to be larger than $0.9999$ and we only require $\gamma$ to satisfy $0.99\le\gamma\le0.9998$. Because the appropriate discount factor $\gamma$ can be different for different problems, we also use a heuristic algorithm to evaluate the frequency of reward signals in each problem so as to determine $\gamma$ for each problem separately. Details concerning this strategy are presented in Ref.~\cite{CDQN} in the appendix. We also normalize the Q functions before training using the evaluated mean and scale of the reward signals, and we follow the techniques in Ref.~\cite{TemporalConsistencyLoss} to transform the Q functions approximately by the square root, so that the Q functions always have appropriate orders of magnitude. 

\begin{figure}[tb]
	\centering
	\!\!\!\!\!\!\!\!\!\!\!\!\!\!\includegraphics[width=1.1\linewidth]{"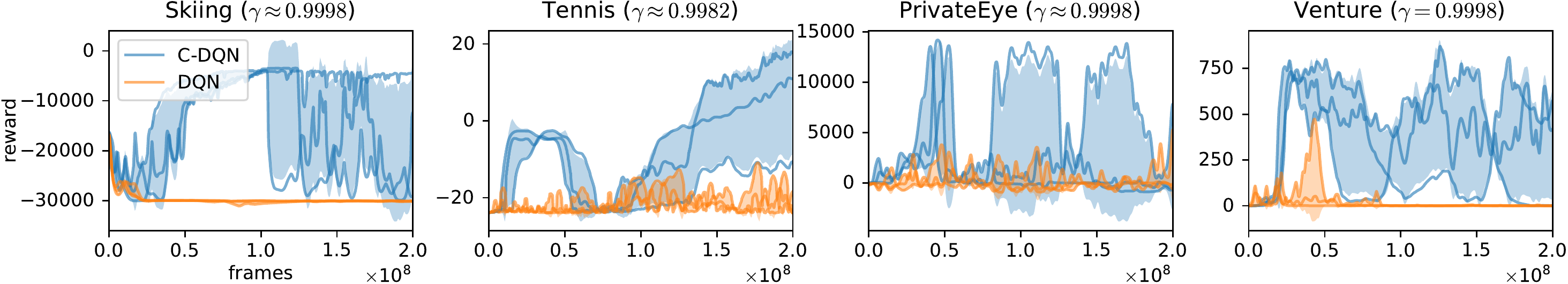"}
	\caption{\label{difficult atari games}Training performance for several difficult games in the \textit{Atari 2600} benchmark. Each line shows the performance in a single repetition of the experiment and the shaded regions show the standard deviation. The discount factors $\gamma$ are shown in the titles. The DQN algorithm fails to learn these tasks and shows significant instability, and the DQN loss increases up to around $10^{5}\sim10^{8}$ in these experiments, while the C-DQN loss stays below $1$.}
\end{figure}

With the C-DQN algorithm and large values of $\gamma$, several difficult problems which could not be solved by simple variants of the DQN algorithm can now be solved, as shown in Fig.~\ref{difficult atari games}. Especially for \textit{Atari 2600} games \textit{Skiing}, \textit{Private Eye} and \textit{Venture}, the agent significantly benefits from large values of $\gamma$ and achieves a higher best performance during training, despite the fact that the games \textit{Private Eye} and \textit{Venture} are only partially observable problems and therefore not fully learnable, which results in unstable performance.

In the following, we compare the test performance of the C-DQN algorithm with other works. To evaluate the test performance, we pick the best-performing agent during training and test its performance using 400 trials, using the $\epsilon$-greedy policy with $\epsilon=0.01$ and no-op starts\footnote{No-op starts mean that whenever an episode begins, the \textit{no-operation} action is executed randomly for $1$ to $30$ frames, so that the agent does not always starts at exactly the same state.} \cite{DQN}. The average of the test performances of the 3 different repetitions of our experiments are shown in Table~\ref{test performance} with the standard error, compared with existing works and the human performance.
\begin{table}
	\caption{Test performance on difficult \textit{Atari 2600} games. The results for the DQN algorithm are obtained by us using the same experimental settings as the C-DQN algorithm. Human results and results for Agent57 are due to Ref.~\cite{Agent57}, and results for Rainbow DQN are due to Ref.~\cite{RainbowDQN}. Note that the human results only correspond to reasonably adequate performance and they are not the highest possible performance of human.}
	\label{test performance}
	\centering
	\begin{tabular}{cccccc}
		\toprule
		Task     & C-DQN & DQN & Human & Rainbow DQN  & Agent57 (SOTA) \\
		\midrule
		Skiing & \textbf{-3697 $\pm$ 157} & -29751 $\pm$ 224 & -4337 & -12958 &  -4203 $\pm$ 608  \\
		Tennis & 10.9 $\pm$ 6.3
		& -2.6 $\pm$ 1.4 & -8.3 & 0.0 &  23.8 $\pm$ 0.1 \\
		Private Eye & 14730 $\pm$ 37 & 7948 $\pm$ 749 & 69571 & 4234 & 79716 $\pm$ 29545 \\
		Venture & 893 $\pm$ 51 & 386 $\pm$ 85 & 1188 & 5.5 & 2624 $\pm$ 442 \\
		\bottomrule
	\end{tabular}
\end{table}

As we have followed the standard procedure of training the DQN agent on the \textit{Atari 2600} benchmark as in Ref.~\cite{DQN} and Ref.~\cite{RainbowDQN}, the performance obtained by our C-DQN agent allows for a fair comparison with the results of the Rainbow DQN algorithm in Ref.~\cite{RainbowDQN}.\footnote{Precisely speaking, a fair comparison with the Rainbow DQN algorithm cannot be made on the game \textit{Skiing}. This is because the reward clipping strategy adopted by the Rainbow DQN algorithm does not permit learning the game \textit{Skiing}. However, this does not affect our conclusion.} In Table~\ref{test performance}, we see that the Rainbow DQN algorithm fails to make progress in learning for these four difficult \textit{Atari 2600} games, and the C-DQN algorithm can achieve performance higher than the Rainbow DQN algorithm and it has non-trivial learning behaviour. The results of the Agent57 algorithm are only for reference \cite{Agent57}, representing the currently known highest overall performance on the \textit{Atari 2600} benchmark. The results of the Agent57 algorithm do not allow for a fair comparison with the C-DQN algorithm, because it requires considerably more computation, sophisticated methods, and larger and more complicated neural networks. Notably, we find that our result on the game \textit{Skiing} is exceptional, where the C-DQN algorithm achieves the state-of-the-art performance despite its simplicity, which is discussed in following. 

\subsubsection{The Atari Game \textit{Skiing}}\label{section skiing}
\footnotetext{\url{https://github.com/mgbellemare/Arcade-Learning-Environment}}In Table~\ref{test performance}, an exceptional result is that the C-DQN algorithm achieves a performance that is higher than the Agent57 algorithm on the game \textit{Skiing}, in fact, utilizing an amount of computation budget less than $0.1\%$ of that of the Agent57 algorithm. We find that this is the highest performance so far and therefore achieves the state-of-the-art (SOTA) for this task. To elucidate the reason, we describe the game in the following. 
\begin{figure}[tb]
	\centering
	\begin{minipage}{0.39\linewidth}
		\centering
		\includegraphics[width=0.8\linewidth]{"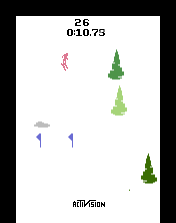"}
		\captionof{figure}{\label{skiing screenshot}A screenshot of game \textit{Skiing} in the \textit{Atari 2600} benchmark. The program is provided under the GNU General Public License v2.0 in Ref.~\cite{AtariEnvironment}.\protect\footnotemark}
	\end{minipage}\qquad%
	\begin{minipage}{0.55\linewidth}
		\centering
		\includegraphics[width=0.9\linewidth]{"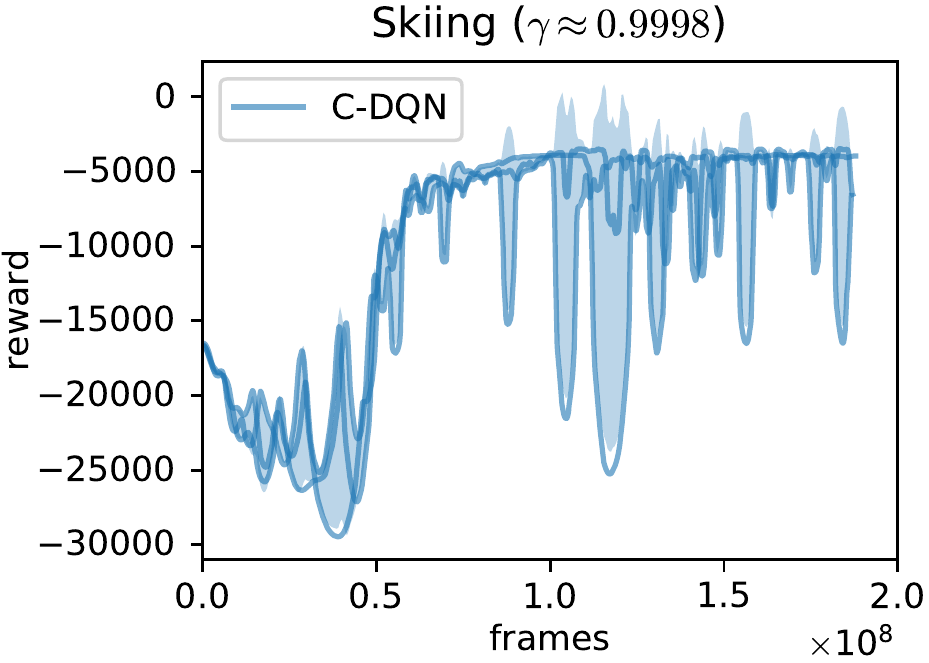"}
		\captionof{figure}{\label{slower skiing CDQN}Training performance of the C-DQN algorithm on \textit{Atari 2600} game \textit{Skiing} when the learning rate is reduced by half, following the same experimental procedure as in Fig.~\ref{difficult atari games}. The standard deviation is shown as the shaded region.}
	\end{minipage}
	\vspace{-5mm}
\end{figure}

A screenshot of \textit{Atari 2600} game \textit{Skiing} is shown in Fig.~\ref{skiing screenshot}. This game is basically a racing game, where the player needs to go downhill as fast as possible, and the time elapsed before reaching the goal is regarded as the minus reward. For each time step, the agent receives a small amount of minus reward which represents the time elapsed, until it reaches the goal and the game terminates. Additionally, the player is required to pass through the gates along the way, which are shown by the two small blue flags in Fig.~\ref{skiing screenshot}. If the player fails to pass through a gate, at the moment when the player reaches the goal, a 5-second penalty is added to the elapsed time. The number of gates that are yet to pass are shown on the top of the screen in the game. With the standard setting in Ref.~\cite{DQN}, the number of time steps for an episode in this game is $\sim1300$ for the random policy, is $\sim4500$ if the player slows down, and is $\sim500$ if the policy is near-optimal. 

Because the penalty for not passing through a gate is given to the agent only at the end of the game, it is necessary for the agent to learn the relation between the penalty at the end of the game and the events that occur early in the game. Therefore, the time horizon for planning must be long enough, and the discount factor $\gamma$ should be at least $1-\frac{1}{500}$ to allow for learning. However, learning may stagnate if $\gamma$ is not even larger than $1-\frac{1}{500}$, because if $\gamma$ is small, the agent would prefer spending longer time before reaching the goal, so that the penalty at the end of the game is delayed and the Q values for the states in the early game are increased, which will further increase the number of time steps for an episode and make learning difficult. Therefore, we have tuned our hyperparameter setting so that we have $\gamma\approx1-\frac{1}{5000}$ on this game. The C-DQN agent learns with this value of $\gamma$ successfully and produces a new record on this task. The large fluctuations on the learning curves shown in Fig.~\ref{difficult atari games} are mainly due to noise coming from the large learning rate, which have been confirmed in Fig.~\ref{slower skiing CDQN} by repeating the experiment using a smaller learning rate. However, we find that with the small learning rate, the policy can easily get trapped in local optima and the test performance is actually worse, and therefore we still use the large learning rate in our experiments. It is worth noting that in fact, we cannot compare this result fairly with that of the Agent57 algorithm, because our hyperparameters have been tuned so that the discount factor $\gamma$ is suitable for this task, while the Agent57 algorithm adopts a bandit algorithm to dynamically determine $\gamma$, which is more general.

\section{Conclusions and Outlook}\label{CDQN outlook}
In this chapter, we have discussed the inefficiency issues of the RG algorithm, and proposed a convergent DQN (C-DQN) algorithm to address the long-standing problem of non-convergence of Q-learning. We have discussed the properties of the C-DQN algorithm and demonstrated the effectiveness of the C-DQN algorithm on the standard \textit{Atari 2600} benchmark for deep reinforcement learning. With the stability of the C-DQN algorithm, we can tune the discount factor $\gamma$ freely without sacrificing stability, and we can consider the possibility of only learning important pieces of data to improve efficiency. The C-DQN algorithm has a better stability and convergence property than the DQN algorithm, and it can be applied to difficult tasks for which the DQN algorithm fails due to instability. The idea of the C-DQN algorithm can be combined with other reinforcement learning strategies involving target networks and potentially improve their stability. 

Many outstanding issues exist concerning the C-DQN algorithm. The C-DQN loss is non-smooth, and it is still not clear how the non-smoothness affects the optimization and the learning process. When the task is stochastic, the MSBE loss $L_{\textit{MSBE}}$ used in the C-DQN algorithm does not converge exactly to the optimal Q function, and therefore, it would be desirable if the C-DQN algorithm can be improved so that stochastic tasks can be learned correctly without bias. It would be interesting to investigate how the C-DQN loss interplays with several other DQN extensions such as distributional DQN and soft Q-learning \cite{DistributionalDQN, SoftQLearning} and how the gradient descent optimization algorithm affects the learning dynamics of the C-DQN algorithm. It is also an interesting problem how the target network in the C-DQN algorithm can be updated smoothly as in the DDPG algorithm \cite{DDPG}.
	\chapter{Control of Continuous Quantum Systems with Convergent Deep Q-Learning}\label{control rigid body}
	\section{Introduction}
In this chapter, we apply our convergent deep Q network (C-DQN) algorithm \cite{CDQN} to quantum measurement-feedback control problems in continuous space, namely, measurement-feedback cooling of a quantum-mechanical quartic oscillator and measurement-feedback cooling of a trapped quantum-mechanical rigid body in numerical simulation, and we compare the results obtained using C-DQN with those obtained using the conventional DQN algorithm to demonstrate the advantages of C-DQN. In Section \ref{quartic oscillator control section}, we present our model of the controlled quantum quartic oscillator, and show that the C-DQN algorithm performs significantly more stably compared with the conventional DQN algorithm, which suffers much less randomness in the final results. In Section \ref{rigid body control section}, we first introduce the background of the quantum-mechanical rigid body and review the derivation of the Hamiltonian following Ref.~\cite{quantumMechanicalRigidBody} in Section \ref{rigid body Hamiltonian review}, and then, we present our model of the trapped and controlled quantum-mechanical rigid body and derive its Hamiltonian and the time-evolution equation in Section \ref{rigid body model derivation}. We then apply the C-DQN and the DQN algorithms to the cooling problem of this system, and compare them with the standard linear-quadratic-Gaussian control strategy that involves approximation in Section \ref{rigid body results}. 

\section{Cooling of a Quantum Quartic Oscillator}\label{quartic oscillator control section}
In this section, we consider the problem of measurement-feedback cooling of a one-dimensional quantum quartic oscillator, which is a minimal model of a nonlinear quantum system in continuous space. We consider a situation in which the system is subject to continuous position measurement, and we consider a controllable external force for use in feedback control to reduce the energy of the system, thereby cooling the system. We show that while the conventional DQN algorithm exhibits instability and a large variance in learning the task, the C-DQN algorithm is much more stable. We have mostly followed our previous work \cite{MasterThesis} concerning the settings of the quantum system.

\subsection{Significance of the Quartic System}
In contrast to a quantum quartic oscillator, a quantum harmonic oscillator is simple and its optimal control strategy can be analytically obtained \cite{HarmonicOscillatorControl}. The Hamiltonian of a harmonic oscillator is given by
\begin{equation}
	\hat{H} = \frac{\hat{p}^{2}}{2m}+\frac{k}{2}\hat{x}^{2},
\end{equation}
where $\hat{p}$ is the momentum operator and $\hat{x}$ is the position operator. The Hamiltonian is therefore quadratic with respect to the operators $\hat{x}$ and $\hat{p}$, and the time-evolution equations of the expectation values $\langle\hat{x}\rangle$ and $\langle\hat{x}\rangle$ are linear and given by
\begin{equation}
	d\langle\hat{x}\rangle = \frac{1}{m}\langle\hat{p}\rangle\,dt,\qquad d\langle\hat{p}\rangle = k\langle\hat{x}\rangle\,dt.
\end{equation}
When the position of the system is continuously measured as discussed in Section \ref{continuous measurement subsection}, the time evolution of the position $\langle\hat{x}\rangle$ and the momentum $\langle\hat{p}\rangle$ is subject to a Gaussian noise, and the standard linear-quadratic-Gaussian (LQG) control is applicable. The LQG control regards $\langle\hat{x}\rangle$ and $\langle\hat{p}\rangle$ as the system variables and it effectively minimizes the term
\begin{equation}
	\mathbb{E}\left[\int\left( \frac{1}{2m}\langle\hat{p}\rangle^{2}+\frac{k}{2}\langle\hat{x}\rangle^{2}\right) dt\right ],
\end{equation}
which corresponds to the minimization of the expectation of the total energy
\begin{equation}
	\mathbb{E}\left[\int\langle\hat{H}\rangle\,dt\right ]=\mathbb{E}\left[\int\left( \frac{1}{2m}\langle\hat{p}\rangle^{2}+\frac{k}{2}\langle\hat{x}\rangle^{2} + \frac{1}{2m}(\langle\hat{p}^{2}\rangle-\langle\hat{p}\rangle^{2})+\frac{k}{2}(\langle\hat{x}^{2}\rangle-\langle\hat{x}\rangle^{2})\right)\,dt\right ],
\end{equation}
since the variances $\langle\hat{p}^{2}\rangle-\langle\hat{p}\rangle^{2}$ and $\langle\hat{x}^{2}\rangle-\langle\hat{x}\rangle^{2}$ in the above equation are known to converge to steady values under continuous measurement \cite{HarmonicOscillatorControl}, and the state becomes a Gaussian state. Therefore, the quantum harmonic oscillator under continuous measurement is effectively classical, in the sense that the position and momentum variables $\langle\hat{x}\rangle$ and $\langle\hat{p}\rangle$ are sufficient to describe the state in the time evolution, and the control strategy can be conveniently derived in the same way as for classical systems. The control force of the LQG control is linear with respect to the variables $\langle\hat{x}\rangle$ and $\langle\hat{p}\rangle$.

\begin{figure}[htb!]
	\centering
	\includegraphics[width=0.44\linewidth]{"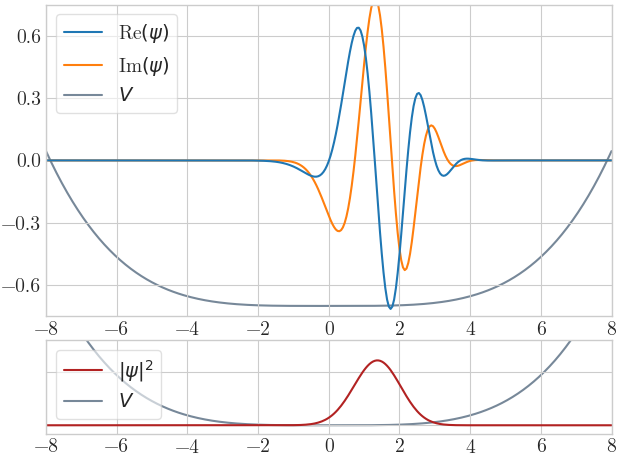"}\qquad
	\includegraphics[width=0.44\linewidth]{"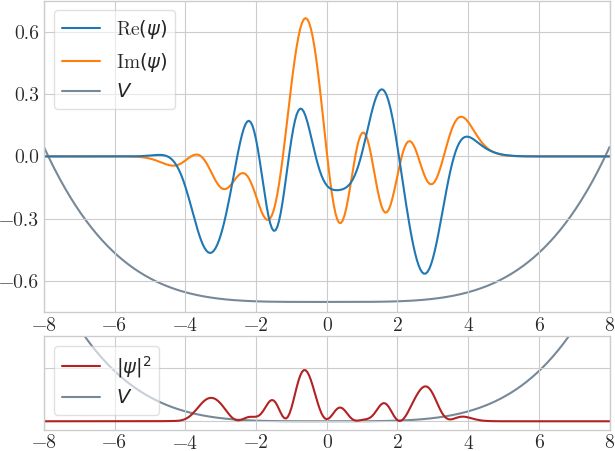"}
	\caption{Snapshots of the time evolution of a state in a quartic potential. The left panels shows the initial state which is Gaussian, and after several oscillations in the quartic potential, the state becomes non-Gaussian and it is shown in the right panel. The blue and the orange curves show the real and the imaginary parts of the wave functions, and the red curves show the probability densities. The grey curves show the potential, and the scale of the potential is arbitrary. }
	\label{quartic evolution example}
\end{figure}

\begin{figure}[htb!]
	\centering
	\includegraphics[width=0.47\linewidth]{"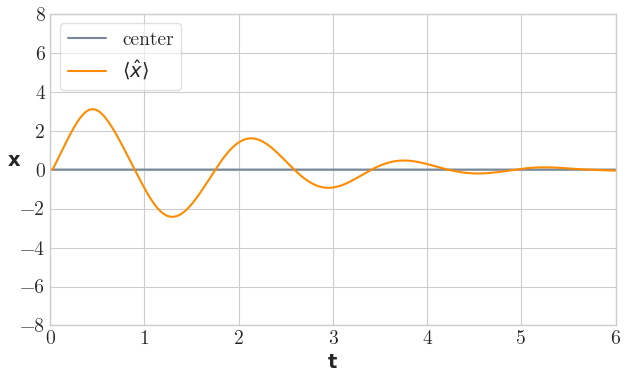"}
	\caption{The time evolution of the expectation value of the position $\langle x\rangle$ of the state evolving in the quartic potential as in Fig.~\ref{quartic evolution example}. It can be seen that the expectation value of the position ceases to oscillate, while the energy of the system remains high.}
	\label{quartic position example}
\end{figure}

Nevertheless, for a quartic oscillator of which the Hamiltonian is given by
\begin{equation}
	\hat{H} = \frac{\hat{p}^{2}}{2m}+\lambda\hat{x}^{4},
\end{equation}
the time evolution of $\langle\hat{x}\rangle$ and $\langle\hat{p}\rangle$ is given by
\begin{equation}
	d\langle\hat{x}\rangle=\frac{\langle\hat{p}\rangle}{m},\qquad d\langle\hat{p}\rangle=-4\lambda\langle\hat{x}^3\rangle,
\end{equation}
which involves the cubic term $\langle\hat{x}^3\rangle$. Therefore, the skewness $\left \langle\left (\hat{x}-\langle\hat{x}\rangle\right )^3\right \rangle$ is also relevant in the dynamics. The time evolution of the skewness is given by \cite{MasterThesis}
\begin{equation}
	d\left \langle\left (\hat{x}-\langle\hat{x}\rangle\right )^3\right \rangle = \frac{3\left \langle(\hat{x}-\langle\hat{x}\rangle)(\hat{p}-\langle\hat{p}\rangle)(\hat{x}-\langle\hat{x}\rangle)\right\rangle}{m} dt
\end{equation}
and the time evolution of $\left\langle(\hat{x}-\langle\hat{x}\rangle)(\hat{p}-\langle\hat{p}\rangle)(\hat{x}-\langle\hat{x}\rangle)\right\rangle$ is given by
\begin{equation}
	\begin{split}	
	d\left \langle(\hat{x}-\langle\hat{x}\rangle)(\hat{p}-\langle\hat{p}\rangle)(\hat{x}-\langle\hat{x}\rangle)\right\rangle&=\frac{2\left \langle(\hat{p}-\langle\hat{p}\rangle)(\hat{x}-\langle\hat{x}\rangle)(\hat{p}-\langle\hat{p}\rangle)\right\rangle }{m}dt\\
	&\quad-4\lambda\left \langle(\hat{x}^3-\langle\hat{x}^3\rangle)(\hat{x}-\langle\hat{x}\rangle)^2\right \rangle dt.
	\end{split}
\end{equation}
For a Gaussian state, of which the skewness, the odd central moments and the excess kurtosis are all zero, the term $\left\langle(\hat{x}^3-\langle\hat{x}^3\rangle)(\hat{x}-\langle\hat{x}\rangle)^2\right \rangle$ is found to be \cite{MasterThesis}
\begin{equation}
\begin{split}
	\left \langle(\hat{x}^3-\langle\hat{x}^3\rangle)(\hat{x}-\langle\hat{x}\rangle)^2\right \rangle= 9\left(\langle\hat{x}^{2}\rangle-\langle\hat{x}\rangle^{2}\right) \langle\hat{x}\rangle,
\end{split}
\end{equation}
which implies that a state that is initially Gaussian in a quartic potential gradually develops nonzero skewness. The dynamics of $\langle\hat{x}\rangle$ and that of $\langle\hat{p}\rangle$ are affected by the profile of the wave function, and therefore, the state cannot be merely characterized by the expectation values $\langle\hat{x}\rangle$ and $\langle\hat{p}\rangle$ and the variances, and the dynamics exhibits genuinely quantum-mechanical effects, as shown in Figs.~\ref{quartic evolution example} and \ref{quartic position example}.

It has been known that a one-dimensional quartic oscillator corresponds to the one-dimensional $\phi^4$ theory \cite{anharmonicphi4}, and the system is difficult to analyse. The appropriate control strategy of this system is unknown, and therefore, we consider cooling of the quartic oscillator as our first nontrivial example of continuous quantum control, and we compare the result obtained using the conventional DQN algorithm with the result obtained using the C-DQN algorithm.

\subsection{Model}
Following Section~\ref{continuous measurement subsection}, the stochastic time-evolution equation of the state subject to continuous position measurement is given by
\begin{gather}\label{quartic time evolution equation}
	d|\psi\rangle=\left[\left(-\frac{i}{\hbar}\hat{H}-\frac{\gamma}{4}(\hat{x}-\langle\hat{x}\rangle)^2\right)dt+\sqrt{\dfrac{\gamma}{2}}(\hat{x}-\langle\hat{x}\rangle)dW\right]|\psi\rangle,\\
	\hat{H} = \frac{\hat{p}^{2}}{2m}+\lambda\hat{x}^{4}-F_{\text{con}}\hat{x},
\end{gather}
where $\gamma$ is the measurement strength and $dW$ is a Wiener increment, which is a Gaussian random variable satisfying $\mathbb{E}[dW]=0$ and $\mathbb{E}[dW^{2}]=dt$, and $F_{\text{con}}$ is the external control force. Due to the Gaussian property of the position measurement, the system tends to behave like a classical particle and to have a Gaussian profile when the measurement strength is large. Therefore, we choose a sufficiently small measurement strength so that quantum effects can be significant. The system parameters that we use in our numerical simulation are given in Table~\ref{quartic oscillator parameter table}.
\begin{table}[h]
	\centering
	\begin{tabular}{p{4.3em}ccccc}\toprule
		& $m$ ($m_c$) & $\lambda$ $\left (\frac{m^2_c\omega^3_c}{\hbar}\right )$ & $\gamma$ ($\frac{m_c\omega_c^2}{\hbar}$) & $F_{\text{max}}$ ($\sqrt{\hbar m_c\omega_c^{3}}$) & $x_{\text{max}}$ ($\sqrt{\frac{\hbar}{m_c\omega_c}}$)\\[8pt] 
		\hline &&&&&\\[-11pt]
		quartic oscillator & $\dfrac{1}{\pi}$ & $\dfrac{\pi}{25}$ & $\dfrac{\pi}{100}$ & $3\pi$ & 8.5\\  \bottomrule
	\end{tabular}
	\caption{\label{quartic oscillator parameter table}System parameters of the quartic oscillator used in our numerical experiments, in terms of a reference angular momentum $\omega_c$ and a reference mass $m_c$. $F_{\text{max}}$ is the maximum of the control force $F_{\text{con}}$ that we allow. $x_{\text{max}}$ is the boundary of the 1D space that we simulate, or the maximal distance away from the center of the potential, in our numerical simulation.}
\end{table}

We assume that the measurement efficiency is unity. As the measurement can purify an arbitrary mixed state and the state is continuously measured, we assume that the state, or the wave function, is already known by the external observer and is therefore available. Every $\frac{1}{18\omega_c}$ time, the controller determines the control force $F_{\text{con}}$ on the basis of the information about the instantaneous wave function $|\psi\rangle$, and the force $F_{\text{con}}$ is kept constant during a time of $\frac{1}{18\omega_c}$. The control loss, or the minus reward in the setting of reinforcement learning, is the energy of the state, given by $\left \langle\frac{\hat{p}^{2}}{2m}+\lambda\hat{x}^{4}\right \rangle$. 

To meet the requirement of deep Q-learning, the space of actions, namely, the space of possible choices of the control force $F_{\text{con}}$, must be discrete. We therefore discretize the continuous interval $\left [-F_{\text{max}},+F_{\text{max}} \right ]$ into 21 points, and the set of control actions is given by
\begin{equation}\label{set of control actions}
	\{F_{\text{con}}\ |\ F_{\text{con}}=n\times0.3\pi\sqrt{\hbar m_c\omega_c^{3}},\quad -10\le n\le 10,\quad n\in\mathbb{Z}\}.
\end{equation}
In the setting of reinforcement learning, the agent, or the controller, determines its action $F_{\text{con}}$ based on the current state $|\psi\rangle$. After an evolution time of $\frac{1}{18\omega_c}$ of the state, the agent experiences its next time step and it observes the state again to make decision on the control action. The reward is given by the minus energy of the state at each time step.   

The deep neural network $Q_\theta$ used in reinforcement learning as discussed in Chapter \ref{deep reinforcement learning} takes information of the state $|\psi\rangle$ as its input, and it outputs the Q values for each $F_{\text{con}}$ in the action set in Eq.~(\ref{set of control actions}). The neural network has 4 layers in total, with hidden units 512, 512, and 256. The neural network is trained using the Adam optimizer \cite{Adam} with a minibatch size 512, learning each experience data for 8 times on average, and the update period of the target network is set to be 300 gradient descent steps. The discount factor $\gamma$ in Q-learning discussed in Chapter \ref{deep reinforcement learning} is set to be $0.99$.

To efficiently give the necessary information of the state to the neural network, we use distribution moments of the Wigner quasiprobability distribution of the state as the input of the neural network. Specifically, we include $\langle\hat{x}\rangle$, $\langle\hat{p}\rangle$, $\left \langle\left( \hat{x}-\langle\hat{x}\rangle\right)^{2}\right \rangle$, $\left \langle\left( \hat{p}-\langle\hat{p}\rangle\right)^{2}\right \rangle$, $\text{Re}\left[\left \langle\left( \hat{x}-\langle\hat{x}\rangle\right)\left( \hat{p}-\langle\hat{p}\rangle\right)\right \rangle \right] $, $\left \langle\left( \hat{x}-\langle\hat{x}\rangle\right)^{3}\right \rangle$, and $\left \langle\left( \hat{x}-\langle\hat{x}\rangle\right)\left( \hat{p}-\langle\hat{p}\rangle\right)\left( \hat{x}-\langle\hat{x}\rangle\right)\right \rangle$, etc.~in the input of the neural network, totally up to all fifth central distribution moments with respect to $\hat{x}$ and $\hat{p}$. The motivation for using distribution moments as the input is that the LQG control only needs $\langle\hat{x}\rangle$ and $\langle\hat{p}\rangle$ to find the optimal control force for a harmonic oscillator, and naturally by including higher distribution moments, the controller may become aware of additional nontrivial details of the state and better control strategies may be found. These distribution moments are also physically relevant, as they are physical observables.

The quantum state is approximately simulated in discrete space, and the spacing between adjacent discrete sites is set to be $0.1\sqrt{\frac{\hbar}{m_c\omega_c}}$, and the time step used in the numerical integration of the time-evolution equation is set to be $\frac{1}{1440\omega_c}$. To numerically evaluate the term $\dfrac{\partial}{\partial x}$ by finite difference methods, we use the formula
\begin{equation}\label{finite difference 1st differential}
	\begin{split}
			\frac{\partial}{\partial x}f(x_0) = &\frac{3f(x_0-4h)-32f(x_0-3h)+168f(x_0-2h)-672f(x_0-h)}{840h}\\
				&+\frac{672f(x_0+h)-168(x_0+2h)+32(x_0+3h)-3(x_0+4h)}{840h} + O(h^{8})
	\end{split}
\end{equation}
and to evaluate the term $\dfrac{\partial^{2}}{\partial x^{2}}$, we use 
\begin{equation}\label{finite difference 2nd differential}
	\begin{split}
		\frac{\partial^{2}}{\partial x^{2}}f(x_0) = &\frac{-9f(x_0-4h)+128f(x_0-3h)-1008f(x_0-2h)+8064f(x_0-h)-14350f(x_0)}{5040h}\\
		&+\frac{8064f(x_0+h)-1008(x_0+2h)+128(x_0+3h)-9(x_0+4h)}{5040h} + O(h^{8}),
	\end{split}
\end{equation}
so that we can obtain accurate results efficiently in our numerical simulation. We use the implicit 1.5 order strong scheme \cite{NumericalSimulationofStochasticDE} for numerical integration of the stochastic differential equation (\ref{quartic time evolution equation}), and additionally, in order to prevent numerical divergence due to the high energy part of the quartic potential, we include high-order terms of the time evolution of the Hamiltonian in our numerical integration, including additional terms $\frac{(-\frac{i}{\hbar}\,dt\,\hat{H})^{n}}{n!}$ up to $n=6$. The programming codes of our numerical experiments have been publicized for reference,\footnote{\url{https://github.com/Z-T-WANG/PhDThesis/tree/main/quartic\%20oscillator}} where all details can be found. For discussion concerning numerical integration of stochastic differential equations, see appendix in Ref.~\cite{MasterThesis}.
\subsection{Evaluation and Results}\label{quartic cooling results}
After establishing the model, we train the reinforcement learning agent to reduce the energy of the state using the control force $F_{\text{con}}$ through trial and error. In the following, we describe how we have set the task for the agent and how we evaluate the performance of the agent, and the results are presented, comparing the DQN algorithm with the C-DQN algorithm.

\subsubsection{Initialization}
To make sure that the initial state at the beginning of the control is a typical non-Gaussian state in the quartic potential, at the initialization, we first initialize the state as a Gaussian wave packet at the center of the potential with a random momentum ranging from $-0.3\pi\sqrt{\hbar m_c \omega_c}$ to $+0.3\pi\sqrt{\hbar m_c \omega_c}$, and then, we let the state evolve during a time which changes randomly from $\dfrac{15}{\omega_c}$ to $\dfrac{20}{\omega_c}$ with $F_{\text{con}}=0$, and lastly we use the resulting state as the initial state for the reinforcement learning agent to start to control. This procedure ensures that the initial state from the perspective of the controller is sufficiently non-Gaussian, and the initial energy of the state approximately lies in $5\hbar\omega_c$ and $7\hbar\omega_c$. We re-initialize the state if its energy exceeds $7\hbar\omega_c$. 

\subsubsection{Setting of Episodes}
Following convention in reinforcement learning, we set an \textit{episode} of the control to be $\dfrac{100}{\omega_c}$ time in simulation. In other words, we maximally allow the state to evolve for a time interval of $\dfrac{100}{\omega_c}$ under control, and after that evolution, we stop the simulation and re-initialize the state to start over again. To evaluate the performance of the controller, we calculate the average energy of the controlled state starting from time $\frac{30}{\omega_c}$ to $\frac{100}{\omega_c}$, so that transient behaviour of the control when the state is initialized is ignored.

In order to make sure that the state does not move out of the boundary of the space that is simulated, we stop the simulation and end the episode whenever the probability distribution around the boundary of the space is not negligible. Also, regarding the energy, we end the episode whenever the energy of the state exceeds $12\hbar\omega_c$. In other words, $12\hbar\omega_c$ is the maximal energy that we allow during the control, beyond which we stop the control and regard the control as having failed in this episode. In these two cases of failure, the evaluated performance, i.e.~the average energy, is considered to be $12\hbar\omega_c$. If an episode ends with failure, we set the Q value that is to be learned at the end of the episode to be the final energy of the state divided by $\frac{1}{1-\gamma}$, which is a large value, so that the reinforcement learning agent should learn to avoid getting close to these failure cases.

At the beginning of training, the control of the agent fails quickly and frequently, and as the learning makes progress, the state is kept at a low energy for a longer period of time, and finally the state can be stabilized and cooled. In training, we first train the agent until it can stabilize the state for a time interval of $\frac{100}{\omega_c}$, i.e., until it completes a whole episode without failure, and then, we train the agent for 11000 more episodes, during which we gradually reduce the learning rate and the $\epsilon$ hyperparameter of the $\epsilon$-greedy policy in Q-learning.

\subsubsection{Results}
\begin{figure}[htb]
	\centering
	\hspace*{-1.5cm}\includegraphics[width=1.16\linewidth]{"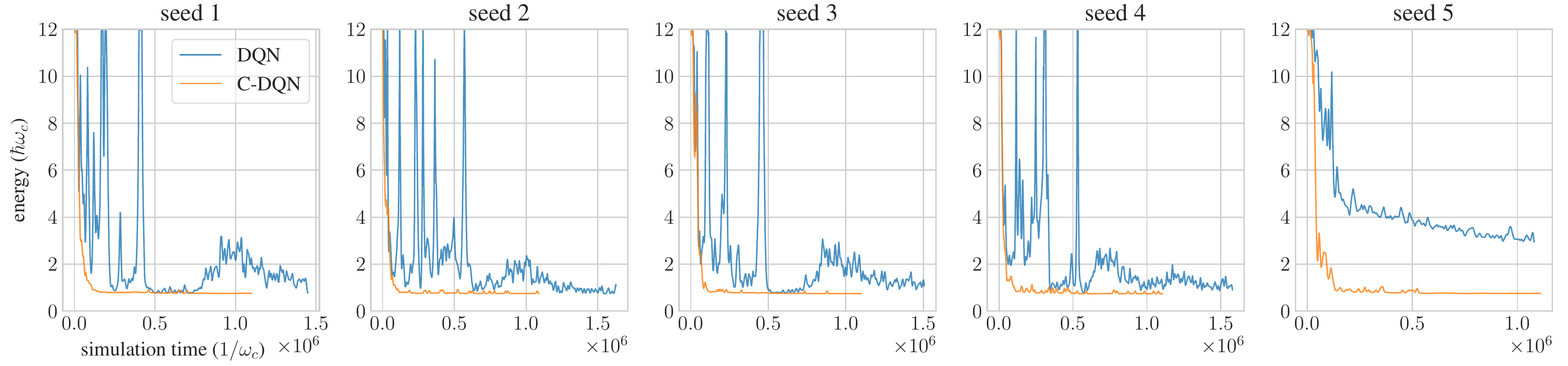"}
	\caption{Learning curves of the quartic cooling problem for the DQN algorithm and the C-DQN algorithm, with the same experimental settings, for 5 different random seeds. The abscissa shows the simulated time of the evolution of the quartic oscillator, which represents the number of learned data and the training time. The ordinate shows the average energy, which shows the performance of the controller, and a smaller value represents a better performance. Gaussian smoothing with the standard deviation of 40 is applied to smooth the performance data. The failure of stabilization and control corresponds to an energy of 12.}
	\label{quartic cooling separate 40}
\end{figure}
\begin{figure}[tb!]
	\centering
	\hspace*{-1.5cm}\includegraphics[width=1.16\linewidth]{"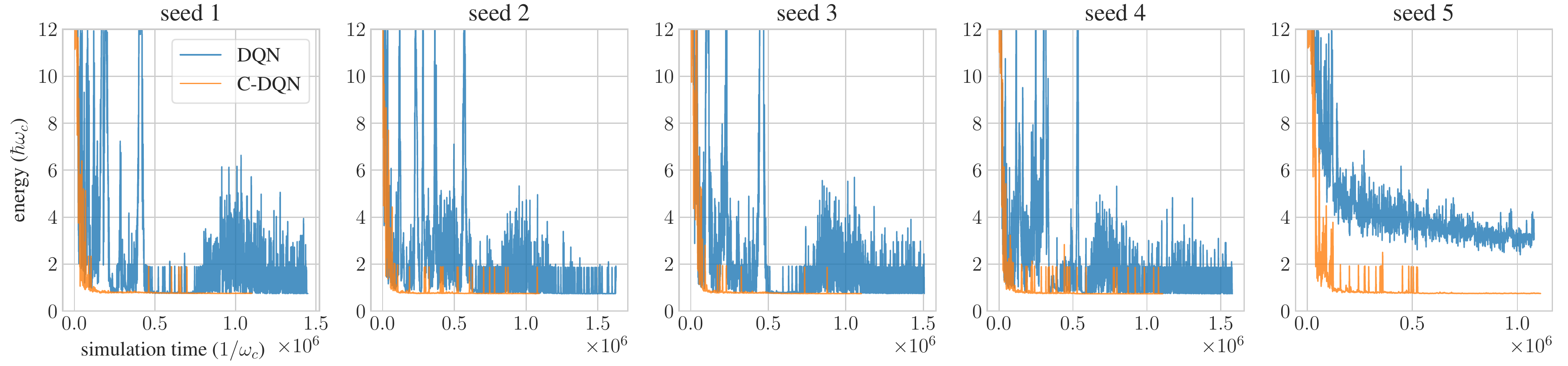"}
	\caption{Learning curves of the quartic cooling problem for the DQN algorithm and the C-DQN algorithm, for 5 different random seeds, which shows the same data as in Fig.~\ref{quartic cooling separate 40} but using Gaussian smoothing with the standard deviation of 4 to show the fluctuations in performance.}
	\label{quartic cooling separate 4}
\end{figure}
\begin{figure}[tb!]
	\centering
	\hspace*{-1.5cm}\includegraphics[width=1.16\linewidth]{"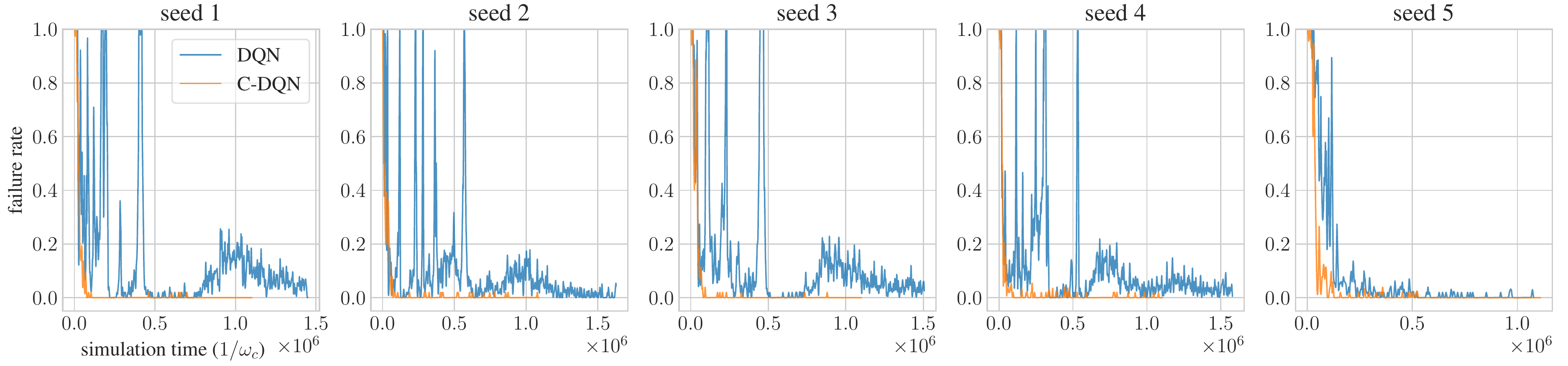"}
	\caption{Failure rate of the controller learned by the DQN algorithm and the C-DQN algorithm in the process of training, for 5 different random seeds, evaluated using a Gaussian window with the standard deviation of 20.}
	\label{quartic cooling separate failure rate}
\end{figure}
\begin{figure}[htb!]
	\centering
	\includegraphics[width=0.47\linewidth]{"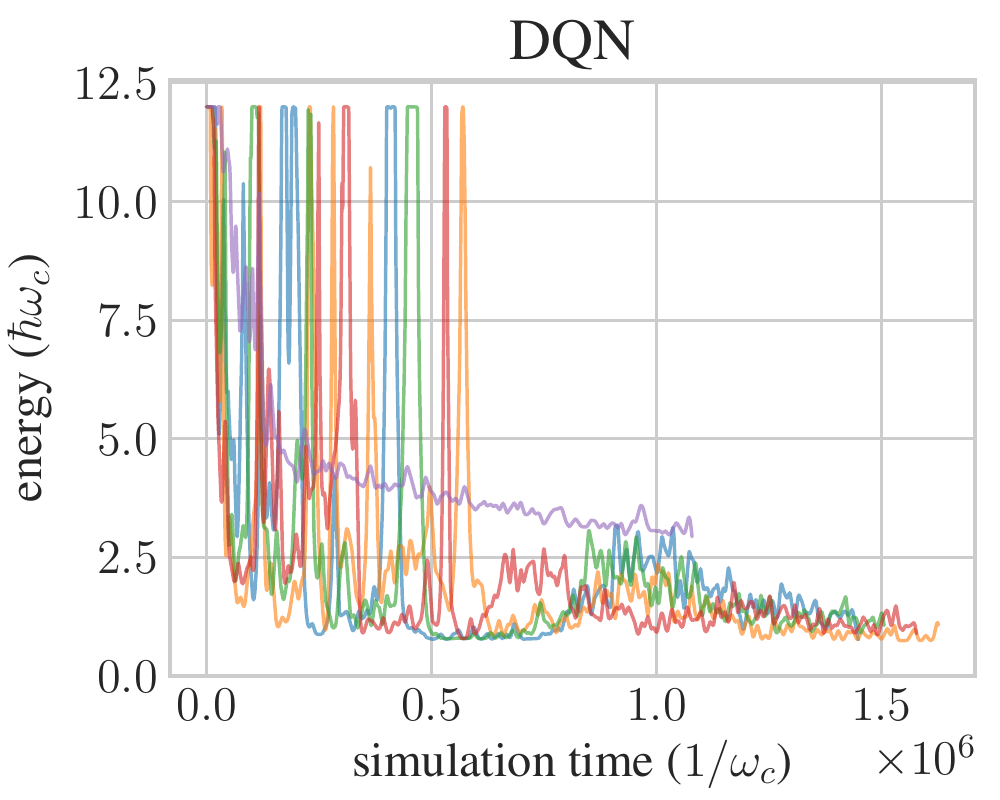"}\quad
	\includegraphics[width=0.47\linewidth]{"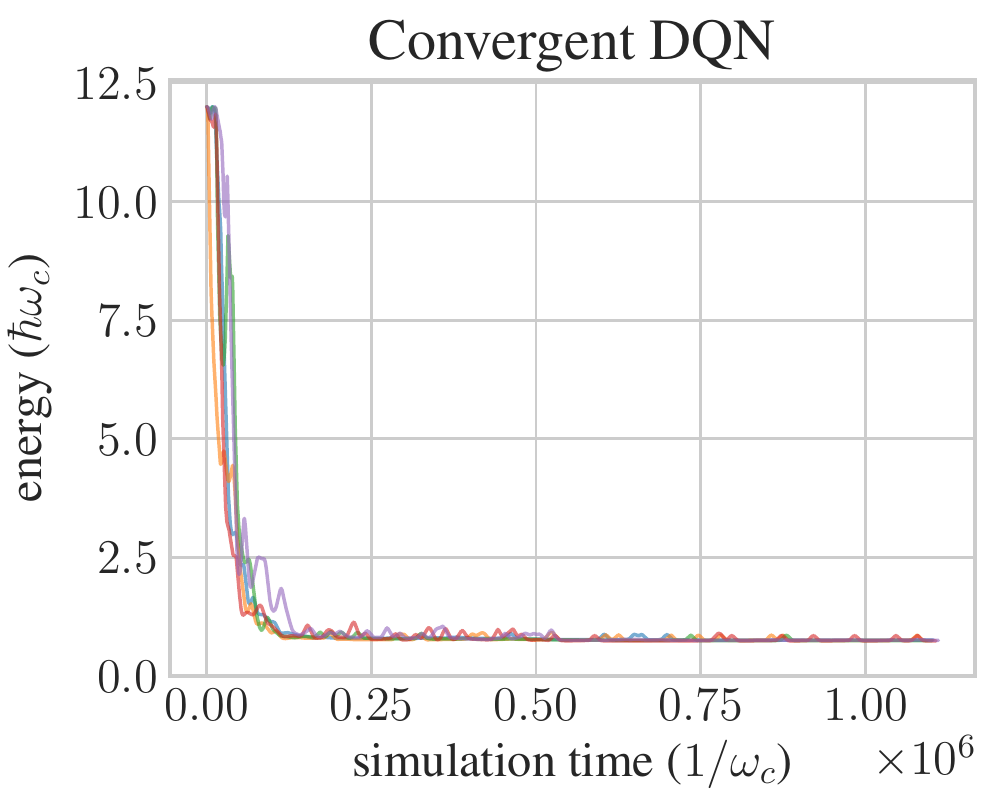"}
	\caption{Learning curves of the quartic cooling problem for the DQN algorithm (left) and the C-DQN algorithm (right). Each curve represent a repetition of the experiment with a different random seed, summarizing the results in Fig.~\ref{quartic cooling separate 40}. }
	\label{quartic cooling DQN CDQN}
\end{figure}

We repeat the experiment for 5 times using 5 different random seeds, and the learning curves of the DQN algorithm and the C-DQN algorithm in the training process are shown in Fig.~\ref{quartic cooling separate 40}, \ref{quartic cooling separate 4} and \ref{quartic cooling separate failure rate}, for each of the random seeds separately, and summarized in Fig.~\ref{quartic cooling DQN CDQN}.

In Fig.~\ref{quartic cooling separate 40}, we see that although both the DQN and the C-DQN algorithm can learn to reduce the energy of the system, the performance of C-DQN is consistently better than DQN during training. In addition, whereas the performance of C-DQN improves steadily and consistently during training, the performance of DQN does not improve consistently, and the performance fluctuates. Especially, the performance can get worse after the agent has learned how to cool the system well. To see more details, we show the performance data with less smoothing in Fig.~\ref{quartic cooling separate failure rate}. In Fig.~\ref{quartic cooling separate failure rate}, it is clear that the performance of C-DQN is very stable, while the performance of DQN strongly fluctuates in training when it does not perform well. Specifically, if we check the failure rate of the controller in training, i.e.~the probability of an episode to end with control failure, we see that the deterioration of the performance of DQN is strongly correlated with control failure, as shown in Fig.~\ref{quartic cooling separate failure rate}, except for the case of the random seed 5. Therefore, we see that the DQN algorithm does not perform stably throughout the training process, and it tends to get unstable and occasionally fails even at a later stage of training, and such instability is hard to remove and does not completely disappear after an extended period of training. 

Meanwhile, by comparing the results of the repetitions of the experiment with different random seeds, we see that the performance of DQN has a large variance with respect to the random seeds, while that of C-DQN almost has no variance, as shown in Fig.~\ref{quartic cooling DQN CDQN}. The learning curves of DQN can be qualitatively different for different random seeds, and especially, the results for the random seed 5 are qualitatively different compared with the others, as can be seen in Fig.~\ref{quartic cooling separate 40}, \ref{quartic cooling separate 4} and \ref{quartic cooling separate failure rate}. The agent does not cool the system to an energy that is as low as the others, but it performs stably throughout training and does not encounter many failures, which implies that the agent has learns differently from the others. These results show that the instability of DQN increases the randomness in its results, and therefore, reduces the reproducibility and reliability of the results, especially if the task is relatively difficult. The significant randomness in results considerably increases the difficulty of improving and fine-tuning the AI algorithm, and requires experimenters to repeat the experiments many times in order to confirm a result, and the result may change qualitatively due to minor details which are supposed to only slightly perturb the training \cite{deepReinforcementLearningReproducibility}. Also, in practical scenarios, as one usually wants a controller that is trained to have a performance as high as possible, one usually needs to run the experiments many times in order to pick the best trained controller, and therefore, the obtained performance eventually would highly depend on the number of repetitions of the experiment.

Compared with the DQN algorithm, using exactly the same experimental setting, the C-DQN algorithm does not encounter any difficulty encountered by the DQN algorithm, and the C-DQN algorithm steadily approaches a satisfactory performance without instability and with little randomness, as shown in Fig.~\ref{quartic cooling DQN CDQN}. This clearly demonstrates the stability of C-DQN and the consistency of its results, which translates into the reliability of the results of C-DQN in solving scientific problems. 

As the satisfactory stability and consistency of C-DQN have been confirmed as above for the nontrivial quantum control problem, we achieve our purpose of developing a stable and reliable deep reinforcement learning algorithm for physical control problems. In the following, we apply the C-DQN algorithm to a more complicated system which is of realistic relevance.

\section{Cooling of a Quantum-Mechanical Rigid Body}\label{rigid body control section}
With rapid development in synthesis and manipulation of nanoparticles, nanorotors have recently been experimentally realized and studied \cite{quantumRotorNature}. When the nanoparticles are sufficiently isolated from their environment and sufficiently cooled, quantum-mechanical effects of rotations are expected to be observed, and these quantum-mechanical rotors are expected to find their applications in sensing devices and fundamental tests of physical principles, because they allow an unprecedented accuracy of sensing using their rotational degrees of freedom. However, in experiments, the rotational degrees of freedom have yet to be successfully cooled to approach a quantum regime so far, and only the center-of-mass degree of freedom has been successfully cooled to approach the ground state \cite{nanoparticleGroundStateCooling1, nanoparticleGroundStateCooling2}. The dynamics of rotation is essentially nonlinear, and it is much harder to manipulate compared to the dynamics in position space. 

In the following, we consider the problem of measurement-feedback cooling of a quantum-mechanical rigid body, which is both of realistic significance and of theoretical interest. We derive the Hamiltonian of the system, describe our model, and present the results of the control learned by the DQN and the C-DQN algorithms, and compare the results with those obtained using the standard LQG control which employs a linear approximation. 
\subsection{Derivation of the Rotational Hamiltonian}\label{rigid body Hamiltonian review}
In this section, we derive the Hamiltonian of a free quantum rigid body in terms of the angles which define the orientation of the rigid body, following Ref.~\cite{quantumMechanicalRigidBody}.
\subsubsection{Euler Angles}
A standard representation of the orientation of a rigid body is given by the Euler angles $(\alpha,\beta,\gamma)$. There are several different conventions on the Euler angles, and here we follow Ref.~\cite{quantumMechanicalRigidBody} and use the z-x-z convention for the Euler rotation (see Eq.~(\ref{Euler rotation operator}) below). In this case, the rotation $(\alpha,\beta,\gamma)$ represents first rotating the rigid body around the $z$ axis of the laboratory frame through angle $\gamma$, then rotating around the $x$ axis of the laboratory frame through angle $\beta$, and finally rotating around the $z$ axis of the laboratory frame through angle $\alpha$. In terms of angular momentum operators $\hat{L}_x, \hat{L}_y, \hat{L}_z$, this Euler rotation is given by
\begin{equation}\label{Euler rotation operator}
	e^{-\frac{i}{\hbar}\alpha\hat{L}_{z}}e^{-\frac{i}{\hbar}\beta\hat{L}_{x}}e^{-\frac{i}{\hbar}\gamma\hat{L}_{z}}.
\end{equation}
If we assume that before the rotation, the rigid body is at the default position and the local $x$, $y$ and $z$ axes of the rigid body align with the $x$, $y$ and $z$ axes of the laboratory frame, then, the rotation operator $e^{-\frac{i}{\hbar}\alpha\hat{L}_{z}}e^{-\frac{i}{\hbar}\beta\hat{L}_{x}}e^{-\frac{i}{\hbar}\gamma\hat{L}_{z}}$ also represents the orientation of the rigid body in the laboratory frame after the rotation. The angle $\beta$ is called the polar angle or the zenith angle, the angle $\alpha$ is called the azimuthal angle, and the angle $\gamma$ corresponds to self-rotation of the object around its local $z$ axis.

Although the representation of Euler angles is physically straightforward, it suffers from the well-known problem called gimbal lock, or, coordinate singularity, when $\beta$ is equal to $0$ and $\pi$. The cases of $\beta$ being equal to $0$ and $\pi$ correspond to the north pole and the south pole of a spherical coordinate. When we have $\beta=0$, we have
\begin{equation}
	e^{-\frac{i}{\hbar}\alpha\hat{L}_{z}}e^{-\frac{i}{\hbar}\beta\hat{L}_{x}}e^{-\frac{i}{\hbar}\gamma\hat{L}_{z}}=e^{-\frac{i}{\hbar}(\alpha+\gamma)\hat{L}_{z}},
\end{equation}
and therefore $\alpha$ and $\gamma$ become equivalent and essentially one degree of freedom disappears. A similar situation occurs for $\beta=\pi$, in which case we have 
\begin{equation}
	e^{-\frac{i}{\hbar}\alpha\hat{L}_{z}}e^{-\frac{i}{\hbar}\beta\hat{L}_{x}}e^{-\frac{i}{\hbar}\gamma\hat{L}_{z}}=e^{-\frac{i}{\hbar}(\alpha-\gamma)\hat{L}_{z}}e^{-\frac{i}{\hbar}\beta\hat{L}_{x}},
\end{equation}
and the remaining degree of freedom becomes $\alpha-\gamma$. This issue leads to singularity when one attempts to find the wave function of a quantum state in the space of Euler angles and attempts to differentiate with respect to $\alpha$, $\beta$ and $\gamma$. Also, if one considers a smooth movement across the point $\beta=0$, it is clear that the angles $\alpha$ and $\gamma$ flip and change discontinuously, which shows that a continuous physical movement does not correspond to a continuous change in the coordinate representation using Euler angles.
\subsubsection{Quaternions}
To overcome the problem of coordinate singularity, we use quaternions in the following. The quaternion variables are related to the Euler angles by
\begin{equation}\label{quaternion angles}
	\begin{split}
		\xi=\cos\dfrac{\alpha-\gamma}{2}\sin\frac{\beta}{2},\\
		\eta=\sin\frac{\alpha-\gamma}{2}\sin\frac{\beta}{2},\\
		\zeta=\sin\frac{\alpha+\gamma}{2}\cos\frac{\beta}{2},\\
		\chi=\cos\frac{\alpha+\gamma}{2}\cos\frac{\beta}{2}.
	\end{split}
\end{equation}
It can be checked that at $\beta=0$, we have $\sin\frac{\beta}{2}=0$ and $\xi=\eta=0$, and the remaining degree of freedom becomes $\alpha+\gamma$; at $\beta=\pi$, we have $\cos\frac{\beta}{2}=0$ and $\zeta=\chi=0$, and the remaining degree of freedom becomes $\alpha-\gamma$. All coordinate singularities are removed by the quaternion representation. Nevertheless, there is a redundant degree of freedom, and we have $\xi^{2}+\eta^{2}+\zeta^{2}+\chi^{2}\equiv1$, and the rotation of $(\xi, \eta, \zeta, \chi)$ and that of $(-\xi, -\eta, -\zeta, -\chi)$ represent the same physical rotation. 

As an extension of complex numbers, in quaternion computation, the quaternions are conventionally written as
\begin{equation}
	\chi + \xi\, i + \eta\, j + \zeta\, k,
\end{equation}
where $\chi$ represents a real number, and the calculation rules are given by
\begin{equation}
	i^{2}=j^{2}=k^{2}=-1, 
\end{equation}
and
\begin{equation}
	i\times j =-j\times i = k, \quad j\times k =-k\times j = i, \quad k\times i =-i\times k = j.
\end{equation}
The inverse of a quaternion is given by
\begin{equation}
	\frac{1}{\chi + \xi\, i + \eta\, j + \zeta\, k}=\frac{\chi - \xi\, i - \eta\, j - \zeta\, k}{\xi^{2}+\eta^{2}+\zeta^{2}+\chi^{2}}.
\end{equation}

The quaternions are intimately related to rotations in three-dimensional space. For a vector $\vec{r}=(r_x,r_y,r_z)$, after rotating around a unit vector $\vec{a}=(a_x,a_y,a_z)$ by angle $\theta$, the resulting vector can be represented by
\begin{equation}
	\begin{split}
			&\left(\cos\frac{\theta}{2}+\sin\frac{\theta}{2}a_x i + \sin\frac{\theta}{2}a_y j + \sin\frac{\theta}{2}a_z k\right) \times \left(r_x i + r_y j + r_z k\right)\\
			&\times\left(\cos\frac{\theta}{2}-\sin\frac{\theta}{2}a_x i - \sin\frac{\theta}{2}a_y j - \sin\frac{\theta}{2}a_z k\right)\\
			={}& \left[\left (\cos\theta+a_x^{2}(1-\cos\theta)\right )r_x+\left(a_x a_y(1-\cos\theta)-a_z\sin\theta\right)r_y+\left(a_z a_x(1-\cos\theta)+a_y\sin\theta\right)r_z  \right] i \\
			&+\left[\left (a_x a_y(1-\cos\theta)+a_z\sin\theta\right )r_x+\left(\cos\theta+a_y^{2}(1-\cos\theta)\right)r_y+\left(a_y a_z(1-\cos\theta)-a_x\sin\theta\right)r_z  \right] j \\
			&+\left[\left (a_z a_x(1-\cos\theta)-a_y\sin\theta\right )r_x+\left(a_y a_z(1-\cos\theta)+a_x\sin\theta\right)r_y+\left(\cos\theta+a_z^{2}(1-\cos\theta)\right)r_z  \right] k,
	\end{split}
\end{equation}
where the factors before $i$, $j$ and $k$ are the $x$, $y$ and $z$ coordinates of the vector, respectively. Therefore, quaternions can be used to compute a combination of rotations. The result of rotation 2, $\chi_2 + \xi_2\, i + \eta_2\, j + \zeta_2\, k$, after rotation 1, $\chi_1 + \xi_1\, i + \eta_1\, j + \zeta_1\, k$, can be computed as
\begin{equation}
	(\chi_2 + \xi_2\, i + \eta_2\, j + \zeta_2\, k)\times(\chi_1 + \xi_1\, i + \eta_1\, j + \zeta_1\, k),
\end{equation}
since the vector $\vec{r}$ after the rotation 1 and rotation 2 is given by
\begin{equation}
		\begin{split}
			&(\chi_2 + \xi_2\, i + \eta_2\, j + \zeta_2\, k)\times(\chi_1 + \xi_1\, i + \eta_1\, j + \zeta_1\, k) \times \left(r_x i + r_y j + r_z k\right)\\
			&\times(\chi_1 - \xi_1\, i - \eta_1\, j - \zeta_1\, k)\times(\chi_2 - \xi_2\, i - \eta_2\, j - \zeta_2\, k).
		\end{split}
\end{equation}
\subsubsection{Angular Momentum Operators}
Consider an infinitesimal rotation around the $z$ axis of the laboratory frame by $\epsilon$, denoted by $\hat{R}_{z;\epsilon}$. In the quaternion representation, the rotation is given by
\begin{equation}
	\hat{R}_{z;\epsilon}=1+\sin\frac{\epsilon}{2}\,k+O(\epsilon^{2}).
\end{equation}
Therefore, given an initial orientation $\hat{R}_{0}=\chi_{0} +\xi_{0}i+ \eta_{0}j+ \zeta_{0}k$, after rotation $\hat{R}_{z;\epsilon}$, the orientation becomes
\begin{equation}
	\begin{split}
		\hat{R}_{z;\epsilon}\hat{R}_{0}= (\chi_{0}-\frac{\epsilon}{2}\zeta_{0})+(\xi_{0}-\frac{\epsilon}{2}\eta_{0})i+(\eta_{0}+\frac{\epsilon}{2}\xi_{0})j+(\zeta_{0}+\frac{\epsilon}{2}\chi_{0})k.
	\end{split}
\end{equation}
That is, the rotation $\hat{R}_{z;\epsilon}$, or in terms of the angular momentum operator, $e^{-\frac{i}{\hbar}\epsilon\hat{L}_{z}}$, shifts the quaternion variables $(\xi, \eta, \zeta, \chi)$ by an amount of $\frac{\epsilon}{2}$. Therefore, for functions defined in the space of the quaternion variables $(\xi, \eta, \zeta, \chi)$, the angular momentum operator $\hat{L}_{z}$ that generates the rotations around the $z$ axis is given by
\begin{equation}
	\hat{L}_{z}=-\frac{i\hbar}{2}\left (-\eta\frac{\partial}{\partial\xi}+\xi\frac{\partial}{\partial\eta}+\chi\frac{\partial}{\partial\zeta}-\zeta\frac{\partial}{\partial\chi}\right ).
\end{equation}
Similarly, the other angular momentum operators are given by
\begin{equation}
	\hat{L}_{x}=-\frac{i\hbar}{2}\left (+\chi\frac{\partial}{\partial\xi}-\zeta\frac{\partial}{\partial\eta}+\eta\frac{\partial}{\partial\zeta}-\xi\frac{\partial}{\partial\chi}\right ),
\end{equation}
\begin{equation}
	\hat{L}_{y}=-\frac{i\hbar}{2}\left (+\zeta\frac{\partial}{\partial\xi}+\chi\frac{\partial}{\partial\eta}-\xi\frac{\partial}{\partial\zeta}-\eta\frac{\partial}{\partial\chi}\right ).
\end{equation}
To confirm the commutation relations among the angular momentum operators, using the bilinearity of the commutator, we can easily obtain
\begin{equation}\label{L commutation1}
	\begin{split}
		[\hat{L}_{x},\hat{L}_{y}]&=\hat{L}_{x}\hat{L}_{y}-\hat{L}_{y}\hat{L}_{x}\\
		&=-\frac{\hbar^2}{2}\left (-\chi\frac{\partial}{\partial\zeta} +\zeta\frac{\partial}{\partial\chi}+\eta\frac{\partial}{\partial\xi}-\xi\frac{\partial}{\partial\eta} \right )\\
		&=i\hbar \hat{L}_{z},
	\end{split}
\end{equation}
where $[\cdot,\cdot]$ is the commutator, defined as $[A,B]:=AB-BA$. Similarly we have 
\begin{equation}\label{L commutation2}
	[\hat{L}_{y},\hat{L}_{z}]=i\hbar \hat{L}_{x},\qquad [\hat{L}_{z},\hat{L}_{x}]=i\hbar \hat{L}_{y},
\end{equation}
which are the standard commutation relations of the angular momentum operators.

Defining the total angular momentum operator $\hat{L}^{2}$ by
\begin{equation}
	\hat{L}^{2}:=\hat{L}_{x}^{2}+\hat{L}_{y}^{2}+\hat{L}_{z}^{2},
\end{equation}
we have
\begin{equation}
	[\hat{L}_{x},\hat{L}^{2}]=[\hat{L}_{y},\hat{L}^{2}]=[\hat{L}_{z},\hat{L}^{2}]=0.
\end{equation}

Besides the operators $\hat{L}_{x}$, $\hat{L}_{y}$ and $\hat{L}_{z}$, there are several other important operators. A rigid body has body-fixed principal axes which rotate together with the rigid body, and these local axes of the rigid body are associated with the principal moments of inertia $I_{x}, I_{y}, I_{z}$. To find the rotational energy of the rigid body, we need the angular momenta projected onto these principal axes, and we consider these principal axes to be the $x$, $y$ and $z$ axes of the local frame of the rigid body.

The angular momentum operators in the directions of the body-fixed local axes are associated with rotations around those local axes. Given an orientation $\hat{R}_{0}$, the orientation $\hat{R}_{0}$ can be interpreted as the orientation after rotating by $\hat{R}_{0}$ from the default position $(\xi=0, \eta=0, \zeta=0, \chi=1)$, i.e., $\alpha=\beta=\gamma=0$, at which point the body-fixed local axes align with the laboratory axes. Therefore, rotating around the local $z$ axis of a rigid body by $\epsilon$ results in the orientation $\hat{R}_{0}\hat{R}_{z;\epsilon}$. This is because the action of rotating around the local $z$ axis of a rigid body is denoted by $\hat{R}_{0}\hat{R}_{z;\epsilon}\hat{R}_{0}^{-1}$, which represents the rotation around the rotated $z$ axis, rotated by $\hat{R}_{0}$, and we have $\hat{R}_{0}\hat{R}_{z;\epsilon}\hat{R}_{0}^{-1}\hat{R}_{0}=\hat{R}_{0}\hat{R}_{z;\epsilon}$. The result means that it is the same for the body to first rotate by $\hat{R}_{0}$ and then rotate around its local $z$ axis, and to first rotate around the laboratory $z$ axis at the default position and then rotate by $\hat{R}_{0}$. The orientation after the rotation is given by
\begin{equation}
	\begin{split}
		\hat{R}_{0}\hat{R}_{z;\epsilon}&=(\xi_{0}, \eta_{0}, \zeta_{0}, \chi_{0})\cdot(0,0,\frac{\epsilon}{2},1)\\
		&= (\xi_{0}+\frac{\epsilon}{2}\eta_{0},\  \eta_{0}-\frac{\epsilon}{2}\xi_{0},\  \zeta_{0}+\frac{\epsilon}{2}\chi_{0},\  \chi_{0}-\frac{\epsilon}{2}\zeta_{0}).
	\end{split}
\end{equation}
Therefore, the corresponding angular momentum operator $\hat{Q}_{z}$ which generates the rotation around the local $z$ axis is given by
\begin{equation}
	\hat{Q}_{z}=-\frac{i\hbar}{2}\left (\eta\frac{\partial}{\partial\xi}-\xi\frac{\partial}{\partial\eta}+\chi\frac{\partial}{\partial\zeta}-\zeta\frac{\partial}{\partial\chi}\right ),
\end{equation}
and the other angular momentum operators are given by
\begin{equation}
	\hat{Q}_{x}=-\frac{i\hbar}{2}\left (\chi\frac{\partial}{\partial\xi}+\zeta\frac{\partial}{\partial\eta}-\eta\frac{\partial}{\partial\zeta}-\xi\frac{\partial}{\partial\chi}\right ),
\end{equation}
\begin{equation}
	\hat{Q}_{y}=-\frac{i\hbar}{2}\left (-\zeta\frac{\partial}{\partial\xi}+\chi\frac{\partial}{\partial\eta}+\xi\frac{\partial}{\partial\zeta}-\eta\frac{\partial}{\partial\chi}\right ).
\end{equation}
The commutation relations among these operators are given by
\begin{equation}\label{Q commutation1}
	\begin{split}
		[\hat{Q}_{x},\hat{Q}_{y}]&=\hat{Q}_{x}\hat{Q}_{y}-\hat{Q}_{y}\hat{Q}_{x}\\
		&=-\frac{\hbar^2}{2}\left (+\chi\frac{\partial}{\partial\zeta} -\zeta\frac{\partial}{\partial\chi}+\eta\frac{\partial}{\partial\xi}-\xi\frac{\partial}{\partial\eta} \right )\\
		&=-i\hbar \hat{Q}_{z},
	\end{split}
\end{equation}
and
\begin{gather}\label{Q commutation2}
		[\hat{Q}_{y},\hat{Q}_{z}]=-i\hbar \hat{Q}_{x},\qquad [\hat{Q}_{z},\hat{Q}_{x}]=-i\hbar \hat{Q}_{y},\\
		\hat{Q}^{2}:=\hat{Q}_{x}^{2}+\hat{Q}_{y}^{2}+\hat{Q}_{z}^{2}\equiv\hat{L}^{2},\\
		[\hat{Q}_{x},\hat{Q}^{2}]=[\hat{Q}_{y},\hat{Q}^{2}]=[\hat{Q}_{z},\hat{Q}^{2}]=0.
\end{gather}
We have the relation $\hat{Q}^{2}\equiv\hat{L}^{2}$, indicating that the total angular momentum is the same for both the local frame and the laboratory frame. \\

Comparing Eqs.~(\ref{L commutation1}) and (\ref{L commutation2}) with Eqs.~(\ref{Q commutation1}) and (\ref{Q commutation2}), one notices that the angular momentum operators $\hat{Q}_{x}$, $\hat{Q}_{y}$ and $\hat{Q}_{z}$ have a different sign in the commutation rules compared with the operators $\hat{L}_{x}$, $\hat{L}_{y}$ and $\hat{L}_{z}$. This is because the rotations generated by $\hat{Q}_{x}$, $\hat{Q}_{y}$ and $\hat{Q}_{z}$ are combined on the right of $\hat{R}_{0}$, while the rotations generated by $\hat{L}_{x}$, $\hat{L}_{y}$ and $\hat{L}_{z}$ are combined on the left. Other important results are
\begin{equation}
	\begin{split}
		[\hat{L}_{z},\hat{Q}_{z}]
		&=-\frac{\hbar^2}{4}\left (\left [-\eta\frac{\partial}{\partial\xi}+\xi\frac{\partial}{\partial\eta},\, +\eta\frac{\partial}{\partial\xi}-\xi\frac{\partial}{\partial\eta}\right ]+\left [\chi\frac{\partial}{\partial\zeta}-\zeta\frac{\partial}{\partial\chi},\,\chi\frac{\partial}{\partial\zeta}-\zeta\frac{\partial}{\partial\chi}\right ] \right )\\
		&=0,
	\end{split}
\end{equation}
and
\begin{equation}
	[\hat{Q}_{x},\hat{L}_{y}]=0,
\end{equation}
that is,
\begin{equation}
	[\hat{Q}_{x/y/z},\hat{L}_{x/y/z}]=0,
\end{equation}
and any of $\hat{Q}_{x}$, $\hat{Q}_{y}$ and $\hat{Q}_{z}$ commutes with all of $\hat{L}_{x}$, $\hat{L}_{y}$ and $\hat{L}_{z}$. 
Therefore, the state of a quantum-mechanical rigid body can be characterized by three quantum numbers $(\hat{L}^{2}, \hat{L}_{z}, \hat{Q}_{z})$. 
\paragraph{Physical Interpretation}Unlike a shapeless particle, e.g., an elementary particle, which can be specified by $(\hat{L}^{2}, \hat{L}_{z})$, the state of a rigid body needs three quantum numbers $(\hat{L}^{2}, \hat{L}_{z}, \hat{Q}_{z})$ to be specified because besides the angular momentum of the rotational motion, there are remaining degrees of freedom concerning how the rotation aligns with the principal axes, i.e., the shape, of the rigid body. For example, a rod can rotate around its axis of symmetry and align with the laboratory $z$ axis, having both large $\hat{L}_{z}$ and $\hat{Q}_{z}$ angular momenta; however, the axis of symmetry of the rod may rotate on the $x$-$y$ plane of the laboratory frame as well, so that the rod can have a large $\hat{L}_{z}$ angular momentum while having a zero $\hat{Q}_{z}$ angular momentum. Therefore, it is necessary to specify both the angular momentum in the laboratory frame and how the angular momentum aligns with the shape, i.e., the local axes, of the rigid body. This explains the reason why $\hat{L}_{z}$ and $\hat{Q}_{z}$ commute but they are not independent: they result from the same angular momentum projected onto the laboratory $z$ axis and onto the rigid body $z$ axis, and therefore both of them are constrained by the total angular momentum $\hat{L}^{2}$ and have the same dimension in their matrix representations.

\subsubsection{Symmetry and Conservation of Angular Momenta}
The rotational energy of a classical rigid body is given by
\begin{equation}
	\frac{1}{2}\left (I_{x}\omega_{x}^{2}+I_{y}\omega_{y}^{2}+I_{z}\omega_{z}^{2}\right ),
\end{equation}
where $\omega_{x}$, $\omega_{y}$ and $\omega_{z}$ are the angular velocities projected onto the local $x$, $y$ and $z$ axes of the rigid body, and therefore the Hamiltonian of a quantum-mechanical rigid body is given by
\begin{equation}
	\hat{H}=\frac{\hat{Q}_{x}^{2}}{2I_{x}}+\frac{\hat{Q}_{y}^{2}}{2I_{y}}+\frac{\hat{Q}_{z}^{2}}{2I_{z}}.
\end{equation}
As $\hat{H}$ commutes with $\hat{L}_{x}$, $\hat{L}_{y}$ and $\hat{L}_{z}$, the angular momenta in the laboratory frame and the total angular momentum $\hat{L}^{2}=\hat{Q}^{2}$ are conserved.

The conservation of $\hat{Q}_{x}$, $\hat{Q}_{y}$ and $\hat{Q}_{z}$ depends on the symmetry of the rigid body, i.e., the relations among $I_{x}$, $I_{y}$ and $I_{z}$. For a spherically symmetric rigid body, we have $I:=I_{x}=I_{y}=I_{z}$ and therefore
\begin{equation}
	\hat{H}=\dfrac{\hat{Q}^{2}}{2I}=\dfrac{\hat{L}^{2}}{2I}
\end{equation}
and all angular momenta are conserved. 

For an axially symmetric rigid body, we have $I_{\perp}:=I_{x}=I_{y}$ and $I_{\parallel}:=I_{z}$ and
\begin{equation}\label{axially symmetric H}
	\begin{split}
		\hat{H}&=\frac{\hat{Q}_{x}^{2}+\hat{Q}_{y}^{2}}{2I_{\perp}}+\frac{\hat{Q}_{z}^{2}}{2I_{\parallel}}\\
		&=\frac{1}{2}\left(\frac{1}{I_{\parallel}}-\frac{1}{I_{\perp}}\right)\hat{Q}_{z}^{2}+\frac{\hat{Q}^{2}}{2I_{\perp}},
	\end{split}
\end{equation}
and therefore we have 
\begin{equation}
	[\hat{Q}_{z},\hat{H}]=\left [\hat{Q}_{z},\,\frac{1}{2}\left(\frac{1}{I_{\parallel}}-\frac{1}{I_{\perp}}\right)\hat{Q}_{z}^{2}\right ] + \left[\hat{Q}_{z} ,\,\frac{\hat{Q}^{2}}{2I_{\perp}} \right]=0,
\end{equation}
showing that the angular momentum $\hat{Q}_{z}$ is conserved. In this case, the angular momentum operators $\hat{L}^{2}$, $\hat{L}_{z}$ and $\hat{Q}_{z}$ fully diagonalize the Hamiltonian. For a generic asymmetric rigid body, $\hat{H}$ does not commute with $\hat{Q}_{x}$, $\hat{Q}_{y}$ or $\hat{Q}_{z}$ and none of the angular momenta $\hat{Q}_{x}$, $\hat{Q}_{y}$ and $\hat{Q}_{z}$ is conserved. These results are consistent with the case for a classical rigid body.

\subsubsection{Expressions of the Angular Momentum Operators and the Hamiltonian in terms of Angles}
To find the expressions in terms of straightforwardly physically relevant variables, we need to take the results from the coordinates of $(\xi, \eta, \zeta, \chi)$ to the coordinates of angles and distances. First, to compensate for the redundant degree of freedom in the quaternion parametrization, we introduce an additional degree of freedom $r$ and define
\begin{equation}
	\begin{split}
		\xi&=r\sin\frac{\beta}{2}\cos\dfrac{\alpha-\gamma}{2}\\
		\eta&=r\sin\frac{\beta}{2}\sin\frac{\alpha-\gamma}{2}\\
		\zeta&=r\cos\frac{\beta}{2}\sin\frac{\alpha+\gamma}{2}\\
		\chi&=r\cos\frac{\beta}{2}\cos\frac{\alpha+\gamma}{2},
	\end{split}
\end{equation}
so that the total number of degrees of freedom becomes 4, and we have\linebreak $\xi^{2}+\eta^{2}+\zeta^{2}+\chi^{2}=r^{2}$. Next, for simplicity, we define
\begin{equation}
	\mu:=\frac{\alpha-\gamma}{2},\quad
	\nu:=\frac{\alpha+\gamma}{2},
\end{equation}
and we have 
\begin{equation}
	\alpha\equiv \nu+\mu,\quad \gamma\equiv\nu-\mu,
\end{equation}
and the relations between coordinates $(\xi, \eta, \zeta, \chi)$ and $(r,\beta, \mu, \nu)$ are given by
\begin{equation}
	\begin{split}
		\xi&=r\sin\frac{\beta}{2}\cos\mu,\\
		\eta&=r\sin\frac{\beta}{2}\sin\mu,\\
		\zeta&=r\cos\frac{\beta}{2}\sin\nu,\\
		\chi&=r\cos\frac{\beta}{2}\cos\nu,
	\end{split}
\end{equation}
and
\begin{equation}
	\begin{split}
		r&=\sqrt{\xi^{2}+\eta^{2}+\zeta^{2}+\chi^{2}},\\
		\beta&=2\arcsin\sqrt{\frac{\xi^{2}+\eta^{2}}{\xi^{2}+\eta^{2}+\zeta^{2}+\chi^{2}}}, \\
		\mu &= \arctan\frac{\eta}{\xi},\\
		\nu &= \arctan\frac{\zeta}{\chi}.
	\end{split}
\end{equation}
To transform from the coordinates from $(\xi, \eta, \zeta, \chi)$ to $(r,\beta, \mu, \nu)$, we need to replace differentials like $\dfrac{\partial}{\partial\xi}$ with $\dfrac{\partial r}{\partial \xi}\dfrac{\partial}{\partial r}+\dfrac{\partial \beta}{\partial \xi}\dfrac{\partial}{\partial \beta}+\dfrac{\partial \mu}{\partial \xi}\dfrac{\partial}{\partial \mu}+\dfrac{\partial \nu}{\partial \xi}\dfrac{\partial}{\partial \nu}$ by utilizing the following relations:
\begin{equation}
	\begin{split}
		\dfrac{\partial r}{\partial\xi}=\frac{\xi}{r},\qquad \dfrac{\partial\beta}{\partial\xi}&=\frac{2}{\tan\frac{\beta}{2}}\frac{\xi}{r^{2}},\qquad\ \ 
		\dfrac{\partial\mu}{\partial\xi}=-\frac{\sin\mu}{r\sin\frac{\beta}{2}},\quad
		\dfrac{\partial\nu}{\partial\xi}=0,\\
		\dfrac{\partial r}{\partial\eta}=\frac{\eta}{r},\qquad \dfrac{\partial\beta}{\partial\eta}&=\frac{2}{\tan\frac{\beta}{2}}\frac{\eta}{r^{2}},\qquad\ \
		\dfrac{\partial\mu}{\partial\eta}=\frac{\cos\mu}{r\sin\frac{\beta}{2}},\quad\ \ 
		\dfrac{\partial\nu}{\partial\eta}=0,\\
		\dfrac{\partial r}{\partial\zeta}=\frac{\zeta}{r},\qquad \dfrac{\partial\beta}{\partial\zeta}&=-2\tan\frac{\beta}{2}\,\frac{\zeta}{r^{2}},\quad
		\dfrac{\partial\mu}{\partial\zeta}=0,\qquad\qquad
		\dfrac{\partial\nu}{\partial\zeta}=\frac{\cos\nu}{r\cos\frac{\beta}{2}},\\
		\dfrac{\partial r}{\partial\chi}=\frac{\chi}{r},\qquad \dfrac{\partial\beta}{\partial\chi}&=-2\tan\frac{\beta}{2}\,\frac{\chi}{r^{2}},\quad
		\dfrac{\partial\mu}{\partial\chi}=0,\qquad\qquad
		\dfrac{\partial\nu}{\partial\chi}=-\frac{\sin\nu}{r\cos\frac{\beta}{2}},
	\end{split}
\end{equation}
where for simplicity, $\xi$, $\eta$, $\zeta$, and $\chi$ are considered as functions of $(r,\beta, \mu, \nu)$ on the right-hand sides of the equations.\\

Then, we can expand the expression of the angular momentum operator $\hat{Q}_{z}$
\begin{equation}
	\begin{split}
		\hat{Q}_{z}= &-\frac{i\hbar}{2}\left (+\eta\frac{\partial}{\partial\xi}-\xi\frac{\partial}{\partial\eta}+\chi\frac{\partial}{\partial\zeta}-\zeta\frac{\partial}{\partial\chi}\right ),\\
		= &-\frac{i\hbar}{2}\left[\eta\left(\dfrac{\partial r}{\partial \xi}\dfrac{\partial}{\partial r}+\dfrac{\partial \beta}{\partial \xi}\dfrac{\partial}{\partial \beta}+\dfrac{\partial \mu}{\partial \xi}\dfrac{\partial}{\partial \mu}+\dfrac{\partial \nu}{\partial \xi}\dfrac{\partial}{\partial \nu}\right)\right . \\
		&-\xi\left(\dfrac{\partial r}{\partial \eta}\dfrac{\partial}{\partial r}+\dfrac{\partial \beta}{\partial \eta}\dfrac{\partial}{\partial \beta}+\dfrac{\partial \mu}{\partial \eta}\dfrac{\partial}{\partial \mu}+\dfrac{\partial \nu}{\partial \eta}\dfrac{\partial}{\partial \nu}\right)\\
		&+\chi\left(\dfrac{\partial r}{\partial \zeta}\dfrac{\partial}{\partial r}+\dfrac{\partial \beta}{\partial \zeta}\dfrac{\partial}{\partial \beta}+\dfrac{\partial \mu}{\partial \zeta}\dfrac{\partial}{\partial \mu}+\dfrac{\partial \nu}{\partial \zeta}\dfrac{\partial}{\partial \nu}\right)\\
		&\left. -\zeta\left(\dfrac{\partial r}{\partial \chi}\dfrac{\partial}{\partial r}+\dfrac{\partial \beta}{\partial \chi}\dfrac{\partial}{\partial \beta}+\dfrac{\partial \mu}{\partial \chi}\dfrac{\partial}{\partial \mu}+\dfrac{\partial \nu}{\partial \chi}\dfrac{\partial}{\partial \nu}\right)\right].
	\end{split}
\end{equation}
The terms involving $\dfrac{\partial}{\partial r}$ in the expression of $\hat{Q}_{z}$ are found to be the following which cancel out
\begin{equation}
	\eta\xi\frac{\partial}{\partial r}-\xi\eta\frac{\partial}{\partial r}+\chi\zeta\frac{\partial}{\partial r}-\zeta\chi\frac{\partial}{\partial r}=0,
\end{equation}
and therefore, we can show that the degree of freedom $r$ is irrelevant to rotations. The same holds true for $\hat{Q}_{x}$, $\hat{Q}_{y}$, $\hat{L}_{x}$, $\hat{L}_{y}$ and $\hat{L}_{z}$, all of which are independent of $r$. After calculation, the angular momentum operators $\hat{Q}_{x}$, $\hat{Q}_{y}$ and $\hat{Q}_{z}$ are given by
\begin{equation}
	\begin{split}
		\hat{Q}_{z}= &-\frac{i\hbar}{2}\left[\eta\left(\dfrac{\partial \beta}{\partial \xi}\dfrac{\partial}{\partial \beta}+\dfrac{\partial \mu}{\partial \xi}\dfrac{\partial}{\partial \mu}\right)-\xi\left(\dfrac{\partial \beta}{\partial \eta}\dfrac{\partial}{\partial \beta}+\dfrac{\partial \mu}{\partial \eta}\dfrac{\partial}{\partial \mu}\right)\right .\\
		&\left. +\chi\left(\dfrac{\partial \beta}{\partial \zeta}\dfrac{\partial}{\partial \beta}+\dfrac{\partial \nu}{\partial \zeta}\dfrac{\partial}{\partial \nu}\right)-\zeta\left(\dfrac{\partial \beta}{\partial \chi}\dfrac{\partial}{\partial \beta}+\dfrac{\partial \nu}{\partial \chi}\dfrac{\partial}{\partial \nu}\right)\right]\\
		= &-\frac{i\hbar}{2}\left(-\frac{\partial}{\partial\mu}+\frac{\partial}{\partial\nu} \right),
	\end{split}
\end{equation}
\begin{equation}\label{Qx angles}
	\begin{split}
		\hat{Q}_{x}= &-\frac{i\hbar}{2}\left[\chi\left(\dfrac{\partial \beta}{\partial \xi}\dfrac{\partial}{\partial \beta}+\dfrac{\partial \mu}{\partial \xi}\dfrac{\partial}{\partial \mu}\right)+\zeta\left(\dfrac{\partial \beta}{\partial \eta}\dfrac{\partial}{\partial \beta}+\dfrac{\partial \mu}{\partial \eta}\dfrac{\partial}{\partial \mu}\right)\right .\\
		&\left. -\eta\left(\dfrac{\partial \beta}{\partial \zeta}\dfrac{\partial}{\partial \beta}+\dfrac{\partial \nu}{\partial \zeta}\dfrac{\partial}{\partial \nu}\right)-\xi\left(\dfrac{\partial \beta}{\partial \chi}\dfrac{\partial}{\partial \beta}+\dfrac{\partial \nu}{\partial \chi}\dfrac{\partial}{\partial \nu}\right)\right]\\
		= &-\frac{i\hbar}{2}\left(2\cos(\nu-\mu)\dfrac{\partial}{\partial \beta}+\frac{\sin(\nu-\mu)}{\tan\frac{\beta}{2}}\frac{\partial}{\partial\mu} +\tan\frac{\beta}{2}\,\sin(\nu-\mu)\frac{\partial}{\partial\nu} \right)
	\end{split}
\end{equation}
and
\begin{equation}\label{Qy angles}
	\begin{split}
		\hat{Q}_{y}= &-\frac{i\hbar}{2}\left[-\zeta\left(\dfrac{\partial \beta}{\partial \xi}\dfrac{\partial}{\partial \beta}+\dfrac{\partial \mu}{\partial \xi}\dfrac{\partial}{\partial \mu}\right)+\chi\left(\dfrac{\partial \beta}{\partial \eta}\dfrac{\partial}{\partial \beta}+\dfrac{\partial \mu}{\partial \eta}\dfrac{\partial}{\partial \mu}\right)\right .\\
		&\left. +\xi\left(\dfrac{\partial \beta}{\partial \zeta}\dfrac{\partial}{\partial \beta}+\dfrac{\partial \nu}{\partial \zeta}\dfrac{\partial}{\partial \nu}\right)-\eta\left(\dfrac{\partial \beta}{\partial \chi}\dfrac{\partial}{\partial \beta}+\dfrac{\partial \nu}{\partial \chi}\dfrac{\partial}{\partial \nu}\right)\right]\\
		= &-\frac{i\hbar}{2}\left(-2\sin(\nu-\mu)\dfrac{\partial}{\partial \beta}+\frac{\cos(\nu-\mu)}{\tan\frac{\beta}{2}}\frac{\partial}{\partial\mu} +\tan\frac{\beta}{2}\,\cos(\nu-\mu)\frac{\partial}{\partial\nu} \right).
	\end{split}
\end{equation}
The rotational Hamiltonian is then given by
\begin{equation}\label{rigid body rotational Hamiltonian}
	\hat{H}=\frac{\hat{Q}_{x}^{2}}{2I_{x}}+\frac{\hat{Q}_{y}^{2}}{2I_{y}}+\frac{\hat{Q}_{z}^{2}}{2I_{z}}.
\end{equation}
\subsection{Model}\label{rigid body model derivation}
In this section, we describe the model that we consider in our numerical experiment of cooling of a quantum rigid body, and derive its time-evolution equation.

We consider a axially symmetric trapped nanorod as in Refs.~\cite{rigidBodyMeasurementCooling,rigidBodyMeasurementCooling2, reimann2018ghz}, with $I_x=I_y$, and the axis of symmetry of the rigid body is approximately aligned with the $z$ axis of the laboratory by a trapping potential. The trap we consider is the optical dipole trap, or the optical tweezer \cite{dipoleTrap, rigidBodyMeasurementCooling}, which uses optical fields to trap the nanoparticle, and the potential is given by
\begin{equation}\label{potential dipole of rigid body}
	V\propto-\left(\vec{l}\cdot\vec{E}\right)^{2}, 
\end{equation}
where $\vec{l}$ is the vector of the axis of the rod, and $\vec{E}$ is the electric field.
\subsubsection{Hamiltonian in terms of Locally Flat Coordinates near $\beta=0$}
As the rod is approximately aligned with the $z$ axis of the laboratory frame, the angle $\beta$ is small, and therefore, we consider a set of coordinates that approximately describe the local plane around $\beta=0$, i.e., around the north pole of spherical coordinates. The coordinates that we consider should straightforwardly describe the position of the head of the rod, as well as the rotation of the rod around its axis of symmetry. Our choice of the coordinates is given by the following:
\begin{equation}
	\begin{split}
		x := {}& \beta\sin(\mu+\nu)=\beta\sin\alpha,\\
		y := {}& -\beta\cos(\mu+\nu)=-\beta\cos\alpha,\\
		\theta := {}& 2\nu.
	\end{split}
\end{equation}
We transform from the coordinates $(\mu, \nu, \beta)$ or $(\alpha,\beta,\gamma)$ to the coordinates $(x,y,\theta)$. Consider the local plane at the point of $\beta=0$ of spherical coordinates, as $\beta$ is small, with the azimuthal angle $\alpha$, the angles $\beta$ and $\alpha$ represent the radius and the azimuth of the two-dimensional polar coordinates for the local plane around $\beta=0$, whereas $x$ and $y$ correspond to the usual Euclidean coordinates.

The inverse relations are given by
\begin{equation}
	\begin{split}
		\beta &= \sqrt{x^{2}+y^{2}},\\
		\cos\alpha&=\cos(\mu+\nu) = -\frac{y}{\sqrt{x^{2}+y^{2}}},\\
		\sin\alpha&=\sin(\mu+\nu) = \frac{x}{\sqrt{x^{2}+y^{2}}},\\
		\nu&=\frac{\theta}{2},\\
		\gamma&=\nu-\mu=\theta+\arctan\dfrac{x}{y}.
	\end{split}
\end{equation}
To evaluate the partial differential operators, we use the following relations:
\begin{equation}\label{Jacobian mu nu to x y}
	\begin{split}
		\dfrac{\partial x}{\partial\beta}&=\sin(\mu+\nu),\quad\quad
		\dfrac{\partial y}{\partial\beta}=-\cos(\mu+\nu),\ \ 
		\dfrac{\partial\theta}{\partial\beta}=0,\\
		\dfrac{\partial x}{\partial\mu}&=\beta\cos(\mu+\nu),\quad
		\dfrac{\partial y}{\partial\mu}=\beta\sin(\mu+\nu),\quad
		\dfrac{\partial\theta}{\partial\mu}=0,\\
		\dfrac{\partial x}{\partial\nu}&=\beta\cos(\mu+\nu),\quad
		\dfrac{\partial y}{\partial\nu}=\beta\sin(\mu+\nu),\quad
		\dfrac{\partial\theta}{\partial\nu}=2,
	\end{split}
\end{equation}
where $\mu$, $\nu$ and $\beta$ on the right-hand sides of the equations are considered as functions of $x$, $y$ and $\theta$.

Then, we can calculate
\begin{equation}
	\begin{split}
		\hat{Q}_{x}= &-{i\hbar}\left  [\cos(\nu-\mu)\frac{\partial}{\partial\beta}+\frac{\sin(\nu-\mu)}{2\tan\frac{\beta}{2}}\frac{\partial}{\partial\mu} +\frac{\tan\frac{\beta}{2}\,\sin(\nu-\mu)}{2} \frac{\partial}{\partial\nu}\right ]\\
		= &-{i\hbar}\left [\cos(\nu-\mu)\left(\sin(\mu+\nu)\frac{\partial}{\partial x}-\cos(\mu+\nu)\frac{\partial}{\partial y} \right)\right .\\
		&+\frac{\sin(\nu-\mu)}{2\tan\frac{\beta}{2}}\left(\beta\cos(\mu+\nu)\frac{\partial}{\partial x}+\beta\sin(\mu+\nu)\frac{\partial}{\partial y} \right)\\
		&\left . +\frac{\tan\frac{\beta}{2}\,\sin(\nu-\mu)}{2}\left(\beta\cos(\mu+\nu)\frac{\partial}{\partial x}+\beta\sin(\mu+\nu)\frac{\partial}{\partial y}+2\frac{\partial}{\partial \theta} \right) \right ]\\
		= &-{i\hbar}\left [\cos\gamma\left(\frac{x}{\sqrt{x^{2}+y^{2}}}\frac{\partial}{\partial x}+\frac{y}{\sqrt{x^{2}+y^{2}}}\frac{\partial}{\partial y} \right)+\sin\gamma\left(-\frac{y}{\sqrt{x^{2}+y^{2}}}\frac{\partial}{\partial x}+\frac{x}{\sqrt{x^{2}+y^{2}}}\frac{\partial}{\partial y}\right)\right .\\
		&\left .+\left(\frac{1}{2\tan\frac{\beta}{2}}-\frac{1}{\beta}+\frac{\tan\frac{\beta}{2}}{2} \right) \sin\gamma\left(-y\frac{\partial}{\partial x}+x\frac{\partial}{\partial y} \right) +\tan\frac{\beta}{2} \sin\gamma\frac{\partial}{\partial \theta}  \right ].
	\end{split}
\end{equation}
For simplicity, we define
\begin{equation}
	g(\beta) := \frac{1}{2\tan\frac{\beta}{2}}-\frac{1}{\beta}+\frac{\tan\frac{\beta}{2}}{2},
\end{equation}
and use
\begin{equation}
	\beta\equiv \sqrt{x^{2}+y^{2}},
\end{equation}
and we have
\begin{equation}
	\begin{split}
		\hat{Q}_{x}= &-{i\hbar}\left [\cos\gamma\left(\frac{x}{\sqrt{x^{2}+y^{2}}}\frac{\partial}{\partial x}+\frac{y}{\sqrt{x^{2}+y^{2}}}\frac{\partial}{\partial y} \right)+\sin\gamma\left(-\frac{y}{\sqrt{x^{2}+y^{2}}}\frac{\partial}{\partial x}+\frac{x}{\sqrt{x^{2}+y^{2}}}\frac{\partial}{\partial y}\right)\right .\\
		&\left .+g(\beta) \sin\gamma\left(-y\frac{\partial}{\partial x}+x\frac{\partial}{\partial y} \right) +\tan\frac{\beta}{2} \sin\gamma\frac{\partial}{\partial \theta}  \right ].\\
		= &-{i\hbar}\left [\left(\frac{x\cos\gamma-y\sin\gamma}{\sqrt{x^{2}+y^{2}}}-g(\beta)y \sin\gamma\right)\frac{\partial}{\partial x} +\left(\frac{x\sin\gamma+y\cos\gamma}{\sqrt{x^{2}+y^{2}}}+g(\beta)x\sin\gamma \right)\frac{\partial}{\partial y}\right .\\
		&\left . +\tan\frac{\beta}{2}\sin\gamma\frac{\partial}{\partial \theta} \right ]\\
		= {}&{i\hbar}\left \{\left[\sin\theta+y(x\cos\theta+y\sin\theta)\frac{g(\beta)}{\beta}\right]\frac{\partial}{\partial x} -\left[\cos\theta+x(x\cos\theta+y\sin\theta)\frac{g(\beta)}{\beta} \right]\frac{\partial}{\partial y}\right .\\
		&\left . -(x\cos\theta+y\sin\theta)\frac{\tan\frac{\beta}{2}}{\beta}\frac{\partial}{\partial \theta} \right \}.
	\end{split}
\end{equation}
The angular momentum operator $\hat{Q}_{y}$ is given by
\begin{equation}
	\begin{split}
		\hat{Q}_{y}	= {}&{i\hbar}\left \{\left[\cos\theta+y(y\cos\theta-x\sin\theta)\frac{g(\beta)}{\beta}\right]\frac{\partial}{\partial x} +\left[\sin\theta-x(y\cos\theta-x\sin\theta)\frac{g(\beta)}{\beta} \right]\frac{\partial}{\partial y}\right .\\
		&\left . -(y\cos\theta-x\sin\theta)\frac{\tan\frac{\beta}{2}}{\beta}\frac{\partial}{\partial \theta} \right \},
	\end{split}
\end{equation}
whereas $\hat{Q}_{z}$ is simply given by
\begin{equation}\label{physical Qz}
	\begin{split}
		\hat{Q}_{z}=&-{i\hbar}\frac{\partial}{\partial\theta}.
	\end{split}
\end{equation}
It can be confirmed that the commutation relations
\begin{equation}
		[\hat{Q}_{x},\hat{Q}_{y}]=-i\hbar \hat{Q}_{z},\qquad[\hat{Q}_{y},\hat{Q}_{z}]=-i\hbar \hat{Q}_{x},\qquad [\hat{Q}_{z},\hat{Q}_{x}]=-i\hbar \hat{Q}_{y}
\end{equation}
indeed hold.

To calculate the Hamiltonian, we need to evaluate the squared operators $\hat{Q}_{x}^{2}$, $\hat{Q}_{y}^{2}$ and $\hat{Q}_{z}^{2}$ which involve multiple differentiations. To proceed, for simplicity, we define
\begin{equation}
	h:= x\cos\theta+y\sin\theta,\qquad k:= y\cos\theta-x\sin\theta,
\end{equation}
and
\begin{equation}
	g_1 := \frac{1}{x}\left(\frac{\partial}{\partial x} \left( \frac{g}{\beta}\right)\right) \equiv \frac{1}{y}\left(\frac{\partial}{\partial y} \left( \frac{g}{\beta}\right)\right),	\qquad g_2 := \frac{1}{x}\left(\frac{\partial}{\partial x} \left(\frac{\tan\frac{\beta}{2}}{\beta}\right)\right) \equiv \frac{1}{y}\left(\frac{\partial}{\partial y} \left( \frac{\tan\frac{\beta}{2}}{\beta}\right)\right),
\end{equation}
and the expressions for $\hat{Q}_{x}^{2}$ and $\hat{Q}_{y}^{2}$ are explicitly given by
\begin{equation}
	\begin{split}
		\hat{Q}_{x}^{2}= &-\hbar^{2}\left \{\left[\sin\theta+yh\frac{g}{\beta}\right]^{2}\partial_{x}^{2} +\left[\cos\theta+xh\frac{g}{\beta} \right]^{2}\partial_{y}^{2}
		+ h^{2}\left(\frac{\tan\frac{\beta}{2}}{\beta}\right)^{2}\partial_{\theta}^{2}\right .\\
		&\qquad-2\left[\sin\theta\cos\theta+\left((y^{2}+x^{2})\sin\theta\cos\theta+xy \right)\frac{g}{\beta}+xyh^{2}\left( \frac{g}{\beta}\right)^{2} \right]\partial_{x}\partial_{y}\\
		&\qquad-\left(\sin\theta+yh\frac{g}{\beta}\right)h\frac{2\tan\frac{\beta}{2}}{\beta}\partial_{x}\partial_{\theta}+\left(\cos\theta+xh\frac{g}{\beta} \right)h\frac{2\tan\frac{\beta}{2}}{\beta}\partial_{y}\partial_{\theta}\\
		&\qquad -h\left[\cos\theta\left(\frac{g}{\beta}+\frac{\tan\frac{\beta}{2}}{\beta}\right)+\left(\left(x^{2}-y^{2}\right)\cos\theta+2xy\sin\theta\right)\left(\frac{g}{\beta}\right)^{2}+yk\left(g_1+\frac{\tan\frac{\beta}{2}}{\beta}\frac{g}{\beta}\right)       \right] \partial_x\\
		&\qquad-h\left[\sin\theta\left(\frac{g}{\beta}+\frac{\tan\frac{\beta}{2}}{\beta}\right)-\left(\left(x^{2}-y^{2}\right)\sin\theta-2xy\cos\theta\right)\left(\frac{g}{\beta}\right)^{2}-xk\left(g_1+\frac{\tan\frac{\beta}{2}}{\beta}\frac{g}{\beta}\right)       \right]\partial_y\\
		&\left .\qquad-hk\left[\frac{g\tan\frac{\beta}{2}}{\beta^{2}}-g_2-\left(\frac{\tan\frac{\beta}{2}}{\beta}\right)^{2} \right]\partial_\theta \right\},
	\end{split}
\end{equation}
\begin{equation}
	\begin{split}
		\hat{Q}_{y}^{2}	= &-\hbar^{2}\left \{\left[\cos\theta+yk\frac{g}{\beta}\right]^{2}\partial_{x}^{2} +\left[\sin\theta-xk\frac{g}{\beta} \right]^{2}\partial_{y}^{2}
		+ k^{2}\left(\frac{\tan\frac{\beta}{2}}{\beta}\right)^{2}\partial_{\theta}^{2}\right .\\
		&\qquad+2\left[\sin\theta\cos\theta+\left((y^{2}+x^{2})\sin\theta\cos\theta-xy \right)\frac{g}{\beta}-xyk^{2}\left( \frac{g}{\beta}\right)^{2} \right]\partial_{x}\partial_{y}\\
		&\qquad-\left(\cos\theta+yk\frac{g}{\beta}\right)k\frac{2\tan\frac{\beta}{2}}{\beta}\partial_{x}\partial_{\theta}-\left(\sin\theta-xk\frac{g}{\beta} \right)k\frac{2\tan\frac{\beta}{2}}{\beta}\partial_{y}\partial_{\theta}\\
		&\qquad +k\left[\sin\theta\left(\frac{g}{\beta}+\frac{\tan\frac{\beta}{2}}{\beta}\right)+\left(\left(x^{2}-y^{2}\right)\sin\theta-2xy\cos\theta\right)\left(\frac{g}{\beta}\right)^{2}+yh\left(g_1+\frac{\tan\frac{\beta}{2}}{\beta}\frac{g}{\beta}\right)       \right] \partial_x\\
		&\qquad-k\left[\cos\theta\left(\frac{g}{\beta}+\frac{\tan\frac{\beta}{2}}{\beta}\right)-\left(\left(x^{2}-y^{2}\right)\cos\theta+2xy\sin\theta\right)\left(\frac{g}{\beta}\right)^{2}+xh\left(g_1+\frac{\tan\frac{\beta}{2}}{\beta}\frac{g}{\beta}\right)       \right]\partial_y\\
		&\left .\qquad+hk\left[\frac{g\tan\frac{\beta}{2}}{\beta^{2}}-g_2-\left(\frac{\tan\frac{\beta}{2}}{\beta}\right)^{2} \right]\partial_\theta \right\}.
	\end{split}
\end{equation}

We consider an axially symmetric rigid body, so that we have $I_{\perp}:=I_{x}=I_{y}$ and $I_{\parallel}:=I_{z}$, and the Hamiltonian is given by
\begin{equation}
		\hat{H}=\frac{\hat{Q}_{x}^{2}+\hat{Q}_{y}^{2}}{2I_{\perp}}+\frac{\hat{Q}_{z}^{2}}{2I_{\parallel}} + V.
\end{equation}
Here the term $\hat{Q}_{x}^{2}+\hat{Q}_{y}^{2}$ is calculated to be
\begin{equation}\label{Q_x^2+Q_y^2 exact}
	\begin{split}
		\hat{Q}_{x}^{2}+\hat{Q}_{y}^{2}= &-\hbar^{2}\left \{\left[1+2y^{2}\frac{g}{\beta}+y^{2}g^{2}\right]^{2}\partial_{x}^{2} +\left[1+2x^{2}\frac{g}{\beta}+x^{2}g^{2} \right]^{2}\partial_{y}^{2}
		+ \left(\tan\frac{\beta}{2}\right)^{2}\partial_{\theta}^{2}\right .\\
		&\qquad-2xy\left[\frac{2g}{\beta}+g^{2} \right]\partial_{x}\partial_{y}-y\left(1+g\beta\right)\frac{2\tan\frac{\beta}{2}}{\beta}\partial_{x}\partial_{\theta}+x\left(1+g\beta\right)\frac{2\tan\frac{\beta}{2}}{\beta}\partial_{y}\partial_{\theta}\\
		&\left.\qquad -x\left(\frac{g}{\beta}+\frac{\tan\frac{\beta}{2}}{\beta} +g^{2} \right) \partial_x-y\left(\frac{g}{\beta}+\frac{\tan\frac{\beta}{2}}{\beta} +g^{2} \right)\partial_y \right\},
	\end{split}
\end{equation}
which is the main result of our derivation. The expression does not involve explicit functions of $\theta$, which is consistent with the fact that the Hamiltonian commutes with $\hat{Q}_{z}$.

The last term $\hat{Q}_{z}^{2}$ in the Hamiltonian is given by
\begin{equation}\label{Q_z^2}
	\hat{Q}_{z}^{2}=-\hbar^{2}\partial_\theta^{2}.
\end{equation}

\subsubsection{Properties of the Hamiltonian}
To see the properties of the Hamiltonian, we calculate the leading terms of the Taylor series of $g$ and $\tan\frac{\beta}{2}$. They are given by
\begin{equation}
	g(\beta)=\frac{\beta}{6}+\frac{7}{360}\beta^{3}+\frac{31}{15120}\beta^{5}+\frac{127}{604800}\beta^{7}+\frac{73}{3421440}\beta^{9}+O(\beta^{11}),
\end{equation}
and
\begin{equation}
	\tan\frac{\beta}{2} = \frac{\beta}{2}+\frac{\beta^{3}}{24}+\frac{\beta^{5}}{240} + \frac{17}{40320}\beta^{7}+\frac{31}{725760}\beta^{9}+O(\beta^{11}),
\end{equation}
showing that the high-order terms decay rapidly, as the angle $\beta$ is the polar angle of spherical coordinates, which necessarily satisfies $\beta\in[0,\pi]$. 

Up to the second order of $\beta$, the term $\hat{Q}_{x}^{2}+\hat{Q}_{y}^{2}$ is given by
\begin{equation}
	\begin{split}
		\hat{Q}_{x}^{2}+\hat{Q}_{y}^{2}	= &-\hbar^{2}\left [\left(1+\frac{y^{2}}{3}\right) \frac{\partial^{2}}{\partial x^{2}}+\left(1+\frac{x^{2}}{3}\right)\frac{\partial^{2}}{\partial y^{2}}+ \frac{x^{2}+y^{2}}{4}\frac{\partial^{2}}{\partial \theta^{2}}-\frac{2xy}{3}\frac{\partial^{2}}{\partial x\,\partial y}\right .\\
		&\left .-y\frac{\partial^{2}}{\partial x\,\partial\theta}+x\frac{\partial^{2}}{\partial y\,\partial\theta} -\frac{2x}{3}\frac{\partial}{\partial x}-\frac{2y}{3}\frac{\partial}{\partial y}\right]+O(\beta^{3}),
	\end{split}
\end{equation}
where we have used $x\sim O(\beta)$ and $y\sim O(\beta)$ because of $\beta=\sqrt{x^{2}+y^{2}}$. When $\beta$ is small, the state of the rigid body is localized around $\beta=0$ in a small region, and, therefore, $x$ and $y$ become small, whereas $\partial_x$ and $\partial_y$ can become large. If we keep the terms involving $\partial_\theta$ and ignore the other small terms, the above equation becomes
\begin{equation}\label{Q_x^2+Q_y^2 lowest order}
	\hat{Q}_{x}^{2}+\hat{Q}_{y}^{2}	\approx -\hbar^{2}\left ( \frac{\partial^{2}}{\partial x^{2}}+\frac{\partial^{2}}{\partial y^{2}}+ \frac{x^{2}+y^{2}}{4}\frac{\partial^{2}}{\partial \theta^{2}}-y\frac{\partial^{2}}{\partial x\,\partial\theta}+x\frac{\partial^{2}}{\partial y\,\partial\theta} \right),
\end{equation}
which is equivalent to the Hamiltonian of a charged particle in a magnetic field. As the Hamiltonian commutes with $\hat{Q}_z\equiv-{i\hbar}{\partial_\theta} $, we assume that the state we investigate is an eigenstate of $\hat{Q}_z$ so that we can regard $\hat{Q}_z$ as a constant. Then, we map Eq.~(\ref{Q_x^2+Q_y^2 lowest order}) to the Hamiltonian of a charged particle in a magnetic field. Since the Hamiltonian of a charged particle in a magnetic field is given by
\begin{equation}
	H=\frac{1}{2m}\left(\vec{p}-\frac{Q}{c}\vec{A}\right)^{2},   
\end{equation}
the term $\frac{Q}{c}\vec{A}$ in our case should be
\begin{equation}
	\frac{Q}{c}\vec{A}=\left(-\frac{y}{2}\hat{Q}_z,\,+\frac{x}{2}\hat{Q}_z,\,0\right), 
\end{equation}
and we have
\begin{equation}
	\nabla\times\frac{Q}{c}\vec{A}=\left(0,\,0,\,\hat{Q}_z\right), 
\end{equation}
which corresponds to a magnetic field that drives the particle to rotate counter-clockwise, given a positive value of $\hat{Q}_z$. Therefore, up to the lowest order with respect to a small $\beta$, the system of the quantum-mechanical rigid body behaves in the same way as a charged particle in a magnetic field, and the strength of the magnetic field is determined by the angular momentum $\hat{Q}_z$. 

\subsubsection{Metric, Hermiticity and Momentum Operators}
The space of Euler angles is non-Euclidean, and when we integrate a function $f$ over the space, we should include the factor resulting from the metric, and the integration is given by
\begin{equation}
	\int_{0}^{2\pi}\int_{0}^{2\pi}\int_{0}^{\pi}f\sin\beta\,d\beta\,d\alpha\,d\gamma.
\end{equation}
In terms of the coordinates $(\beta,\mu,\nu)$, the integration is given by
\begin{equation}
	\int_{0}^{\pi}\int_{\mu}^{\mu+2\pi}\int_{0}^{\pi}f\cdot2\sin\beta\,d\beta\,d\nu\,d\mu.
\end{equation}
The determinant of the Jacobian matrix given by Eq.~(\ref{Jacobian mu nu to x y}) is equal to $2\beta$, and therefore, in terms of the coordinates $(x,y,\theta)$, the integration is given by
\begin{equation}
	\int_{0}^{2\pi}\left (\iint_{\sqrt{x^{2}+y^{2}}\le\pi}f\cdot\frac{\sin\sqrt{x^{2}+y^{2}}}{\sqrt{x^{2}+y^{2}}}\,dx\,dy\right )\,d\theta,
\end{equation}
and we see that the metric term $\frac{\sin\sqrt{x^{2}+y^{2}}}{\sqrt{x^{2}+y^{2}}}$ is approximately equal to $1$ when both $x$ and $y$ are small, showing that the space is indeed approximately flat near $\beta=0$. 

Due to the nontrivial metric term that one needs to take into account when one performs the integration, the inner product $\langle\phi|\psi\rangle$ should also be calculated using
\begin{equation}
	\int_{0}^{2\pi}\iint_{\sqrt{x^{2}+y^{2}}\le\pi}\phi^{*}\psi\cdot\frac{\sin\sqrt{x^{2}+y^{2}}}{\sqrt{x^{2}+y^{2}}}\,dx\,dy\,d\theta, 
\end{equation}
and, therefore, usual operators like $-{i}{\hbar}\frac{\partial}{\partial x}$ and $-{i}{\hbar}\frac{\partial}{\partial y}$ are no longer hermitian, as one generally has
\begin{equation}
	\begin{split}
			\int_{x_0}^{x_1} \phi^{*}\left (\frac{\partial}{\partial x}\psi\right )\cdot\frac{\sin\sqrt{x^{2}+y^{2}}}{\sqrt{x^{2}+y^{2}}}\,dx \ne -\int_{x_0}^{x_1} \left (\frac{\partial}{\partial x}\phi^{*}\right )\psi\cdot\frac{\sin\sqrt{x^{2}+y^{2}}}{\sqrt{x^{2}+y^{2}}}\,dx
	\end{split}
\end{equation}
even if $\phi$ and $\psi$ vanish at $x_0$ and $x_1$, subject to the vanishing boundary condition. Instead, integration by parts yields
\begin{equation}
	\int_{x_0}^{x_1} \phi^{*}\left (\frac{\partial}{\partial x}\psi\right )\cdot\frac{\sin\sqrt{x^{2}+y^{2}}}{\sqrt{x^{2}+y^{2}}}\,dx = -\int_{x_0}^{x_1} \left (\frac{\partial}{\partial x}\left (\phi^{*}\frac{\sin\sqrt{x^{2}+y^{2}}}{\sqrt{x^{2}+y^{2}}}\right )\right )\psi\,dx,
\end{equation}
and an additional term $\phi^{*}\left (\frac{\partial}{\partial x}\frac{\sin\sqrt{x^{2}+y^{2}}}{\sqrt{x^{2}+y^{2}}}\right )$ appears.

With the nontrivial metric, the hermitian momentum operators conjugate to the position operators $x$ and $y$ can be given by
\begin{equation}\label{momentum operator rigid body x}
	\hat{p}_x := -i\hbar\left(\frac{\partial}{\partial x} + \frac{1}{2}\left(\frac{\partial}{\partial x} \frac{\sin\sqrt{x^{2}+y^{2}}}{\sqrt{x^{2}+y^{2}}}\right ) \right),
\end{equation}
\begin{equation}\label{momentum operator rigid body y}
	\hat{p}_y := -i\hbar\left(\frac{\partial}{\partial y} + \frac{1}{2}\left(\frac{\partial}{\partial y} \frac{\sin\sqrt{x^{2}+y^{2}}}{\sqrt{x^{2}+y^{2}}}\right ) \right).
\end{equation}
To confirm the hermiticity, using the vanishing boundary condition and integration by parts, we calculate
\begin{equation}
	\begin{split}
		&\int_{x_0}^{x_1} \phi^{*}\left (\hat{p}_x\psi\right )\cdot\frac{\sin\sqrt{x^{2}+y^{2}}}{\sqrt{x^{2}+y^{2}}}\,dx\\
		=&\int_{x_0}^{x_1} \phi^{*}\left (-i\hbar\left(\frac{\partial}{\partial x}\psi + \frac{1}{2}\left(\frac{\partial}{\partial x} \frac{\sin\sqrt{x^{2}+y^{2}}}{\sqrt{x^{2}+y^{2}}}\right )\psi \right)\right )\cdot\frac{\sin\sqrt{x^{2}+y^{2}}}{\sqrt{x^{2}+y^{2}}}\,dx \\
		=&\int_{x_0}^{x_1} i\hbar\left(\frac{\partial}{\partial x}\left( \phi^{*}\frac{\sin\sqrt{x^{2}+y^{2}}}{\sqrt{x^{2}+y^{2}}}\right)\right)\psi\,dx -\frac{i\hbar}{2}\int_{x_0}^{x_1} \phi^{*}\psi\left ( \frac{\partial}{\partial x} \frac{\sin\sqrt{x^{2}+y^{2}}}{\sqrt{x^{2}+y^{2}}} \right)\cdot\frac{\sin\sqrt{x^{2}+y^{2}}}{\sqrt{x^{2}+y^{2}}}\,dx\\
		=&\int_{x_0}^{x_1} i\hbar\left(\frac{\partial}{\partial x} \phi^{*}\right)\psi\cdot\frac{\sin\sqrt{x^{2}+y^{2}}}{\sqrt{x^{2}+y^{2}}}\,dx +\frac{i\hbar}{2}\int_{x_0}^{x_1} \phi^{*}\psi\left ( \frac{\partial}{\partial x} \frac{\sin\sqrt{x^{2}+y^{2}}}{\sqrt{x^{2}+y^{2}}} \right)\cdot\frac{\sin\sqrt{x^{2}+y^{2}}}{\sqrt{x^{2}+y^{2}}}\,dx\\
		=&\int_{x_0}^{x_1} i\hbar\left(\frac{\partial}{\partial x} \phi^{*}+\frac{1}{2}\left(\frac{\partial}{\partial x}\frac{\sin\sqrt{x^{2}+y^{2}}}{\sqrt{x^{2}+y^{2}}}\right)\phi^{*}\right)\psi\cdot\frac{\sin\sqrt{x^{2}+y^{2}}}{\sqrt{x^{2}+y^{2}}}\,dx\\
		=&\int_{x_0}^{x_1} \left (\hat{p}_x\phi\right)^{*} \psi\cdot\frac{\sin\sqrt{x^{2}+y^{2}}}{\sqrt{x^{2}+y^{2}}}\,dx,
	\end{split}
\end{equation}
which confirms the hermiticity.

The momentum operators $\hat{p}_x$ and $\hat{p}_y$ satisfy the usual commutation relations
\begin{equation}
	\left[\hat{p}_x, \hat{p}_y\right] = 0,\quad 	\left[x, \hat{p}_x\right] = i\hbar,\quad \left[y, \hat{p}_y\right] = i\hbar, \quad \left[x, \hat{p}_y\right] =\left[y, \hat{p}_x\right] = 0.
\end{equation}
We use these operators to compute the distribution moments of the wave function in our numerical algorithm in the following sections.

\subsubsection{Control Fields}
In the following, we proceed to describe the model of the trapped and controlled rigid body that we consider. 

We consider a nanorod trapped by an electromagnetic field through dipole-induced dipole interaction using a laser, and the potential is given by Eq.~(\ref{potential dipole of rigid body}), i.e. 
\begin{equation}
	V\propto-\left(\vec{l}\cdot\vec{E}\right)^{2}, 
\end{equation}
where $\vec{l}$ is the vector of the axis of the rod, and $\vec{E}$ is the electric field. In order to control the rotation of the rigid body, specifically, to control the movement of the rigid body in $x$ and $y$ coordinates discussed above, we consider two additional lasers which can shift the center of the trapping potential in the space spanned by $x$ and $y$. The trapping laser has a fixed intensity $E_z$ with the polarization direction $z$, and the two control lasers have tunable intensities $E_x$ and $E_y$ with the polarization directions $x$ and $y$. The two control lasers are arranged perpendicularly, and the three laser beams are phased-locked and arranged on the same plane, the $x$-$y$ plane, and they intersect at the position of the rigid body, creating a field $\vec{E}=(E_x,E_y,E_y)$ at the rigid body. Then, we use $E_x$ and $E_y$ as the relevant control variables to control the time evolution of the system and cool the system. 

The position of the head of the nanorod relative to the center of the rod is given by
\begin{equation}
	\left (\frac{xl}{2}\frac{\sin\sqrt{x^{2}+y^{2}}}{\sqrt{x^{2}+y^{2}}},\, \frac{yl}{2}\frac{\sin\sqrt{x^{2}+y^{2}}}{\sqrt{x^{2}+y^{2}}},\, \frac{l}{2}\cos\sqrt{x^{2}+y^{2}}\right ),
\end{equation}
where $l$ is the length of the rigid body. Therefore, the potential is given by
\begin{equation}\label{potential exact}
	V=-\left(x\frac{\sin\beta}{\beta}E_x+y\frac{\sin\beta}{\beta}E_y+\cos\beta E_z\right)^{2}, \qquad \beta\equiv \sqrt{x^{2}+y^{2}},
\end{equation}
where the other coefficients have been absorbed into the coefficients $E_x$, $E_y$ and $E_z$. With $E_x=E_y=0$, up to the lowest order in $\beta$, we have 
\begin{equation}\label{potential harmonic}
	V = -E_z^{2}+E_z^{2}\beta^{2}=\frac{k}{2}(x^{2}+y^{2})-E_z^{2}+O(\beta^{4}), \qquad k:=2E_z^{2},
\end{equation}
where we have used $x\sim O(\beta)$ and $y\sim O(\beta)$, and $k$ is the strength of the trapping potential. Therefore, in the neighbourhood of $\beta=0$, the trapping potential is the standard harmonic potential up to a shift of the constant $-E_z^{2}$. 

With nonzero $E_x$ and $E_y$, we have
\begin{equation}\label{potential lowest order with control}
	V=E_z^{2}\left[\left(x-\frac{E_x}{E_z}\right)^{2}+ \left(y-\frac{E_y}{E_z}\right)^{2}\right]-E_z^{2}-E_x^{2}-E_y^{2} +O(\beta^{3}),
\end{equation}
where we have assumed $\dfrac{E_x}{E_z}\sim O(\beta)$ and $\dfrac{E_y}{E_z}\sim O(\beta)$. The result shows that the center of the potential is shifted by $\dfrac{E_x}{E_z}$ and $\dfrac{E_y}{E_z}$ in the $x$ and $y$ directions.

\subsubsection{Measurements}
To our knowledge, so far there is not any widely accepted model of quantum-mechanical measurement for a rigid body system, and in our model, we simply assume that the measurement is Gaussian with respect to the $x$ and $y$ coordinates, which represent the position of the head of the quantum-mechanical rod, and we assume that the measurement efficiency is unity.

The measurement can be regarded as Gaussian if the following assumptions hold true: (1) the polar angle $\beta$ is small so that the space spanned by $x$ and $y$ is isotropic; (2) the measurement is week and repetitive so that it can be regarded as approximately continuous; (3) the state of the rigid body is sufficiently localized and the variance of measurement outcomes is not large, so that one can take the average of measurement outcomes without ambiguity in the space of angles. 

One simple example of the measurement that measures the position of the head of the nanorod is given in Ref.~\cite{rigidBodyMeasurementCooling}, where a probe light shines at the rigid body in the $z$ direction and the direction of the scattered light is probed. Because the rigid body is axially symmetric, the scattered light does not provide information on the angle $\theta$, and it only provides information about $x$ and $y$, which represent the position of the head of the trapped rod. As both $x$ and $y$ are measured, the measurement backaction perturbs the state in both the $x$ and $y$ directions simultaneously, and we regard the perturbations in the $x$ and $y$ directions as independent. 

The model of measurement completes the description of our quantum-mechanical model of the controlled rigid body. The stochastic time-evolution equation of the state is given by
\begin{gather}\label{time evolution equation rigid body}
	d|\psi\rangle=\left[\left(-\frac{i}{\hbar}\hat{H}-\frac{\gamma}{4}(\hat{x}-\langle\hat{x}\rangle)^2-\frac{\gamma}{4}(\hat{y}-\langle\hat{y}\rangle)^2\right)dt+\sqrt{\dfrac{\gamma}{2}}(\hat{x}-\langle\hat{x}\rangle)dW_1+\sqrt{\dfrac{\gamma}{2}}(\hat{y}-\langle\hat{y}\rangle)dW_2\right]|\psi\rangle,\\	\hat{H}=\frac{\hat{Q}_{x}^{2}+\hat{Q}_{y}^{2}}{2I_{\perp}}+\frac{\hat{Q}_{z}^{2}}{2I_{\parallel}} + V,
\end{gather}
where $\gamma$ is the measurement strength, $I_{\perp}$ and $I_{\parallel}$ are the moments of inertia, $dW_1$ and $dW_2$ are independent Wiener increments, i.e.~random variables satisfying $dW_1\sim\mathcal{N}(0,dt)$ and $dW_2\sim\mathcal{N}(0,dt)$, following the convention discussed in Section \ref{continuous measurement subsection}. The terms $\hat{Q}_{x}^{2}+\hat{Q}_{y}^{2}$, $\hat{Q}_{z}^{2}$ and $V$ are given by Eqs.~(\ref{Q_x^2+Q_y^2 exact}), (\ref{Q_z^2}) and (\ref{potential exact}), respectively.
\subsubsection{Settings of the Numerical Algorithm}
We use Eq.~(\ref{time evolution equation rigid body}) in our numerical simulation of the controlled quantum-mechanical nanorod, and we do not use any approximation with respect to the polar angle $\beta$, or, with respect to $x$ and $y$. We assume that the state is an eigenstate of $\hat{Q}_z$ and we set the value of $\hat{Q}_z\equiv-{i\hbar}{\partial_\theta}$ to be a constant, and we simulate the time evolution of the wave function in the two-dimensional space spanned by $x$ and $y$. As in Section \ref{quartic oscillator control section}, we use finite difference methods, discretizing the continuous space and time into discrete sites and time steps, so as to simulate the time evolution of the state.

As discussed in the section of the properties of the Hamiltonian, when $\beta$ is very small, the system is approximately linear, because if we define
\begin{equation}
	\hat{p}_x:=-i\hbar\frac{\partial}{\partial{x}} ,\qquad\hat{p}_y:= -i\hbar\frac{\partial}{\partial{y}},
\end{equation}
we can rewrite Eq.~(\ref{Q_x^2+Q_y^2 lowest order}) into
\begin{equation}
	\begin{split}
		\hat{Q}_{x}^{2}+\hat{Q}_{y}^{2}	&\approx-\hbar^{2}\left ( \frac{\partial^{2}}{\partial x^{2}}+\frac{\partial^{2}}{\partial y^{2}}+ \frac{x^{2}+y^{2}}{4}\frac{\partial^{2}}{\partial \theta^{2}}-y\frac{\partial^{2}}{\partial x\,\partial\theta}+x\frac{\partial^{2}}{\partial y\,\partial\theta} \right)\\
		&= \hat{p}_x^{2} + \hat{p}_y^{2} + \frac{x^2+y^2}{4}\hat{Q}_z^{2}-y\hat{p}_x\hat{Q}_z+x\hat{p}_y\hat{Q}_z,
	\end{split}
\end{equation}
which is quadratic with respect to $x$, $y$, $\hat{p}_x$ and $\hat{p}_y$, and the time evolutions of $x$, $y$, $\hat{p}_x$ and $\hat{p}_y$ are thus linear. Especially, as the high-order terms in Eq.~(\ref{Q_x^2+Q_y^2 exact}) are relatively small as discussed above, we need to ensure that $x$ and $y$ are moderately large so that nonlinear behaviour of the system can be observed. Therefore, we use the parameter regime $x,y\sim 0.5$ in our numerical experiments, and we describe our system parameters in the following. Note that the coordinates $x$ and $y$ are dimensionless because they represent angles.

We simulate the space spanned by $x$ and $y$ in the region of  $-1.29\le x \le1.29$ and $-1.29\le y \le1.29$, which approximately corresponds to a polar angle $\beta$ of $74$ degrees. The spacing between adjacent discrete sites in the simulated discretized space is set to be $0.03$, and therefore, we simulate on the $87\times87$ grid.

Instead of directly setting the system parameters $k$, $I_{\perp}$ and $\gamma$, we set them indirectly through the parameters in Table~\ref{rigid body parameters}.
\begin{table}[h]
	\centering
	\begin{tabular}{p{5em}ccccc}\toprule
		& $\omega$ ($\omega_c$) & $\sigma_g$ & $\dfrac{\gamma}{k}$ & $\hat{Q}_z$ ($\hbar$) & $\dfrac{\max\left( \sqrt{E_x^2+E_y^2}\right)}{E_z}$\\[8pt] 
		\hline &&&&&\\[-6pt]
		rigid body & $\pi$ & $0.1$ & $1,2,3$ & $5,80$ & 1\\  \bottomrule
	\end{tabular}
	\caption{\label{rigid body parameters}System parameters of the quantum-mechanical rigid body that we use in our numerical experiments, in terms of a reference angular momentum $\omega_c$. $\omega$ is the angular frequency of the harmonic potential given by Eq.~(\ref{potential harmonic}) ignoring the high-order terms and $\sigma_g$ is the standard deviation of the probability distribution of the corresponding ground state, which is a Gaussian state, of the harmonic potential. We use $\omega$ and $\sigma_g$ to find the parameters $k$, $E_z$ and $I_{\perp}$ of the system of the rigid body.}
\end{table}

Using the angular frequency $\omega$ of the potential $V$ at the lowest-order approximation given by Eq.~(\ref{potential harmonic}), and the standard deviation $\sigma_g$ of the probability distribution of the corresponding ground-state wave function, as given in Table~\ref{rigid body parameters}, we can find the values of $I_{\perp}$ and $k$, given by
\begin{equation}
	I_{\perp} = \frac{\hbar}{2\omega\sigma_g^{2}}
\end{equation}
and
\begin{equation}
	k = \omega^{2}I_{\perp},
\end{equation}
and we have 
\begin{equation}
	E_z = \sqrt{\frac{k}{2}},
\end{equation}
following Eq.~(\ref{potential harmonic}).

Because $k$ and the measurement strength $\gamma$ are of the same physical dimension, instead of directly setting the value of $\gamma$, we tune the ratio $\frac{\gamma}{k}$, and we try three different values $1,2$ and $3$ in our experiments. The angular momentum $\hat{Q}_z$ is set to be $5$ or $80$. When we have $\hat{Q}_z=5$, assuming $I_\parallel\approx\dfrac{I_{\perp}}{10}$, the energy of the rotation around the local $z$ axis, given by $\dfrac{\hat{Q}_z}{2I_\parallel}$, is comparable to the rest of the energy of the system, given by $\dfrac{\hat{Q}_{x}^{2}+\hat{Q}_{y}^{2}}{2I_{\perp}} + V$; when we have $\hat{Q}_z=80$, the rotational energy around the local $z$ axis is hundreds of times larger than the rest of the energy of the system. We therefore try the two different cases in our experiments.

The control forces we allow are given by the following discrete set
\begin{equation}\label{rigid body control choices}
	\left \{(E_x, E_y)\ |\ \sqrt{E_x^{2}+E_y^{2}}\le E_z, \quad\frac{E_x}{E_z}=0.2 n_x, \quad\frac{E_y}{E_z}=0.2 n_y,\quad n_x,n_y\in\mathbb{Z}\right \}, 
\end{equation}
including 81 different choices of control forces in total.

Whenever the controller determines a choice of control forces, the control forces are kept constant during a time of $\dfrac{1}{5\omega_c}$. After an evolution time of $\dfrac{1}{5\omega_c}$ of the state, the energy $\langle\hat{H}\rangle$ with $E_x=E_y=0$ is calculated as the minus reward for the reinforcement learning agent, and the controller determines a new choice of control forces on the basis of the information about the instantaneous wave function.

Most of the other settings of the numerical algorithm are the same as those in Section \ref{quartic oscillator control section}. The discount factor $\gamma$ in Q-learning is tuned to be 0.96, and we change the numbers of the hidden units in the deep neural network to be 512, 1024, 512. 

The input of the neural network is set to be the distribution moments of the wave function as in our experiments of the case of a quartic oscillator. In the case of a rigid body, the distribution moments are computed with respect to the 4 operators $x$, $y$, $\hat{p}_x$ and $\hat{p}_y$, where $\hat{p}_x$ and $\hat{p}_y$ are given by Eqs.~(\ref{momentum operator rigid body x}) and (\ref{momentum operator rigid body y}), and we compute up to the fifth central moments, totally constituting 125 real numbers, and use them as the input of the neural network. Additionally, we rescale $x$ and $y$ by a factor of $\sqrt{\frac{2}{k}}$ and rescale $\hat{p}_x$ and $\hat{p}_y$ by a factor of $\sqrt{2I_{\perp}}$ to ensure that the results are of the order of $O(1)$.

As in the case of the simulation of a quartic oscillator, we use the finite difference methods given by Eqs.~(\ref{finite difference 1st differential}) and (\ref{finite difference 2nd differential}) to evaluate the partial differential operators. To numerically integrate the stochastic time-evolution equation given by Eq.~(\ref{time evolution equation rigid body}), we use a time step of $0.00125\times\frac{1}{\omega_c}$, and we use the explicit 1.5 order strong scheme \cite{NumericalSimulationofStochasticDE} with several modifications. First, we additionally include terms obtained using the 4th-order Runge-Kutta method to include the high-order terms of the deterministic time evolution of the state; next, we do not include the 1.5 order terms of the stochastic evolution that are contributed by the cross terms of the stochastic noise, $\iiint dW_{1/2}\,dW_{1/2}\,dW_{1/2}$, because they are very hard calculate, and they have been confirmed to contribute little to the time evolution, only making a difference smaller than $0.1\%$ to the state; lastly, we ignore the parts of the wave function that are too far away from the center of the probability distribution, setting the amplitudes to be zero for $|x-\langle x\rangle|>\frac{\pi}{3}$ and $|y-\langle y\rangle|>\frac{\pi}{3}$. The cross term $\iint dW_{1}\,dW_{2}$ which is necessary for the numerical integration of the differential equation is evaluated approximately based on series expansion using the Legendre series, following Ref.~\cite{ItoIntegralApproximations} (pp.~761), up to 500 terms. Our numerical codes have been publicized and all details can be found therein.\footnote{\url{https://github.com/Z-T-WANG/PhDThesis/tree/main/rigid\%20body}}

\subsection{Evaluation and Results}\label{rigid body results}
The settings of episodes and the evaluation of performances are the same as in the case of our experiments of a quartic oscillator, described in Section \ref{quartic cooling results}, except for several minor differences. 

The initial state we use in our numerical experiments is set to be a Gaussian wave packet centered at a position randomly picked in the region $-0.4\le x\le0.4$ and $-0.4\le y\le0.4$, and the momentum of the wave packet is set such that the velocity is zero, i.e.~$\dfrac{d\langle x\rangle}{dt}=0$ and $\dfrac{d\langle y\rangle}{dt}=0$, on the basis of the approximation given by Eq.~(\ref{Q_x^2+Q_y^2 lowest order}). The highest energy we allow in our simulation is $30\hbar\omega$, or, $30\pi\hbar\omega_c$, beyond which we stop the simulation and consider the control as having failed. We also stop the simulation when the wave function gets close to the boundary of the simulated space, when the probability of being around the boundary is larger than $1.5\times10^{-3}$. We use the 4 discrete sites counted from the boundary inwards to calculate this probability. Regarding training, the agent is trained for 9000 episodes after it successfully stabilizes the state for a time evolution of $\dfrac{100}{\omega_c}$. 
\subsubsection{The LQG Control}
To compare the results obtained using the DQN algorithm and the C-DQN algorithm with standard control strategies, we consider the linear-quadratic-Gaussian (LQG) control on the basis of the approximation given by Eqs.~(\ref{Q_x^2+Q_y^2 lowest order}) and (\ref{potential lowest order with control}). The control force is chosen such that after an evolution time of $\dfrac{1}{5\omega_c}$ of the state, the state is expected to satisfy
\begin{equation}
	\frac{v_x}{\langle x\rangle}=-\sqrt{\frac{k}{I_{\perp}}},\qquad \frac{v_y}{\langle y\rangle}=-\sqrt{\frac{k}{I_{\perp}}},
\end{equation}
where 
\begin{equation}
	v_x:=\dfrac{d\langle x\rangle}{dt},\qquad 	v_y:=\dfrac{d\langle y\rangle}{dt},
\end{equation}
and the LQG controller effectively treats the state as a classical particle, only taking the variables $\langle x\rangle$, $\langle y\rangle$, $v_x$ and $v_y$ into consideration. This controller pushes the particle onto the physical trajectory that corresponds to the Lagrangian
\begin{equation}
	L=T-V=\frac{1}{2I_{\perp}}\left (v_x^{2}+v_y^{2}\right ) - \left (-\frac{k}{2}\left (x^{2}+y^{2}\right )\right )=\frac{1}{2I_{\perp}}\left (v_x^{2}+v_y^{2}\right ) +\frac{k}{2}\left (x^{2}+y^{2}\right ),
\end{equation}
which is exactly equal to the energy of the system. Therefore, this control minimizes the energy of the system integrated with respect to time according to the Hamilton principle.

Although the LQG control is provably optimal \cite{LinearQuadraticControl} when the dynamics is linear and Gaussian noise is present, it can fail to stabilize the rigid body due to nonlinearity of the system. Specifically, because the interaction potential is given by
\begin{equation}
	V\propto-\left(\vec{l}\cdot\vec{E}\right)^{2},
\end{equation}
when the direction of the control field and the direction of the controlled rod are perpendicular, the control force becomes $0$, and when the angle between the two directions exceeds $\frac{\pi}{2}$, the interaction becomes repulsive. However, the LQG control replies on the linear approximation and considers that larger values of $E_x$ and $E_y$ should always drive the particle in the $x$ and $y$ directions more strongly, which is not necessarily true, as the particle can move away from the center and the angle between $\vec{l}$ and $\vec{E}$ can become large. Therefore, in order to prevent the systematic failure of the LQG control in controlling the rigid body, we put constraints on $E_x$ and $E_y$ such that we have $\left |\langle x\rangle -\frac{E_x}{E_z}\right |\le \frac{\pi}{4}$ and $\left |\langle y\rangle -\frac{E_y}{E_z}\right |\le \frac{\pi}{4}$, and we do not put these constraints on the AI-based controllers. With these constraints, the LQG control can stabilize the rigid body for a long time in our numerical experiments. For a fair comparison with the AI-based controllers, we also discretize the control forces of the LQG control, mapping the control forces of the LQG control to the nearest choice in the set given by Eq.~(\ref{rigid body control choices}). 

\subsubsection{Data Augmentation}
\begin{figure}[htb!]
	\centering
	\includegraphics[width=0.8\linewidth]{"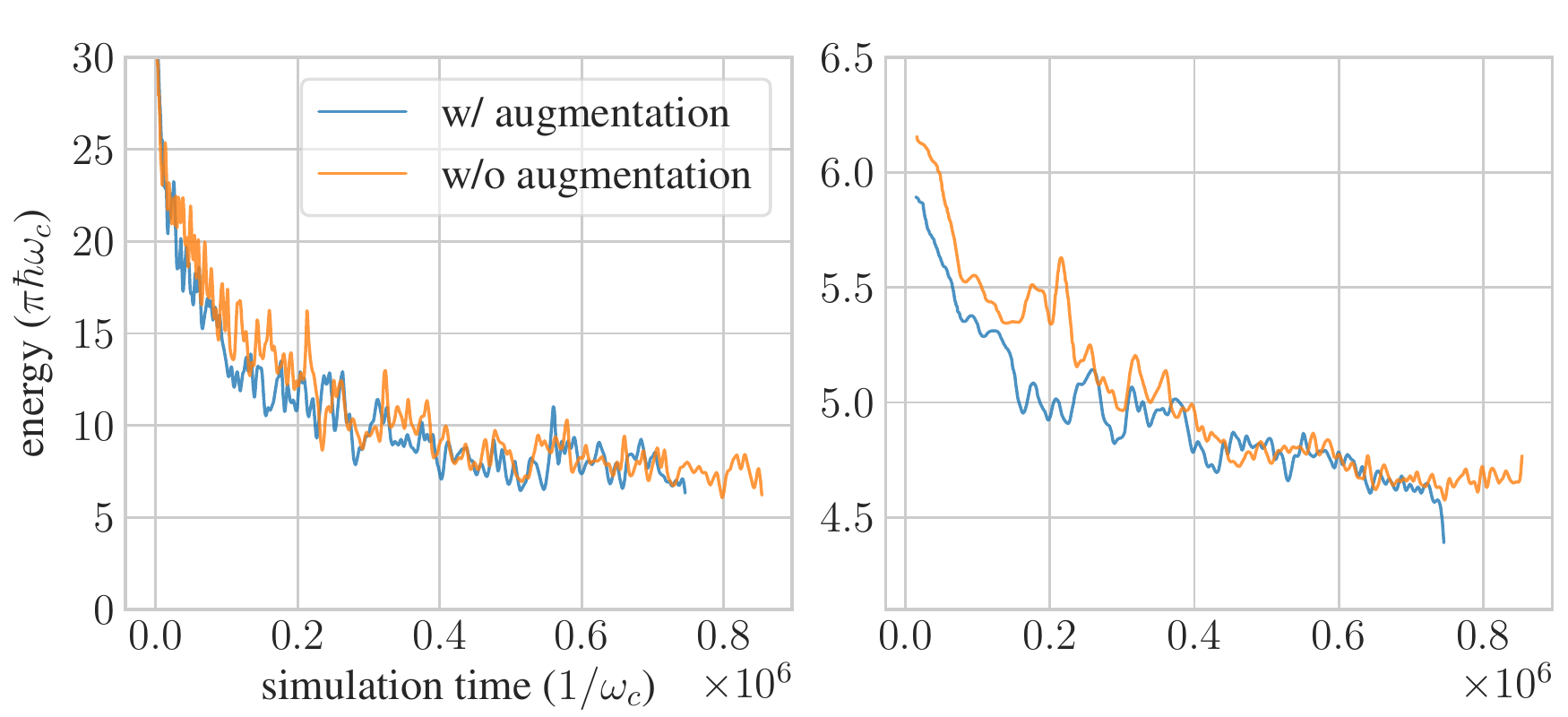"}
	\caption{Learning curves of the cooling problem of a rigid body for the C-DQN algorithm, with and without the data augmentation technique. The ordinate and the abscissa are the same in the two figures. The left panel shows the results including the episodes that end with control failure, applying Gaussian smoothing with the standard deviation of 40, and the right panel shows the results excluding the episodes that end with control failure, applying Gaussian smoothing with the standard deviation of 20. The system parameters are given by $\dfrac{\gamma}{k}=3$ and $\hat{Q}_z=5$. }
	\label{data augmentation rigid body}
\end{figure}

Because the time-evolution equation given by Eq.~(\ref{time evolution equation rigid body}) has central symmetry, given a trajectory of the time evolution of a quantum state, we can flip the $x$ and $y$ directions and obtain an equally valid trajectory, i.e., employing the transformation $(x,y)\rightarrow(-x,-y)$ and $(E_x,E_y)\rightarrow(-E_x,-E_y)$. Therefore, we use both
the data of the directly simulated state and the data of the symmetric counterpart of the simulated state produced by the flip of the $x$ and $y$ directions for the AI to learn. This method is called data augmentation in machine learning literature because it increases the number of data for the AI to learn, and in our case the number of data effectively becomes twice. In our numerical experiments, we randomly flip the $x$ and $y$ directions of the data. The learning curves with and without the data augmentation technique are shown in Fig.~(\ref{data augmentation rigid body}), showing that the data augmentation is indeed beneficial for learning.
\subsubsection{Experimental Results}
\begin{figure}[htb!]
	\centering
	\includegraphics[width=0.8\linewidth]{"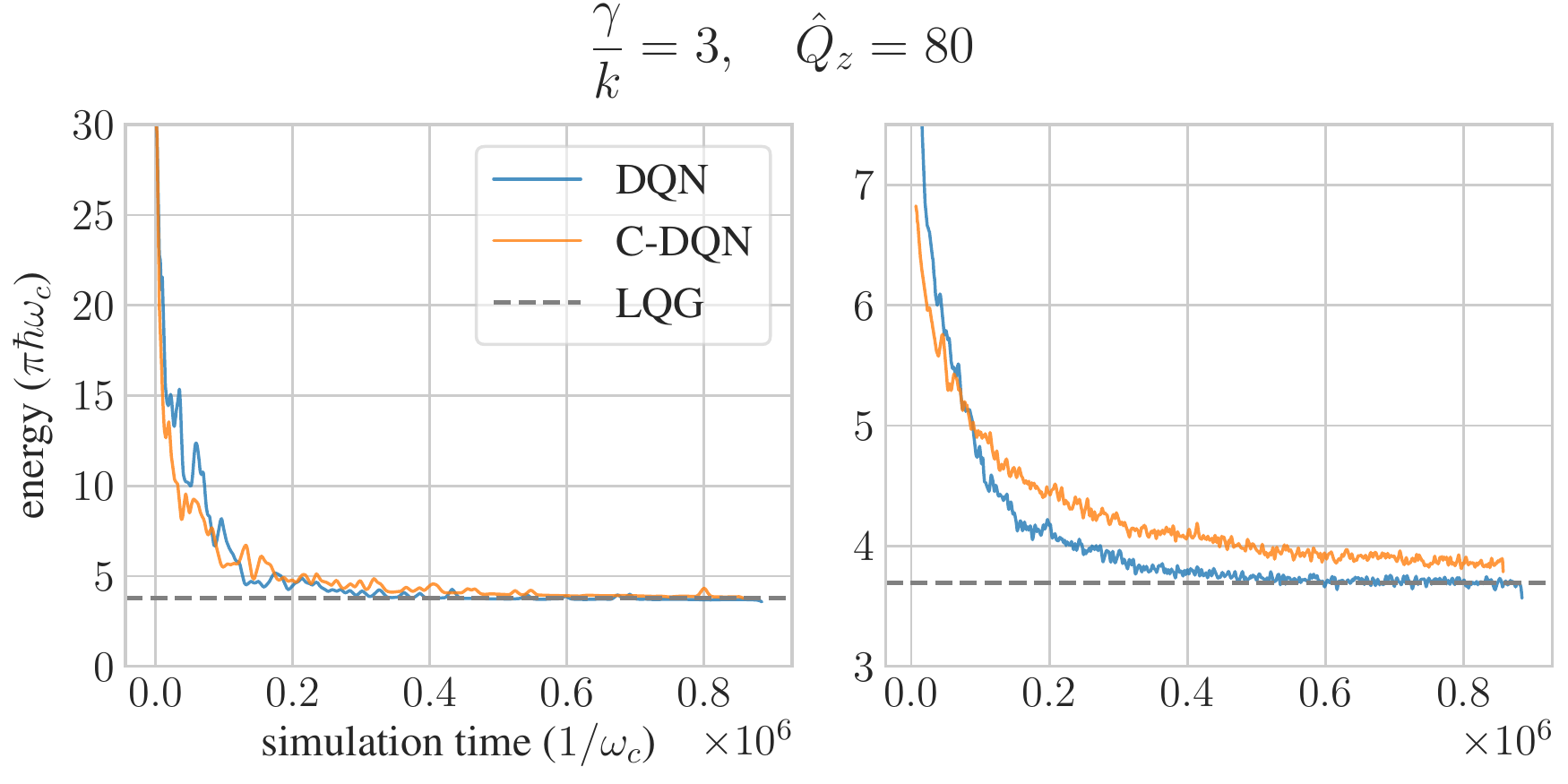"}
	\caption{Learning curves of the cooling problem of a rigid body for the DQN and the C-DQN algorithms, with the system parameters $\dfrac{\gamma}{k}=3$ and $\hat{Q}_z=80$, and the horizontal dashed line shows the performance of the LQG control. The left panel shows the results including the episodes that end with control failure, applying Gaussian smoothing with the standard deviation of 40, and the right panel shows the results excluding the episodes that end with control failure, applying Gaussian smoothing with the standard deviation of 10. The figure in the right panel is enlarged.}
	\label{rigid body 3 80}
\end{figure}

\begin{figure}[htb!]
	\centering
	\includegraphics[width=0.8\linewidth]{"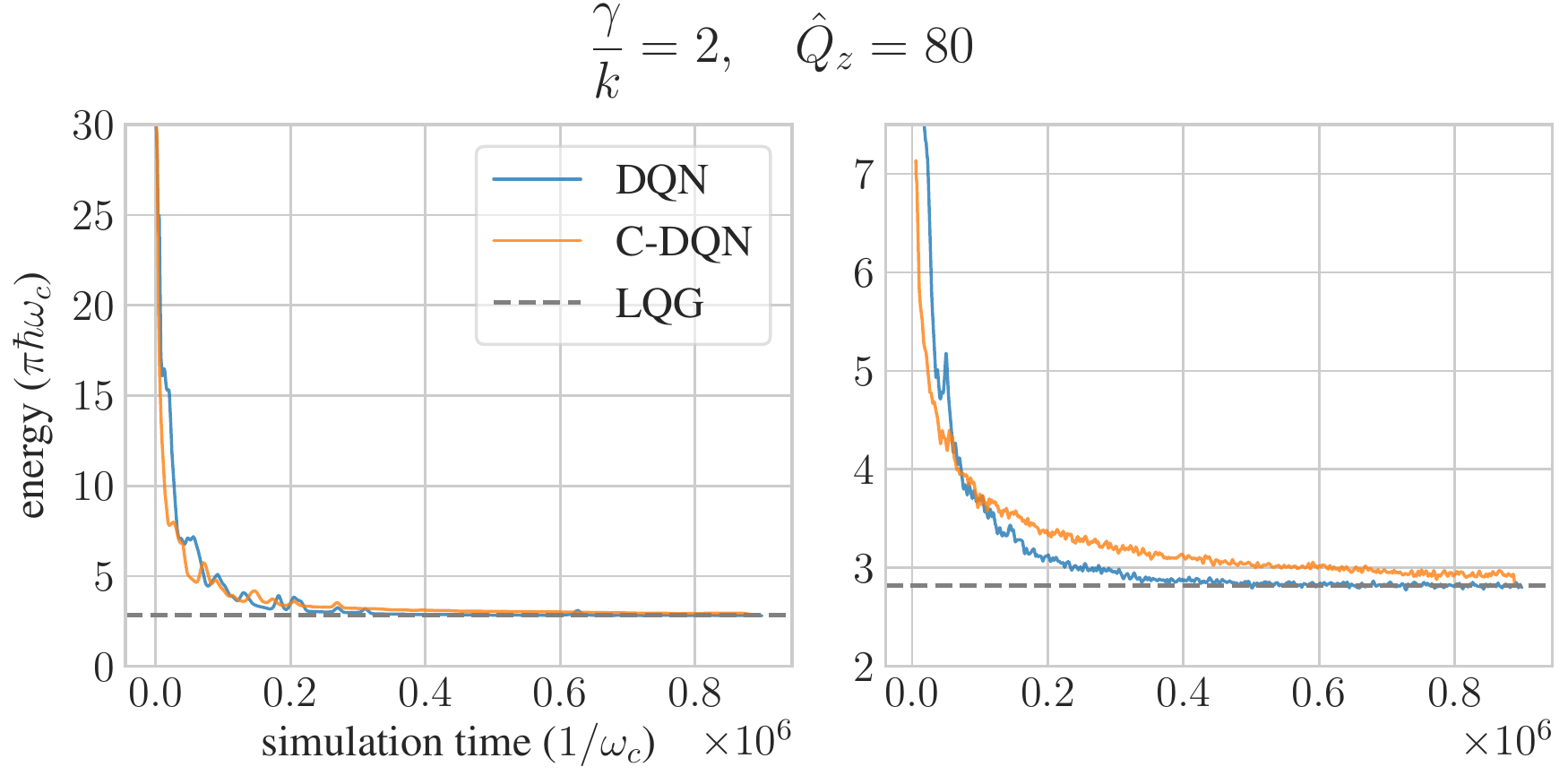"}
	\caption{Learning curves of the cooling problem of a rigid body for the DQN and the C-DQN algorithms as in Fig.~\ref{rigid body 3 80}, with the system parameters $\dfrac{\gamma}{k}=2$ and $\hat{Q}_z=80$.}
	\label{rigid body 2 80}
\end{figure}

\begin{figure}[htb!]
	\centering
	\includegraphics[width=0.8\linewidth]{"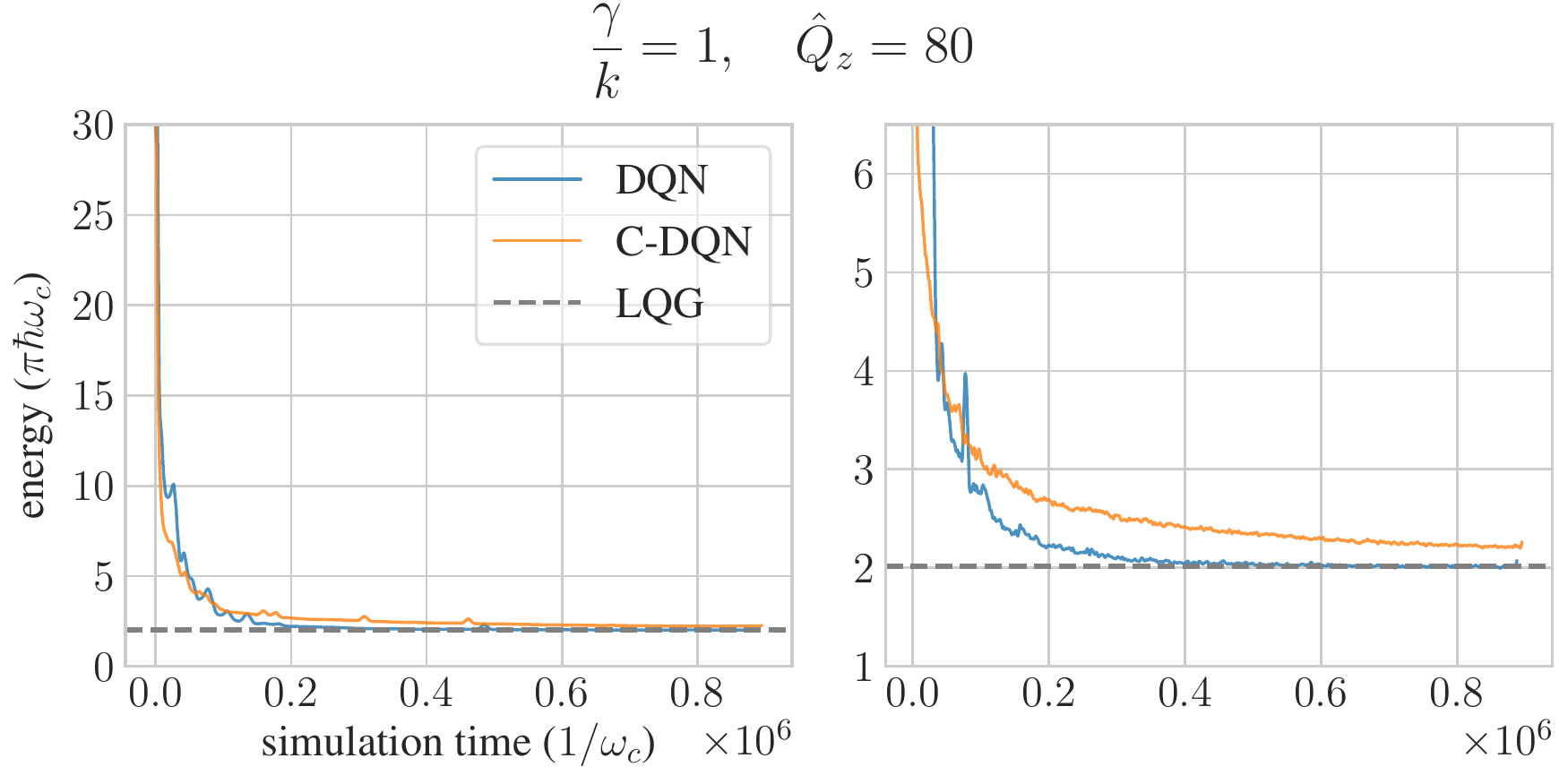"}
	\caption{Learning curves of the cooling problem of a rigid body for the DQN and the C-DQN algorithms as in Fig.~\ref{rigid body 3 80}, with the system parameters $\dfrac{\gamma}{k}=1$ and $\hat{Q}_z=80$.}
	\label{rigid body 1 80}
\end{figure}

\begin{figure}[htb!]
	\centering
	\includegraphics[width=0.8\linewidth]{"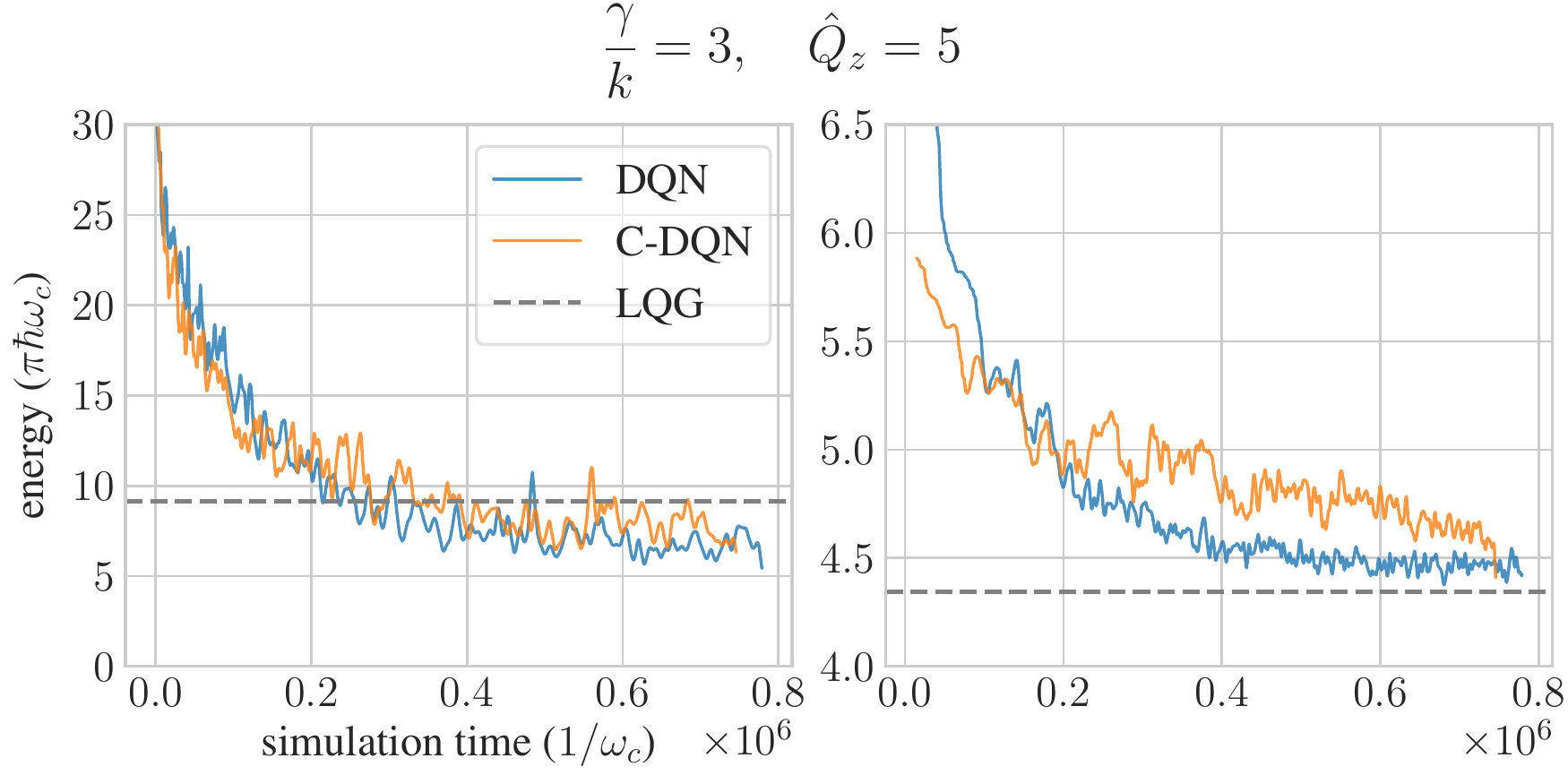"}
	\caption{Learning curves of the cooling problem of a rigid body for the DQN and the C-DQN algorithms as in Fig.~\ref{rigid body 3 80}, with the system parameters $\dfrac{\gamma}{k}=3$ and $\hat{Q}_z=5$.}
	\label{rigid body 3 5}
\end{figure}

\begin{figure}[htb!]
	\centering
	\includegraphics[width=0.8\linewidth]{"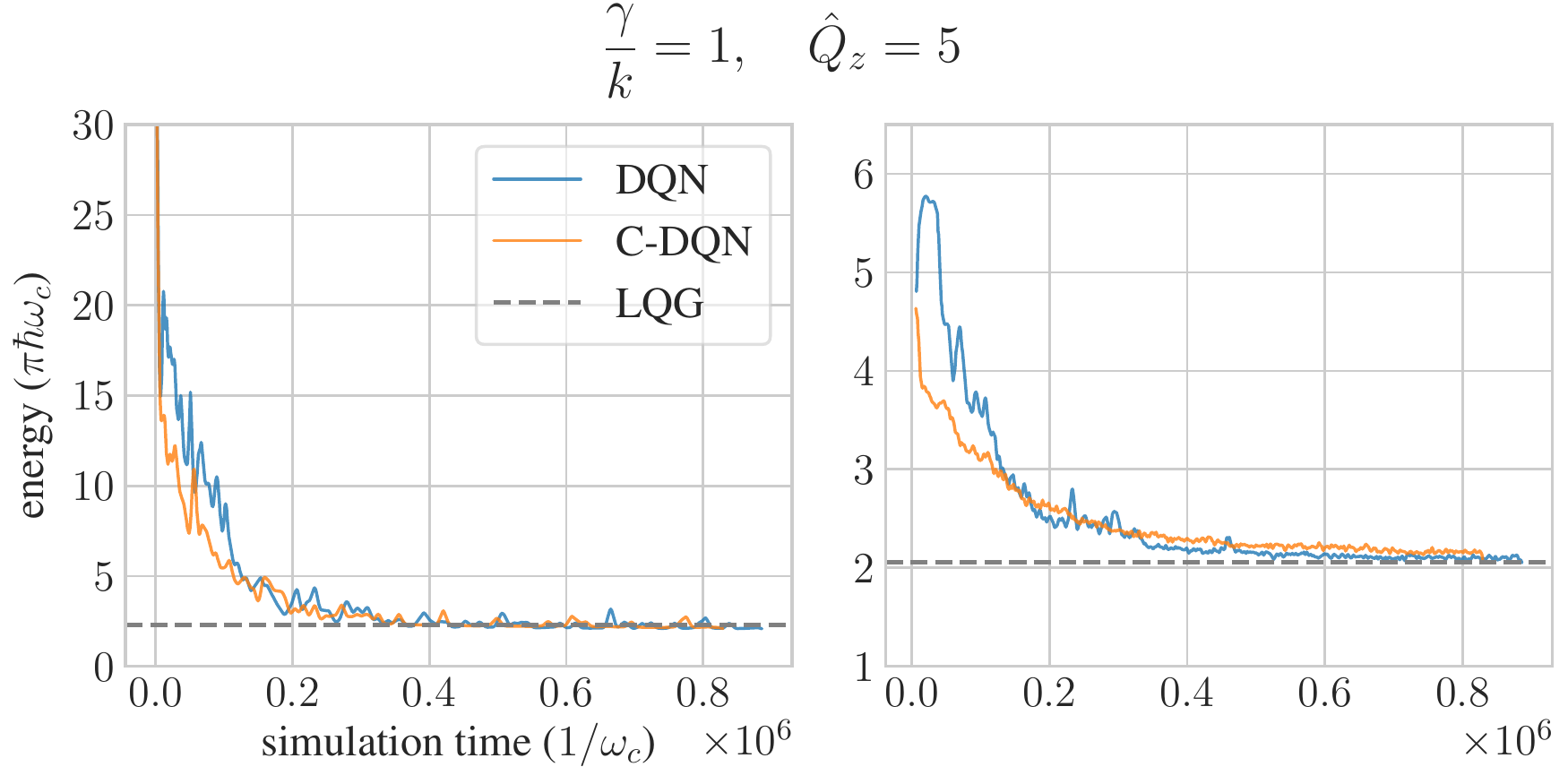"}
	\caption{Learning curves of the cooling problem of a rigid body for the DQN and the C-DQN algorithms as in Fig.~\ref{rigid body 3 80}, with the system parameters $\dfrac{\gamma}{k}=1$ and $\hat{Q}_z=5$.}
	\label{rigid body 1 5}
\end{figure}

As discussed above, we follow the same procedure as in Section \ref{quartic oscillator control section} with several modifications to train the AI using the DQN and the C-DQN algorithms on the task of cooling a quantum-mechanical rigid body. The learning curves for different values of system parameters $\gamma$ and $\hat{Q}_z$ are shown in Figs.~\ref{rigid body 3 80}, \ref{rigid body 2 80}, \ref{rigid body 1 80}, \ref{rigid body 3 5} and \ref{rigid body 1 5}, where the performances of the LQG control are shown for comparison. In these figures, we see that both the DQN and the C-DQN algorithms can learn the cooling problem and their performances are comparable, which are also comparable with that of the LQG control, showing that the AI algorithms can indeed successfully cool the quantum-mechanical rigid body.

\begin{figure}[tbh!]
	\centering
	\includegraphics[width=0.8\linewidth]{"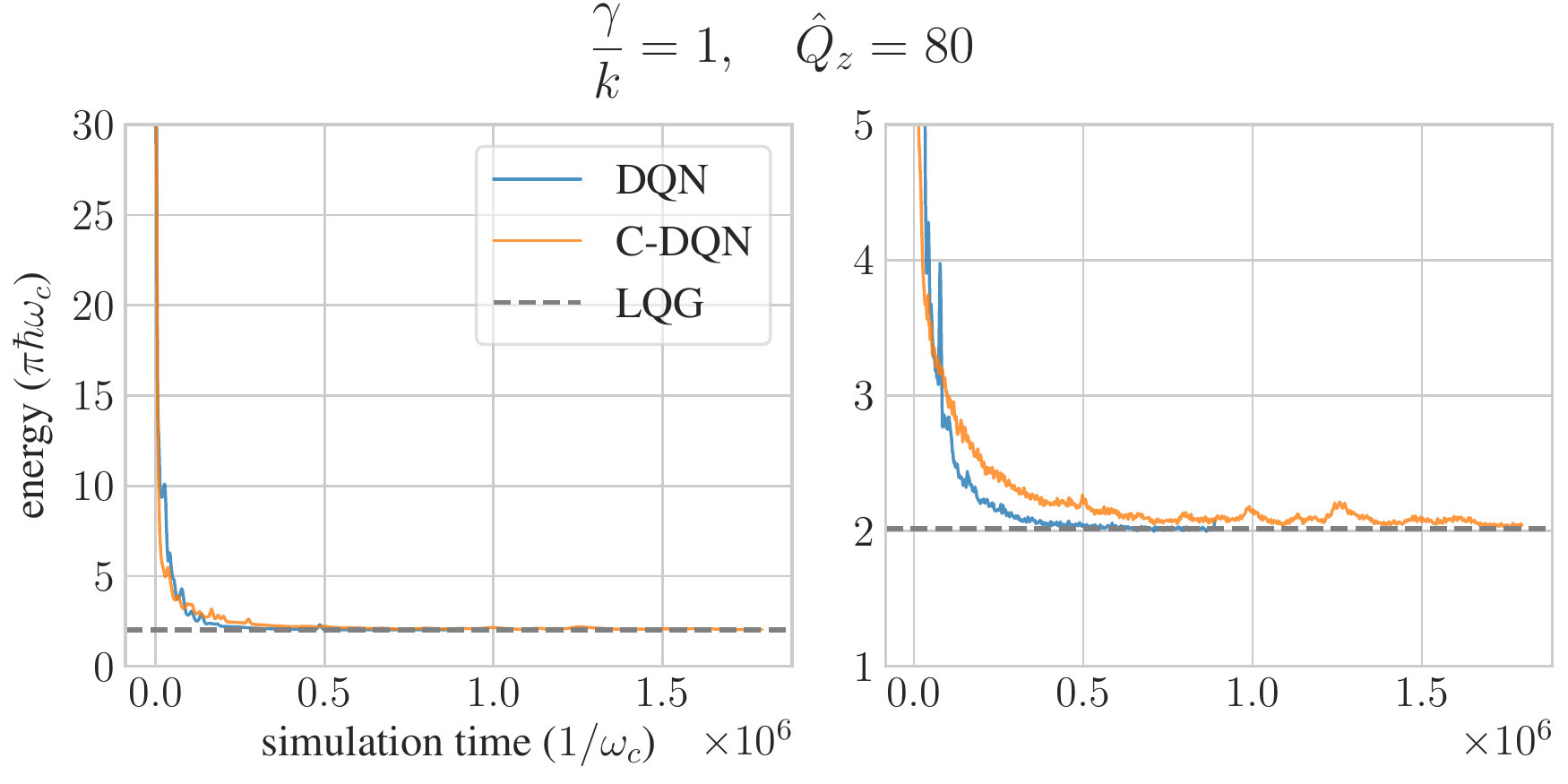"}
	\caption{Learning curves of the cooling problem of a rigid body for the DQN and the C-DQN algorithms as in Fig.~\ref{rigid body 1 80}, with a doubled time of training for the C-DQN algorithm.}
	\label{rigid body doubled CDQN 1 80}
\end{figure}

\begin{figure}[tbh!]
	\centering
	\includegraphics[width=0.8\linewidth]{"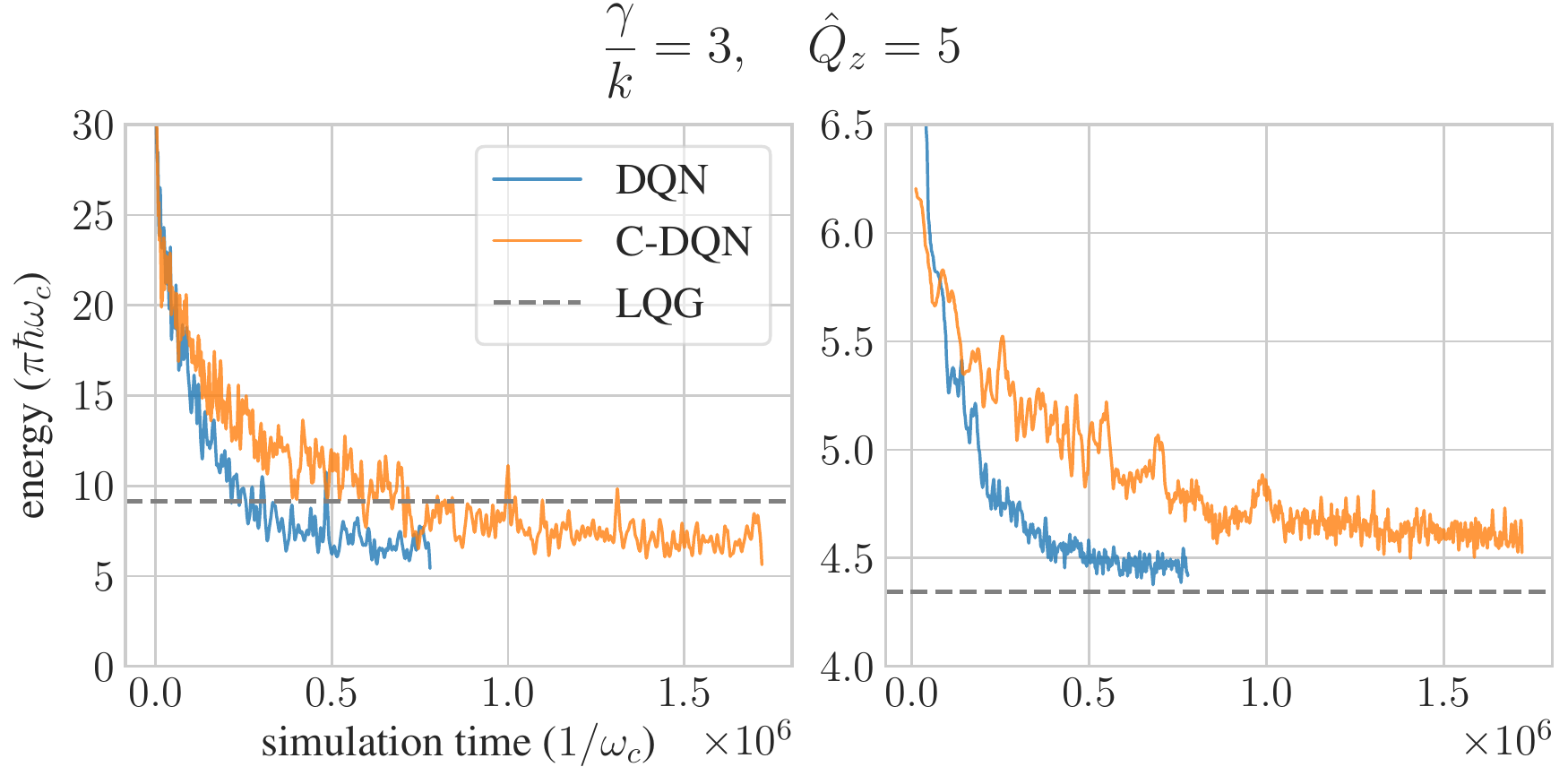"}
	\caption{Learning curves of the cooling problem of a rigid body for the DQN and the C-DQN algorithms as in Fig.~\ref{rigid body 3 5}, with a doubled time of training for the C-DQN algorithm.}
	\label{rigid body doubled CDQN 3 5}
\end{figure}

As shown in the above figures, compared with the DQN algorithm, while the C-DQN algorithm learns faster at the beginning of training when the behaviour of the controller is unstable on this task, it learns more slowly at a later stage. Given an equal amount of time of training, the performances of the C-DQN algorithm can be marginally lower than that of the DQN algorithm, and as we gradually reduce the learning rate throughout training, the DQN algorithm quickly reaches a stable performance while the learning of the C-DQN algorithm is slowed down and does not exactly reach the performance of the DQN algorithm at the end of training. To see the results for a prolonged training period, we double the training time of the C-DQN algorithm for the experiments for the system parameters $\frac{\gamma}{k}=1, \hat{Q}_z=80$ and $\frac{\gamma}{k}=3, \hat{Q}_z=5$, and the results are shown in Figs.~\ref{rigid body doubled CDQN 1 80} and \ref{rigid body doubled CDQN 3 5}. The results show that the performance of the C-DQN algorithm indeed approaches that of the DQN algorithm given a longer time of training, and confirms the effectiveness of the learning of the C-DQN algorithm.

The lower performance of the C-DQN algorithm compared with the DQN algorithm can be attributed to the behaviour of the C-DQN algorithm on the stochasticity of the task. As discussed in Section \ref{stochasticity effect on CDQN discussion} and experimentally shown in Section \ref{wet chicken experiment}, if the time evolution of the system is stochastic, the C-DQN algorithm does not necessarily converge to the optimal solution of the Bellman equation, and it may converge somewhere between the solution of the residual gradient algorithm and the DQN algorithm, and as a result, the performance may be slightly worse than that of the optimal control, which is a disadvantage of the C-DQN algorithm. Nevertheless, as discussed in Section \ref{wet chicken experiment} and shown in Figs.~\ref{rigid body doubled CDQN 1 80} and \ref{rigid body doubled CDQN 3 5} above, after a prolonged period of training, the performance of C-DQN can improve to be approximately equal to that of DQN, and, therefore, the disadvantage of C-DQN is not significant and can easily be remedied on this task. 

Concerning properties of the system of the controlled rigid body, comparing Figs.~\ref{rigid body 3 80}, \ref{rigid body 2 80} and \ref{rigid body 1 80}, we see that with the reduced measurement strength $\gamma$, the fluctuations in performance are reduced and the system becomes more stable, and the controllers achieve lower energies of cooling and the reinforcement learning algorithms learn faster. Comparing Figs.~\ref{rigid body 3 80} and \ref{rigid body 3 5}, when the angular momentum $\hat{Q}_z$ is reduced, with a large measurement strength, the stochasticity of the system increases, and the rate of the failure of control becomes significant for the case of $\frac{\gamma}{k}=3$ and $\hat{Q}_z=5$. This is because a large angular momentum $\hat{Q}_z$ behaves effectively like a magnetic field which localizes the state, and with a localized state, the measurement backaction of the position measurement is weaker, which leads to less stochastic perturbation and better stability of the system. Therefore, the control is easier for the case of $\frac{\gamma}{k}=3$ and $\hat{Q}_z=80$ compared with the control for the case of $\frac{\gamma}{k}=3$ and $\hat{Q}_z=5$. When the measurement strength is relatively weak, as shown in Figs.~\ref{rigid body 1 80} and \ref{rigid body 1 5}, the change of $\hat{Q}_z$ may not make a significant difference.

When the stochasticity is large and the energy is relatively high, as shown in Fig.~\ref{rigid body 3 5} and \ref{rigid body doubled CDQN 3 5}, the rate of the failure of control is non-negligible, and the overall performance of the DQN algorithm and that of the C-DQN algorithm are better than the performance of the LQG control. If we exclude the cases of control failure, the energy of the control of the C-DQN algorithm is higher than that of DQN which is then higher than that of the LQG control. This result implies that although the energy of the system cooled by the LQG controller is often relatively low, the LQG control fails more frequently than the DQN controller and the C-DQN controller, and the DQN controller fails more frequently than the C-DQN controller. This result is consistent with our argument regarding the nonlinearity of the system. When the energy of the state is high and the wave function moves away from the center of the trap, the nonlinearity becomes significant and the LQG control tends to have an inferior performance. The result also shows that C-DQN tends to avoid failure and has a more stable behaviour compared with DQN.
\section{Conclusion and Future Perspectives}
In this section, we summarize and discuss the results, and we present the conclusions and future perspectives.
\subsubsection{Summary and Discussion}
In the experiments of cooling of a quantum quartic oscillator, we confirm that both of the DQN and the C-DQN algorithms can learn to sufficiently cool the nonlinear quantum system, in which case conventional control strategies do not work satisfactorily as the state is non-Gaussian in general \cite{quantumCartpole}. However, the DQN algorithm exhibits significant instability and does not perform stably throughout the course of training, and it has a large variance in its final performance, which is likely due to the complexity of the nonlinear quantum system. On the other hand, the C-DQN algorithm always performs stably for this complicated system, and as a consequence, it achieves a better performance at the end of training and has a negligible variance in its final performance. The results demonstrate the stability and reliability of the C-DQN algorithm for physical control problems.

In the experiments of cooling of a quantum-mechanical rigid body, we confirm that both of the DQN and the C-DQN algorithms can learn to stabilize and cool the system of the trapped quantum-mechanical nanorod, which is a non-Euclidean multi-dimensional system, and that when the linear approximation holds, the constrained conventional LQG controller also cools the system efficiently. Within the regime of the system parameters we have considered, both the DQN and C-DQN algorithms perform comparably with the LQG controller. The final performances of the C-DQN algorithm are marginally lower than those of the DQN algorithm given the same computational budget, because the C-DQN algorithm generally learns more slowly and possibly converges to a solution that is suboptimal when the system is stochastic. Nevertheless, our experimental results show that the performances of the C-DQN algorithm can match those of the DQN algorithm if the AI is trained for a longer period of time. Compared with the results of the cooling of a quartic oscillator, we see that the DQN algorithm may be preferred if it does not encounter any instability, in which case the problem is often relatively easy and the AI learns quickly and converges to an optimal solution. If the problem to be solved is difficult and complicated, the DQN algorithm may suffer from significant instabilities, in which case the C-DQN algorithm clearly performs better. The C-DQN algorithm tends to have more stable and reproducible behaviour in the learning process, which is beneficial for research and fine-tuning in general.
\subsubsection{Future Perspectives}
Concerning future perspectives, as the C-DQN algorithm may not converge to the optimal solution, it is worthwhile to investigate the case where the C-DQN and the DQN algorithms are used in combination. For example, one may train the AI using the C-DQN algorithm at an early stage to avoid instability, and change to the DQN algorithm to improve the final performance at a later stage. It is also worthwhile to consider possible modifications of the C-DQN algorithm so that it can satisfactorily deal with stochasticity.

Regarding the study of a quantum-mechanical rigid body, the model of measurement we considered is only approximate, and models of more convenient and realistic measurement protocols are desired. For example, one may consider the measurement of the position of the head of the nanorod in the three-dimensional space, by attaching particles or charges that are easily measurable at the heads of the nanorod. Although different models of measurement should result in different behaviour of the system, we believe the our results concerning the DQN and the C-DQN algorithms are universal and hold true for any measurement model. As experimental techniques continue developing, the controls that we have investigated may be applied to real experiments to cool the state of a rigid body to a quantum regime.

For simplicity, the rigid body we considered in our numerical experiments is axially symmetric. It is also worthwhile to investigate the more complicated case where the rigid body is asymmetric in general. The nonlinearity of an asymmetric rigid body is more significant, and as $\hat{Q}_z$ does not commute with the Hamiltonian, one would be able to cool all rotational degrees of freedom using control fields in the $x$ and $y$ directions only. However, experimentally, it can be more difficult to measure the orientation of an asymmetric rigid body, and the computational cost is also considerably higher.

	\chapter{Conclusions}\label{conclusion}
	In this thesis, we have reviewed the formulations of continuous quantum measurement and deep reinforcement learning in Chapters \ref{continuous quantum measurement} and \ref{deep reinforcement learning}. We have discussed the non-convergence issue of conventional Q-learning strategies and the inefficiency issues of existing convergent approaches in the reinforcement learning literature, and developed a new convergent deep Q-learning algorithm in Chapter \ref{convergent DQN}, which we call the convergent deep Q network (C-DQN) algorithm, as an alternative to the conventional deep Q network (DQN) algorithm. The C-DQN algorithm is provably convergent, scalable and efficient, and we have demonstrated its effectiveness on standard benchmarks in the reinforcement learning literature, namely, the \textit{Atari 2600} benchmark \cite{AtariEnvironment}. Finally, in Chapter \ref{control rigid body}, we have applied the C-DQN algorithm to the measurement-feedback cooling problems of a quantum-mechanical quartic oscillator and a trapped quantum-mechanical rigid body. We presented the physical models and analysed the properties of the systems, and showed that although both of the DQN and the C-DQN algorithms can learn to cool the systems, the C-DQN algorithm learns stably and has better performances if the DQN algorithm suffers from instability when the task is difficult; however, the C-DQN algorithm learns relatively more slowly when the task is sufficiently simple such that the DQN algorithm can work stably and quickly. Because the performances of the DQN algorithm can have large variances and lack consistency from trial to trial if the underlying task is difficult, the C-DQN algorithm can be a better choice for researches on complicated physical control problems. 

Our contribution is twofold: we have investigated the non-convergence issue of the standard reinforcement learning algorithm, Q-learning, and developed a new convergent algorithm and examined the properties of our algorithm; we have established the quantum-mechanical model of the trapped and controlled rigid body, and demonstrated the effectiveness of our control strategies for the measurement-feedback cooling problem of this system. 

Regarding future directions, we may consider the combination of the DQN and the C-DQN algorithms so that we can obtain both the stability of the C-DQN algorithm and the high performance of the final result of the DQN algorithm. It is also desired if the C-DQN algorithm can be improved to deal with stochasticity satisfactorily and to converge to an optimal solution of the Bellman equation in the presence of stochasticity. Concerning the study of a quantum rigid body, the control strategies we have investigated may be applied to real experiments, using application-specific integrated circuits which embed the control strategies to control the lasers to reduce the energy of the trapped rigid bodies. The control strategies we have considered can help cool the system of a trapped rigid body so that a quantum regime may be realized, which has applications in sensing devices and fundamental physical research \cite{quantumRotorNature}. It is also possible to extend our research to the case of a more complicated asymmetric rigid body, which has highly nonlinear dynamics. 

This thesis contributes to the field of the interdisciplinary study of quantum control and machine learning, and we hope that our work helps the development of the use of machine learning technologies for physical problems and the development of better control strategies in the microscopic quantum world.
	
	\renewcommand\chaptername{Appendix}
		

	\bibliography{references}

\begin{thebibliography}{100}

\bibitem{quantumControlSurvey}
Daoyi Dong and Ian~R Petersen.
\newblock Quantum control theory and applications: a survey.
\newblock {\em IET Control Theory \& Applications}, 4(12):2651--2671, 2010.

\bibitem{quantumControlScience}
Warren~S. Warren, Herschel Rabitz, and Mohammed Dahleh.
\newblock Coherent control of quantum dynamics: The dream is alive.
\newblock {\em Science}, 259(5101):1581--1589, 1993.

\bibitem{ColdAtomQuantumControl}
Steven Chu.
\newblock Cold atoms and quantum control.
\newblock {\em Nature}, 416(6877):206, 2002.

\bibitem{quantumChemistryControl1}
Moshe Shapiro and Paul Brumer.
\newblock {\em Quantum control of molecular processes}.
\newblock John Wiley \& Sons, 2012.

\bibitem{quantumChemistryControl2}
Marcos Dantus and Vadim~V Lozovoy.
\newblock Experimental coherent laser control of physicochemical processes.
\newblock {\em Chemical reviews}, 104(4):1813--1860, 2004.

\bibitem{quantumChemistryControl3}
Vlasta Bonaci{\'c}-Kouteck{\`y} and Roland Mitri{\'c}.
\newblock Theoretical exploration of ultrafast dynamics in atomic clusters:
  Analysis and control.
\newblock {\em Chemical reviews}, 105(1):11--66, 2005.

\bibitem{QuantumDots}
Peter Lodahl, Sahand Mahmoodian, and S\o{}ren Stobbe.
\newblock Interfacing single photons and single quantum dots with photonic
  nanostructures.
\newblock {\em Rev. Mod. Phys.}, 87:347--400, May 2015.

\bibitem{QuantumSpinDots}
Ronald Hanson, Leo~P Kouwenhoven, Jason~R Petta, Seigo Tarucha, and Lieven~MK
  Vandersypen.
\newblock Spins in few-electron quantum dots.
\newblock {\em Reviews of modern physics}, 79(4):1217, 2007.

\bibitem{QuantumSimulation}
Iulia~M Georgescu, Sahel Ashhab, and Franco Nori.
\newblock Quantum simulation.
\newblock {\em Reviews of Modern Physics}, 86(1):153, 2014.

\bibitem{TrappedIon}
H.~Häffner, C.F. Roos, and R.~Blatt.
\newblock Quantum computing with trapped ions.
\newblock {\em Physics Reports}, 469(4):155 -- 203, 2008.

\bibitem{NVCenter}
Lucio Robledo, Lilian Childress, Hannes Bernien, Bas Hensen, Paul~FA Alkemade,
  and Ronald Hanson.
\newblock High-fidelity projective read-out of a solid-state spin quantum
  register.
\newblock {\em Nature}, 477(7366):574, 2011.

\bibitem{Superconducting}
Andreas Wallraff, David~I Schuster, Alexandre Blais, Luigi Frunzio, R-S Huang,
  Johannes Majer, Sameer Kumar, Steven~M Girvin, and Robert~J Schoelkopf.
\newblock Strong coupling of a single photon to a superconducting qubit using
  circuit quantum electrodynamics.
\newblock {\em Nature}, 431(7005):162, 2004.

\bibitem{OpticalQuantumComputation}
Shuntaro Takeda and Akira Furusawa.
\newblock Perspective: Toward large-scale fault-tolerant universal photonic
  quantum computing.
\newblock {\em arXiv preprint arXiv:1904.07390}, 2019.

\bibitem{CavityOptomechanics}
Markus Aspelmeyer, Tobias~J. Kippenberg, and Florian Marquardt.
\newblock Cavity optomechanics.
\newblock {\em Rev. Mod. Phys.}, 86:1391--1452, Dec 2014.

\bibitem{LinearQuadraticControl}
Brian D.~O. Anderson and John~B. Moore.
\newblock {\em Optimal Control: Linear Quadratic Methods}.
\newblock Prentice-Hall, Inc., Upper Saddle River, NJ, USA, 1990.

\bibitem{QuantumOptimalControlTheory}
J~Werschnik and E~K~U Gross.
\newblock Quantum optimal control theory.
\newblock {\em Journal of Physics B: Atomic, Molecular and Optical Physics},
  40(18):R175--R211, sep 2007.

\bibitem{CRAB}
Patrick Doria, Tommaso Calarco, and Simone Montangero.
\newblock Optimal control technique for many-body quantum dynamics.
\newblock {\em Phys. Rev. Lett.}, 106:190501, May 2011.

\bibitem{GRAPE}
Navin Khaneja, Timo Reiss, Cindie Kehlet, Thomas Schulte-Herbrüggen, and
  Steffen~J. Glaser.
\newblock Optimal control of coupled spin dynamics: design of nmr pulse
  sequences by gradient ascent algorithms.
\newblock {\em Journal of Magnetic Resonance}, 172(2):296 -- 305, 2005.

\bibitem{humanOutperformNumericalMethods}
Jens Jakob~WH S{\o}rensen, Mads~Kock Pedersen, Michael Munch, Pinja Haikka,
  Jesper~Halkj{\ae}r Jensen, Tilo Planke, Morten~Ginnerup Andreasen, Miroslav
  Gajdacz, Klaus M{\o}lmer, Andreas Lieberoth, et~al.
\newblock Exploring the quantum speed limit with computer games.
\newblock {\em Nature}, 532(7598):210--213, 2016.

\bibitem{geneticQuantumControl}
Ehsan Zahedinejad, Sophie Schirmer, and Barry~C. Sanders.
\newblock Evolutionary algorithms for hard quantum control.
\newblock {\em Phys. Rev. A}, 90:032310, Sep 2014.

\bibitem{spinOptimal}
Navin Khaneja, Roger Brockett, and Steffen~J. Glaser.
\newblock Time optimal control in spin systems.
\newblock {\em Phys. Rev. A}, 63:032308, Feb 2001.

\bibitem{HarmonicOscillatorControl}
A.~C. Doherty and K.~Jacobs.
\newblock Feedback control of quantum systems using continuous state
  estimation.
\newblock {\em Phys. Rev. A}, 60:2700--2711, Oct 1999.

\bibitem{refereeCitation}
Leigh Martin, Felix Motzoi, Hanhan Li, Mohan Sarovar, and K.~Birgitta Whaley.
\newblock Deterministic generation of remote entanglement with active quantum
  feedback.
\newblock {\em Phys. Rev. A}, 92:062321, Dec 2015.

\bibitem{superconductingQubitAnharmonicityPulseControl}
F.~Motzoi, J.~M. Gambetta, P.~Rebentrost, and F.~K. Wilhelm.
\newblock Simple pulses for elimination of leakage in weakly nonlinear qubits.
\newblock {\em Phys. Rev. Lett.}, 103:110501, Sep 2009.

\bibitem{shortcutsToAdiabaticity}
David Gu{\'e}ry-Odelin, Andreas Ruschhaupt, Anthony Kiely, Erik Torrontegui,
  Sofia Mart{\'\i}nez-Garaot, and Juan~Gonzalo Muga.
\newblock Shortcuts to adiabaticity: concepts, methods, and applications.
\newblock {\em Reviews of Modern Physics}, 91(4):045001, 2019.

\bibitem{SpinEcho}
S.~Meiboom and D.~Gill.
\newblock Modified spin‐echo method for measuring nuclear relaxation times.
\newblock {\em Review of Scientific Instruments}, 29(8):688--691, 1958.

\bibitem{DynamicDecoupling}
Lorenza Viola, Emanuel Knill, and Seth Lloyd.
\newblock Dynamical decoupling of open quantum systems.
\newblock {\em Phys. Rev. Lett.}, 82:2417--2421, Mar 1999.

\bibitem{ImageNetClassification}
Alex Krizhevsky, Ilya Sutskever, and Geoffrey~E. Hinton.
\newblock Imagenet classification with deep convolutional neural networks.
\newblock In {\em Proceedings of the 25th International Conference on Neural
  Information Processing Systems - Volume 1}, NIPS'12, pages 1097--1105, USA,
  2012. Curran Associates Inc.

\bibitem{facialRecognition}
Florian Schroff, Dmitry Kalenichenko, and James Philbin.
\newblock Facenet: A unified embedding for face recognition and clustering.
\newblock In {\em Proceedings of the IEEE conference on computer vision and
  pattern recognition}, pages 815--823, 2015.

\bibitem{ZeroResourceSpeech}
Ewan {Dunbar}, Xuan {Nga Cao}, Juan {Benjumea}, Julien {Karadayi}, Mathieu
  {Bernard}, Laurent {Besacier}, Xavier {Anguera}, and Emmanuel {Dupoux}.
\newblock {The Zero Resource Speech Challenge 2017}.
\newblock {\em arXiv e-prints}, page arXiv:1712.04313, Dec 2017.

\bibitem{transformer}
Ashish Vaswani, Noam Shazeer, Niki Parmar, Jakob Uszkoreit, Llion Jones,
  Aidan~N Gomez, {\L}ukasz Kaiser, and Illia Polosukhin.
\newblock Attention is all you need.
\newblock {\em Advances in neural information processing systems}, 30, 2017.

\bibitem{RLDota2}
Christopher Berner, Greg Brockman, Brooke Chan, Vicki Cheung, Przemys{\l}aw
  D{\k{e}}biak, Christy Dennison, David Farhi, Quirin Fischer, Shariq Hashme,
  Chris Hesse, et~al.
\newblock Dota 2 with large scale deep reinforcement learning.
\newblock {\em arXiv preprint arXiv:1912.06680}, 2019.

\bibitem{Starcraft2}
Oriol {Vinyals}, Timo {Ewalds}, Sergey {Bartunov}, Petko {Georgiev}, Alexander
  {Sasha Vezhnevets}, Michelle {Yeo}, Alireza {Makhzani}, Heinrich
  {K{\"u}ttler}, John {Agapiou}, Julian {Schrittwieser}, John {Quan}, Stephen
  {Gaffney}, Stig {Petersen}, Karen {Simonyan}, Tom {Schaul}, Hado {van
  Hasselt}, David {Silver}, Timothy {Lillicrap}, Kevin {Calderone}, Paul
  {Keet}, Anthony {Brunasso}, David {Lawrence}, Anders {Ekermo}, Jacob {Repp},
  and Rodney {Tsing}.
\newblock {StarCraft II: A New Challenge for Reinforcement Learning}.
\newblock {\em arXiv e-prints}, page arXiv:1708.04782, Aug 2017.

\bibitem{chess-like}
David Silver, Thomas Hubert, Julian Schrittwieser, Ioannis Antonoglou, Matthew
  Lai, Arthur Guez, Marc Lanctot, Laurent Sifre, Dharshan Kumaran, Thore
  Graepel, Timothy Lillicrap, Karen Simonyan, and Demis Hassabis.
\newblock A general reinforcement learning algorithm that masters chess, shogi,
  and go through self-play.
\newblock {\em Science}, 362(6419):1140--1144, 2018.

\bibitem{diffusionNet}
Robin Rombach, Andreas Blattmann, Dominik Lorenz, Patrick Esser, and Bj{\"o}rn
  Ommer.
\newblock High-resolution image synthesis with latent diffusion models.
\newblock In {\em Proceedings of the IEEE/CVF Conference on Computer Vision and
  Pattern Recognition}, pages 10684--10695, 2022.

\bibitem{neutronStar}
Yuki Fujimoto, Kenji Fukushima, and Koichi Murase.
\newblock Methodology study of machine learning for the neutron star equation
  of state.
\newblock {\em Phys. Rev. D}, 98:023019, Jul 2018.

\bibitem{phaseTransition}
Tomohiro Mano and Tomi Ohtsuki.
\newblock Phase diagrams of three-dimensional anderson and quantum percolation
  models using deep three-dimensional convolutional neural network.
\newblock {\em Journal of the Physical Society of Japan}, 86(11):113704, 2017.

\bibitem{tauDecay}
E.~Barberio, B.~Le, E.~Richter-Was, Z.~Was, J.~Zaremba, and D.~Zanzi.
\newblock Deep learning approach to the higgs boson $cp$ measurement in
  $\ensuremath{H}\ensuremath{\rightarrow}\ensuremath{\tau}\ensuremath{\tau}$
  decay and associated systematics.
\newblock {\em Phys. Rev. D}, 96:073002, Oct 2017.

\bibitem{quarticTomographyNN}
Talitha Weiss and Oriol Romero-Isart.
\newblock Quantum motional state tomography with non-quadratic potentials and
  neural networks.
\newblock {\em arXiv preprint arXiv:1906.08133}, 2019.

\bibitem{materialDesign}
Yue Liu, Tianlu Zhao, Wangwei Ju, and Siqi Shi.
\newblock Materials discovery and design using machine learning.
\newblock {\em Journal of Materiomics}, 3(3):159 -- 177, 2017.
\newblock High-throughput Experimental and Modeling Research toward Advanced
  Batteries.

\bibitem{GateOptimization}
Murphy~Yuezhen Niu, Sergio Boixo, Vadim~N. Smelyanskiy, and Hartmut Neven.
\newblock Universal quantum control through deep reinforcement learning.
\newblock {\em npj Quantum Information}, 5(1):33, 2019.

\bibitem{SpinControlDeepLearning}
Marin Bukov, Alexandre G.~R. Day, Dries Sels, Phillip Weinberg, Anatoli
  Polkovnikov, and Pankaj Mehta.
\newblock Reinforcement learning in different phases of quantum control.
\newblock {\em Phys. Rev. X}, 8:031086, Sep 2018.

\bibitem{alphaZeroSearchControl}
Dalgaard Mogens, Motzoi Felix, Jens~Jakob S{\o}rensen, and Sherson Jacob.
\newblock Global optimization of quantum dynamics with alphazero deep
  exploration.
\newblock {\em NPJ Quantum Information}, 6(1), 2020.

\bibitem{ErrorCorrectionDeepLearning}
Thomas F\"osel, Petru Tighineanu, Talitha Weiss, and Florian Marquardt.
\newblock Reinforcement learning with neural networks for quantum feedback.
\newblock {\em Phys. Rev. X}, 8:031084, Sep 2018.

\bibitem{deepReinforcementLearningReproducibility}
Peter Henderson, Riashat Islam, Philip Bachman, Joelle Pineau, Doina Precup,
  and David Meger.
\newblock Deep reinforcement learning that matters.
\newblock In {\em Proceedings of the AAAI conference on artificial
  intelligence}, volume~32, 2018.

\bibitem{deepReinforcementLearningDoesntWorkYet}
Alex Irpan.
\newblock Deep reinforcement learning doesn't work yet.
\newblock \url{https://www.alexirpan.com/2018/02/14/rl-hard.html}, 2018.

\bibitem{quantumRotorNature}
Benjamin~A Stickler, Klaus Hornberger, and MS~Kim.
\newblock Quantum rotations of nanoparticles.
\newblock {\em Nature Reviews Physics}, 3(8):589--597, 2021.

\bibitem{residualNet}
Kaiming He, Xiangyu Zhang, Shaoqing Ren, and Jian Sun.
\newblock Deep residual learning for image recognition.
\newblock In {\em Proceedings of the IEEE conference on computer vision and
  pattern recognition}, pages 770--778, 2016.

\bibitem{cutout}
Terrance DeVries and Graham~W Taylor.
\newblock Improved regularization of convolutional neural networks with cutout.
\newblock {\em arXiv preprint arXiv:1708.04552}, 2017.

\bibitem{AtariEnvironment}
Marc~G Bellemare, Yavar Naddaf, Joel Veness, and Michael Bowling.
\newblock The arcade learning environment: An evaluation platform for general
  agents.
\newblock {\em Journal of Artificial Intelligence Research}, 47:253--279, 2013.

\bibitem{CDQN}
Zhikang~T. Wang and Masahito Ueda.
\newblock Convergent and efficient deep {Q} network algorithm.
\newblock In {\em International Conference on Learning Representations}, 2022.

\bibitem{quantumCartpole}
Zhikang~T Wang, Yuto Ashida, and Masahito Ueda.
\newblock Deep reinforcement learning control of quantum cartpoles.
\newblock {\em Physical Review Letters}, 125(10):100401, 2020.

\bibitem{NielsenChuang}
Michael~A Nielsen and Isaac Chuang.
\newblock Quantum computation and quantum information, 2002.

\bibitem{QuantumMeasurement}
Howard~M. Wiseman and Gerard~J. Milburn.
\newblock {\em Quantum Measurement and Control}.
\newblock Cambridge University Press, 2009.

\bibitem{OpenQuantumSystemsAngelRivas}
Angel Rivas and Susana~F Huelga.
\newblock {\em Open quantum systems}.
\newblock Springer, 2012.

\bibitem{GaussianMeasurement}
G.~C. Ghirardi, A.~Rimini, and T.~Weber.
\newblock Unified dynamics for microscopic and macroscopic systems.
\newblock {\em Phys. Rev. D}, 34:470--491, Jul 1986.

\bibitem{GaussianMeasurement2}
Carlton~M. Caves and G.~J. Milburn.
\newblock Quantum-mechanical model for continuous position measurements.
\newblock {\em Phys. Rev. A}, 36:5543--5555, Dec 1987.

\bibitem{DIOSI1988419}
L.~Diósi.
\newblock Continuous quantum measurement and itô formalism.
\newblock {\em Physics Letters A}, 129(8):419 -- 423, 1988.

\bibitem{MeasurementItoFormalism}
Lajos Di{\'o}si.
\newblock Continuous quantum measurement and it{\^o} formalism.
\newblock {\em Physics Letters A}, 129(8-9):419--423, 1988.

\bibitem{COMmeasurement}
Yuto Ashida and Masahito Ueda.
\newblock Multiparticle quantum dynamics under real-time observation.
\newblock {\em Physical Review A}, 95(2):022124, 2017.

\bibitem{ReinforcementlearningAnintroduction}
Richard~S Sutton and Andrew~G Barto.
\newblock {\em Reinforcement learning: An introduction}.
\newblock MIT press, 2018.

\bibitem{DQN}
Volodymyr Mnih, Koray Kavukcuoglu, David Silver, Andrei~A. Rusu, Joel Veness,
  Marc~G. Bellemare, Alex Graves, Martin Riedmiller, Andreas~K. Fidjeland,
  Georg Ostrovski, Stig Petersen, Charles Beattie, Amir Sadik, Ioannis
  Antonoglou, Helen King, Dharshan Kumaran, Daan Wierstra, Shane Legg, and
  Demis Hassabis.
\newblock Human-level control through deep reinforcement learning.
\newblock {\em Nature}, 518:529 EP --, Feb 2015.

\bibitem{DeepLearningBook}
Ian Goodfellow, Yoshua Bengio, and Aaron Courville.
\newblock {\em Deep Learning}.
\newblock MIT Press, 2016.
\newblock \url{http://www.deeplearningbook.org}.

\bibitem{Adam}
Diederik Kingma and Jimmy Ba.
\newblock Adam: A method for stochastic optimization.
\newblock {\em International Conference on Learning Representations}, 12 2014.

\bibitem{RMSprop}
Tijmen Tieleman and Geoffrey~Everest Hinton.
\newblock Neural networks for machine learning: Lecture 6a -- overview of
  mini-batch gradient descent.
\newblock
  \url{http://www.cs.toronto.edu/~tijmen/csc321/slides/lecture_slides_lec6.pdf},
  2012.

\bibitem{prioritizedSampling}
Tom {Schaul}, John {Quan}, Ioannis {Antonoglou}, and David {Silver}.
\newblock {Prioritized Experience Replay}.
\newblock {\em arXiv e-prints}, page arXiv:1511.05952, Nov 2015.

\bibitem{DoubleDQN}
Hado {van Hasselt}, Arthur {Guez}, and David {Silver}.
\newblock {Deep Reinforcement Learning with Double {Q}-learning}.
\newblock {\em arXiv e-prints}, page arXiv:1509.06461, Sep 2015.

\bibitem{DuelDQN}
Ziyu {Wang}, Tom {Schaul}, Matteo {Hessel}, Hado {van Hasselt}, Marc {Lanctot},
  and Nando {de Freitas}.
\newblock {Dueling Network Architectures for Deep Reinforcement Learning}.
\newblock {\em arXiv e-prints}, page arXiv:1511.06581, Nov 2015.

\bibitem{RainbowDQN}
Matteo Hessel, Joseph Modayil, Hado Van~Hasselt, Tom Schaul, Georg Ostrovski,
  Will Dabney, Dan Horgan, Bilal Piot, Mohammad Azar, and David Silver.
\newblock Rainbow: Combining improvements in deep reinforcement learning.
\newblock In {\em Thirty-Second AAAI Conference on Artificial Intelligence},
  2018.

\bibitem{ResidualLearningInitial}
Leemon Baird.
\newblock Residual algorithms: Reinforcement learning with function
  approximation.
\newblock In {\em Machine Learning Proceedings 1995}, pages 30--37. Elsevier,
  1995.

\bibitem{TsitsiklisNonlinearDivergence}
John~N Tsitsiklis and Benjamin Van~Roy.
\newblock An analysis of temporal-difference learning with function
  approximation.
\newblock {\em IEEE transactions on automatic control}, 42(5):674--690, 1997.

\bibitem{DeadlyTriad}
Hado Van~Hasselt, Yotam Doron, Florian Strub, Matteo Hessel, Nicolas Sonnerat,
  and Joseph Modayil.
\newblock Deep reinforcement learning and the deadly triad.
\newblock {\em arXiv preprint arXiv:1812.02648}, 2018.

\bibitem{TCLoss}
Tobias Pohlen, Bilal Piot, Todd Hester, Mohammad~Gheshlaghi Azar, Dan Horgan,
  David Budden, Gabriel Barth-Maron, Hado~P. van Hasselt, John Quan, Mel
  Vecer{\'i}k, Matteo Hessel, R{\'e}mi Munos, and Olivier Pietquin.
\newblock Observe and look further: Achieving consistent performance on atari.
\newblock {\em CoRR}, abs/1805.11593, 2018.

\bibitem{ConstrainedTDUpdate}
Ishan Durugkar and Peter Stone.
\newblock {TD} learning with constrained gradients.
\newblock In {\em Proceedings of the Deep Reinforcement Learning Symposium,
  NIPS 2017}, Long Beach, CA, USA, December 2017.

\bibitem{GradientTDFirst}
Richard~S Sutton, Csaba Szepesv{\'a}ri, and Hamid~Reza Maei.
\newblock A convergent {O}({N}) temporal-difference algorithm for off-policy
  learning with linear function approximation.
\newblock In {\em NIPS}, 2008.

\bibitem{GradientTDLinear}
Richard~S Sutton, Hamid~Reza Maei, Doina Precup, Shalabh Bhatnagar, David
  Silver, Csaba Szepesv{\'a}ri, and Eric Wiewiora.
\newblock Fast gradient-descent methods for temporal-difference learning with
  linear function approximation.
\newblock In {\em Proceedings of the 26th Annual International Conference on
  Machine Learning}, pages 993--1000, 2009.

\bibitem{GradientTDGeneral}
Shalabh Bhatnagar, Doina Precup, David Silver, Richard~S Sutton, Hamid Maei,
  and Csaba Szepesv{\'a}ri.
\newblock Convergent temporal-difference learning with arbitrary smooth
  function approximation.
\newblock {\em Advances in neural information processing systems},
  22:1204--1212, 2009.

\bibitem{SBEEDResidualLearning}
Bo~Dai, Albert Shaw, Lihong Li, Lin Xiao, Niao He, Zhen Liu, Jianshu Chen, and
  Le~Song.
\newblock Sbeed: Convergent reinforcement learning with nonlinear function
  approximation.
\newblock In {\em International Conference on Machine Learning}, pages
  1125--1134. PMLR, 2018.

\bibitem{GradientTDbyKernelToCorrelateData}
Yihao Feng, Lihong Li, and Qiang Liu.
\newblock A kernel loss for solving the {B}ellman equation.
\newblock {\em arXiv preprint arXiv:1905.10506}, 2019.

\bibitem{GradientTDwithRegularization}
Sina Ghiassian, Andrew Patterson, Shivam Garg, Dhawal Gupta, Adam White, and
  Martha White.
\newblock Gradient temporal-difference learning with regularized corrections.
\newblock In {\em International Conference on Machine Learning}, pages
  3524--3534. PMLR, 2020.

\bibitem{GradientTreeBackup}
Ahmed Touati, Pierre-Luc Bacon, Doina Precup, and Pascal Vincent.
\newblock Convergent tree backup and retrace with function approximation.
\newblock In {\em International Conference on Machine Learning}, pages
  4955--4964. PMLR, 2018.

\bibitem{CharacterizingDivergenceInTD}
Joshua Achiam, Ethan Knight, and Pieter Abbeel.
\newblock Towards characterizing divergence in deep {Q}-learning.
\newblock {\em arXiv preprint arXiv:1903.08894}, 2019.

\bibitem{PolyakMomentum}
Boris~T Polyak.
\newblock Some methods of speeding up the convergence of iteration methods.
\newblock {\em Ussr computational mathematics and mathematical physics},
  4(5):1--17, 1964.

\bibitem{IsBellmanResidualABadProxy}
Matthieu Geist, Bilal Piot, and Olivier Pietquin.
\newblock Is the {B}ellman residual a bad proxy?
\newblock In {\em Advances in Neural Information Processing Systems}, pages
  3205--3214, 2017.

\bibitem{PCL}
Ofir Nachum, Mohammad Norouzi, Kelvin Xu, and Dale Schuurmans.
\newblock Bridging the gap between value and policy based reinforcement
  learning.
\newblock In {\em Advances in Neural Information Processing Systems}, pages
  2775--2785, 2017.

\bibitem{NAC}
Yang Gao, Huazhe Xu, Ji~Lin, Fisher Yu, Sergey Levine, and Trevor Darrell.
\newblock Reinforcement learning from imperfect demonstrations.
\newblock {\em arXiv preprint arXiv:1802.05313}, 2018.

\bibitem{RetraceOperator}
R{\'e}mi Munos, Tom Stepleton, Anna Harutyunyan, and Marc~G Bellemare.
\newblock Safe and efficient off-policy reinforcement learning.
\newblock {\em arXiv preprint arXiv:1606.02647}, 2016.

\bibitem{Agent57}
Adri{\`a}~Puigdom{\`e}nech Badia, Bilal Piot, Steven Kapturowski, Pablo
  Sprechmann, Alex Vitvitskyi, Zhaohan~Daniel Guo, and Charles Blundell.
\newblock Agent57: Outperforming the atari human benchmark.
\newblock In {\em International Conference on Machine Learning}, pages
  507--517. PMLR, 2020.

\bibitem{FirstFittedQIteration}
Damien Ernst, Pierre Geurts, and Louis Wehenkel.
\newblock Tree-based batch mode reinforcement learning.
\newblock {\em Journal of Machine Learning Research}, 6:503--556, 2005.

\bibitem{FittedValueIteration}
R{{\'e}}mi Munos and Csaba Szepesv{{\'a}}ri.
\newblock Finite-time bounds for fitted value iteration.
\newblock {\em Journal of Machine Learning Research}, 9(27):815--857, 2008.

\bibitem{DistributionalDQN}
Marc~G. {Bellemare}, Will {Dabney}, and R{\'e}mi {Munos}.
\newblock {A Distributional Perspective on Reinforcement Learning}.
\newblock {\em arXiv e-prints}, page arXiv:1707.06887, Jul 2017.

\bibitem{SoftQLearning}
Tuomas Haarnoja, Haoran Tang, Pieter Abbeel, and Sergey Levine.
\newblock Reinforcement learning with deep energy-based policies.
\newblock {\em arXiv preprint arXiv:1702.08165}, 2017.

\bibitem{WetChickenOriginal}
Volker Tresp.
\newblock The wet game of chicken.
\newblock {\em Siemens AG, CT IC 4, Technical Report}, 1994.

\bibitem{WetChickenPaper}
Alexander Hans and Steffen Udluft.
\newblock Efficient uncertainty propagation for reinforcement learning with
  limited data.
\newblock In {\em International Conference on Artificial Neural Networks},
  pages 70--79. Springer, 2009.

\bibitem{WetChickenReferee}
Alexander Hans and Steffen Udluft.
\newblock Ensemble usage for more reliable policy identification in
  reinforcement learning.
\newblock In {\em ESANN}, 2011.

\bibitem{NeuralFittedQIteration}
Martin Riedmiller.
\newblock Neural fitted {Q} iteration--first experiences with a data efficient
  neural reinforcement learning method.
\newblock In {\em European Conference on Machine Learning}, pages 317--328.
  Springer, 2005.

\bibitem{DuelingDQN}
Ziyu Wang, Tom Schaul, Matteo Hessel, Hado Hasselt, Marc Lanctot, and Nando
  Freitas.
\newblock Dueling network architectures for deep reinforcement learning.
\newblock In {\em International conference on machine learning}, pages
  1995--2003. PMLR, 2016.

\bibitem{PrioritizedExperienceReplay}
Tom Schaul, John Quan, Ioannis Antonoglou, and David Silver.
\newblock Prioritized experience replay.
\newblock {\em arXiv preprint arXiv:1511.05952}, 2015.

\bibitem{TemporalConsistencyLoss}
Tobias Pohlen, Bilal Piot, Todd Hester, Mohammad~Gheshlaghi Azar, Dan Horgan,
  David Budden, Gabriel Barth-Maron, Hado Van~Hasselt, John Quan, Mel
  Ve{\v{c}}er{\'\i}k, et~al.
\newblock Observe and look further: Achieving consistent performance on atari.
\newblock {\em arXiv preprint arXiv:1805.11593}, 2018.

\bibitem{DDPG}
Timothy~P Lillicrap, Jonathan~J Hunt, Alexander Pritzel, Nicolas Heess, Tom
  Erez, Yuval Tassa, David Silver, and Daan Wierstra.
\newblock Continuous control with deep reinforcement learning.
\newblock {\em arXiv preprint arXiv:1509.02971}, 2015.

\bibitem{quantumMechanicalRigidBody}
Rotation of a rigid body in quantum-mechanics.
\newblock {\em Nature}, 129(3265):780--780, May 1932.

\bibitem{MasterThesis}
Zhikang Wang.
\newblock Quantum control based on deep reinforcement learning, 2022.

\bibitem{anharmonicphi4}
Carl~M. Bender and Tai~Tsun Wu.
\newblock Anharmonic oscillator.
\newblock {\em Phys. Rev.}, 184:1231--1260, Aug 1969.

\bibitem{NumericalSimulationofStochasticDE}
Peter~E Kloeden and Eckhard Platen.
\newblock {\em Numerical solution of stochastic differential equations},
  volume~23.
\newblock Springer Science \& Business Media, 2013.

\bibitem{nanoparticleGroundStateCooling1}
Uro{\v{s}} Deli{\'c}, Manuel Reisenbauer, Kahan Dare, David Grass, Vladan
  Vuleti{\'c}, Nikolai Kiesel, and Markus Aspelmeyer.
\newblock Cooling of a levitated nanoparticle to the motional quantum ground
  state.
\newblock {\em Science}, 367(6480):892--895, 2020.

\bibitem{nanoparticleGroundStateCooling2}
Lorenzo Magrini, Philipp Rosenzweig, Constanze Bach, Andreas Deutschmann-Olek,
  Sebastian~G Hofer, Sungkun Hong, Nikolai Kiesel, Andreas Kugi, and Markus
  Aspelmeyer.
\newblock Real-time optimal quantum control of mechanical motion at room
  temperature.
\newblock {\em Nature}, 595(7867):373--377, 2021.

\bibitem{rigidBodyMeasurementCooling}
Jaehoon Bang, Troy Seberson, Peng Ju, Jonghoon Ahn, Zhujing Xu, Xingyu Gao,
  Francis Robicheaux, and Tongcang Li.
\newblock Five-dimensional cooling and nonlinear dynamics of an optically
  levitated nanodumbbell.
\newblock {\em Physical Review Research}, 2(4):043054, 2020.

\bibitem{rigidBodyMeasurementCooling2}
T~Seberson and F~Robicheaux.
\newblock Parametric feedback cooling of rigid body nanodumbbells in levitated
  optomechanics.
\newblock {\em Physical Review A}, 99(1):013821, 2019.

\bibitem{reimann2018ghz}
Ren{\'e} Reimann, Michael Doderer, Erik Hebestreit, Rozenn Diehl, Martin
  Frimmer, Dominik Windey, Felix Tebbenjohanns, and Lukas Novotny.
\newblock Ghz rotation of an optically trapped nanoparticle in vacuum.
\newblock {\em Physical review letters}, 121(3):033602, 2018.

\bibitem{dipoleTrap}
Rudolf Grimm, Matthias Weidem{\"u}ller, and Yurii~B Ovchinnikov.
\newblock Optical dipole traps for neutral atoms.
\newblock In {\em Advances in atomic, molecular, and optical physics},
  volume~42, pages 95--170. Elsevier, 2000.

\bibitem{ItoIntegralApproximations}
DF~Kuznetsov.
\newblock Strong approximation of iterated ito and stratonovich stochastic
  integrals.
\newblock {\em Theory of Probability and its Applications}, 65(1):142--143,
  2020.

\end{thebibliography}
	\printindex
\end{document}